\documentclass[a4paper,USenglish,cleveref, autoref, thm-restate]{lipics-v2021}
\newcommand{\longversion}[1]{#1}
\newcommand{\shortversion}[1]{}
\usepackage{todonotes}
\usepackage{comment}

\usepackage[utf8]{inputenc}          
\usepackage[T1]{fontenc}             
\usepackage{amsfonts}                
\usepackage{amssymb}                 
\usepackage{amsmath}                 

\longversion{\usepackage[appendix=inline]{apxproof}}
\shortversion{\usepackage{apxproof}}%
\shortversion{}
\usepackage{mathtools}
\usepackage{setspace}
\usepackage{graphicx}
\usepackage{hyperref}
\usepackage{xcolor}
\usepackage{xspace}

\usepackage{algorithm}
\usepackage[noend]{algpseudocode}

\newcommand{\DS}{\textsc{Dominating Set}\xspace}
\newcommand{\EDS}{\textsc{Extension Dominating Set}\xspace}
\newcommand{\ERD}{\longversion{\textsc{Extension Roman Domination}}\shortversion{\textsc{Ext RD}}\xspace}
\newcommand{\RD}{\textsc{Roman Domination}\xspace}
\newcommand{\HS}{\textsc{Hitting Set}\xspace}
\newcommand{\RHS}{\shortversion{\textsc{RHS}}\longversion{\textsc{Roman Hitting Set}}\xspace}
\newcommand{\ExtRHS}{\shortversion{\textsc{Ext RHS}}\longversion{\textsc{Extension Roman Hitting Set}}\xspace}
\newcommand{\RHF}{\shortversion{\textsc{RHF}}\longversion{\textsc{Roman Hitting Function}}\xspace}
\newcommand{\ExtRHF}{\shortversion{\textsc{Ext RHF}}\longversion{\textsc{Extension Roman Hitting Function}}\xspace}
\newcommand{\PTIME}{\textsf{P}\xspace}
\newcommand{\NP}{\textsf{NP}\xspace}
\newcommand{\FPT}{\textsf{FPT}\xspace}
\newcommand{\XP}{\textsf{XP}\xspace}
\newcommand{\W}[1]{\ensuremath{\textsf{W}[#1]}\xspace}
\newcommand{\paraNP}{\textsf{para-NP}\xspace}

\newcommand{\iffl}{if\longversion{ and only i}f\xspace}

\newcommand{\no}{\textsf{no}\xspace}
\newcommand{\yes}{\textsf{yes}\xspace}

\newcommand{\Oh}{\mathcal{O}}
\newcommand{\weightfun}{\omega}
\newcommand{\weightfunction}[1]{\weightfun\left(#1\right)}

\newenvironment{pf}{\begin{proof}}{\hfill
\end{proof}}
\newenvironment{pfclaim}{\begin{proof}[Proof of the claim]}{\hfill$\Diamond$\renewcommand{\qed}{}\end{proof}}

\newtheorem{brarule}{Branching Rule}

\newtheorem{redrule}{Reduction Rule}

\makeatletter
\newcommand{\crossout}[1]{%
  \begingroup
  \sbox\z@{#1}%
  \dimen\z@=\wd\z@
  \dimen\tw@=\ht\z@
  \dimen\z@=.99626\dimen\z@   
  \dimen\tw@=.99626\dimen\tw@ 
  \edef\co@wd{\strip@pt\dimen\z@}
  \edef\co@ht{\strip@pt\dimen\tw@}
  \leavevmode
  \rlap{\pdfliteral{q 1 J 0.4 w 0 0 m \co@wd\space \co@ht\space l S Q}}%
  \rlap{\pdfliteral{q 1 J 0.4 w 0 \co@ht\space m \co@wd\space 0 l S Q}}%
  #1%
  \endgroup
}
\makeatother

\title{Hitting the Romans}
\author{Henning Fernau}
{Universit\"at Trier, Fachbereich IV, Informatikwissenschaften, Germany \and \url{https://www.uni-trier.de/index.php?id=49861}}
{fernau@informatik.uni-trier.de}
{https://orcid.org/0000-0002-4444-3220}
{}

\author{Kevin Mann}
{Universit\"at Trier, Fachbereich IV, Informatikwissenschaften, Germany}
{mann@informatik.uni-trier.de}
{https://orcid.org/0000-0002-0880-2513} 
{}

\authorrunning{H. Fernau and K. Mann}

\Copyright{Henning Fernau and Kevin Mann} 


\category{} 

\relatedversion{} 

\longversion{\nolinenumbers} 

\EventEditors{}
\EventNoEds{1}
\EventLongTitle{}
\EventShortTitle{}
\EventAcronym{}
\EventYear{}
\EventDate{}
\EventLocation{}
\EventLogo{}
\SeriesVolume{}
\ArticleNo{}

\begin{document}

\maketitle

\begin{abstract}
    Roman domination is one of few examples where the related  extension problem is polynomial-time solvable even if the original decision problem is \NP-complete. This is interesting, as it allows to establish polynomial-delay enumeration algorithms for finding minimal Roman dominating functions, while it is open for more than four decades if all minimal dominating sets of a graph or if all hitting sets of a hypergraph can be enumerated with polynomial delay. To find the reason why this is the case, we combine the idea of hitting set with the idea of Roman domination. We  hence obtain and study two new problems, called \textsc{Roman Hitting Function} and \textsc{Roman Hitting Set}, both generalizing \textsc{Roman Domination}. This allows us to delineate the borderline of polynomial-delay enumerability. Here, we assume what we call the \emph{Hitting Set Transversal Thesis}, claiming that it is impossible to enumerate all minimal hitting sets of a hypergraph with polynomial delay. Our first focus is on the extension versions of these problems. While doing this, we find some conditions under which the \textsc{Extension Roman Hitting Function} problem is \NP-complete. We then use parameterized complexity to get a better understanding of why \textsc{Extension Roman Hitting Function} behaves in this way. Furthermore, we  analyze the parameterized and approximation complexity of the underlying optimization problems. We also discuss consequences for Roman variants of other problems like \textsc{Vertex Cover}. 
\end{abstract}
\section{Introduction}
\label{sec:intro}

Defense strategies in the Roman Empire have been studied throughout the centuries, mostly from a historical-military perspective, but then also from a more graph-theoretic viewpoint, commencing with an article by Arquilla\footnote{Arquilla is quite known for his analysis of cyberwar, see \cite{Arq2011}.} and Fredricksen~\cite{ArqFre95}.
The strategy that we are also discussing in this paper is nowadays known as Roman domination and was introduced in~\cite{Ste99} in line with the discussions of Arquilla and Fredricksen, formalizing a military strategic decision going back to Constantine the Great.

The concept of Roman domination is a well-studied variation of domination in graphs, 
certified by its own 45-page chapter in the relatively recent textbook~\cite{HayHedHen2020}, written by Chellali \emph{et al.}
Here, armies are placed in different regions. A region is secured if there is at least one army in this region or there are two armies in one neighbored region. Mathematically, we model the map as a graph, where regions correspond to vertices and connections between regions correspond to edges; moreover, the placement of armies is described by a function which maps each vertex to~0,~1 or~2. Such a function is called \emph{Roman dominating} if each vertex with value~0 has a neighbor with value~2. It is clearly desirable to minimize expenses by finding a Roman dominating function that uses the least number of armies; the corresponding decision problem is called \textsc{Roman Domination}.

Apart from quite a number of combinatorial (graph-theoretic) results that have been obtained for Roman domination, nicely surveyed in~\cite{HayHedHen2020}, the decision problem  \textsc{Roman Domination} has been studied from various aspects. Roughly speaking,  \textsc{Roman Domination} behaves quite similarly to the classical and well-known \textsc{Dominating Set} problem. More specifically, to mention some of these results:
\begin{itemize}
    \item \textsc{Roman Domination} is \NP-complete, even on special graph classes; see~\cite{Dre2000a}. A simpler proof is shown in \cite{Pagetal2002}, from which it is also clear (though not stated there) that subexponential-time algorithms can be excluded assuming ETH (Exponential Time Hypothesis).
    \item \textsc{Minimum Roman Domination} can be approximated up to a logarithmic factor but not any better, unless $\PTIME=\NP$, confer~\cite{Pagetal2002} and also~\cite{PadPal2020}.
    \item \textsc{Roman Domination} under standard parameterization (by an upper-bound~$k$ on the number of armies) is complete for \W{2}; see~\cite{Fer08}. However, the dual parameterization puts \textsc{Roman Domination} in \FPT; this is explicit in \cite{AbuBCF2016} and can be also obtained by combining \cite{BerFer2015} with \cite{BerFerSig2014}.
\end{itemize}

Even though \textsc{Roman Domination} and \textsc{Dominating Set} behave the same in terms of complexity in a variety of settings, this parallelism possibly surprisingly breaks down for two (mutually related) tasks:
\begin{itemize}
    \item Can we enumerate all minimal solutions of a given instance with polynomial delay?
    \item Can we decide, given a certain part of the solution, if there exists a minimal solution that extends the given pre-solution?
\end{itemize}
The first type of question is also known as an \emph{output-sensitive enumeration} problem. It is an open question since four decades if there exists an algorithm that enumerates all inclusion-wise minimal dominating sets with polynomial delay only, i.e., a user has to wait only polynomial time until the first minimal dominating set is shown, and also between any two outputs, and also between the last output solution and the termination of the algorithm.
The corresponding enumeration question for Roman dominating functions (for two natural notions of minimality) can be solved with polynomial delay, as proven in~\cite{AbuFerMan2022b}.
This result on Roman domination is based on another result giving a polynomial-time algorithm for the \emph{extension problem(s)} as described in the second item. Namely, the idea is to call an extension test before diving further into branching. This strategy is well-established in the area of enumeration algorithms, but few concrete examples are known; we only refer to the discussion in~\cite{Str2019}.
This makes \textsc{Extension Roman Domination} one of few examples where the extension problem is polynomial-time solvable, while the original problem is \NP-complete. 
For more details on extension problems, we refer to the survey~\cite{CasFGMS2022} that also suggests a general framework to describe such problems and provides further motivations and applications. 

The simple scientific question that we want to investigate in this paper is ``why'': \emph{What causes Roman domination to be feasible with respect to enumeration and extension?}
To find out why \textsc{Extension Roman Domination} behaves in this peculiar way and what can be seen as a difference to \textsc{Extension Dominating Set}, we want to generalize the concept of Roman domination and try to find the borderline of tractability. By this, we refer to the question if minimal hitting sets can be enumerated  with polynomial delay. This so-called \emph{Hitting Set Transversal Problem} is open for four decades. 
It is possibly time to put it up in the form of a conjecture: The \emph{Hitting Set Transversal Thesis} (or HSTT for short) would claim that there is no polynomial-delay enumeration algorithm for  minimal hitting sets.  This question is equivalent to several enumeration problems in logic, database theory and also to enumerating minimal dominating sets in graphs, see \cite{CreKPSV2019,EitGot95,GaiVer2017,KanLMN2014}.
 Our paper can be read as trying to understand for which problems polynomial-delay enumeration is not possible, assuming HSTT, and furthermore, to describe situations when polynomial-delay enumeration algorithms exist. 
Previous research on enumeration algorithms for minimal dominating sets often tried to look into special graph classes where polynomial-delay enumeration could be exhibited, or not, assuming HSTT. For instance, in~\cite{KanLMN2014} it is shown that minimal dominating sets can be enumerated with polynomial delay in split graphs,\footnote{Kanté \emph{et al.} use a direct argument for this polynomial delay enumeration result. It can be also shown that the extension problem to minimal dominating sets is solvable in polynomial time in split graphs.} while minimal connected dominating sets cannot, assuming HSTT, also see \cite{KanLMN2011}. In fact, it is shown in~\cite{KanLMN2014} that several graph enumeration problems are equivalent to enumerating minimal hitting sets when it comes to the question of polynomial delay. This approach can be seen as specializing a known (HSTT-hard) enumeration problem by studying special graph classes.
Our approach is different, as we come from a domination-type problem with a known polynomial-delay enumeration algorithm for general graphs and we try to stretch this result by generalization to understand when this enumeration task becomes infeasible, assuming HSTT.

It is well-known that \textsc{Hitting Set} can be viewed as a generalization of \textsc{Dominating Set} by modelling a graph by the closed-neighborhood hypergraph. Although \textsc{Hitting Set} and \textsc{Roman Domination} are both established concepts that generalize \textsc{Dominating Set}, it seems that there is no combination of both concepts published in the literature. Actually, trying to define such a combination comes with some problems.  
 If we want a \textsc{Hitting set} instance to represent a \textsc{Dominating Set} instance, the vertex set of the given graph is the vertex set of the hypergraph that we have to construct, also known as the ``universe'', and the set of all closed neighborhoods is the (hyper)edge set. Ignoring twins, this implies a bijection between the universe and the (hyper)edge set. But in general hypergraphs, the number of hyperedges and the number of elements in the universe does not necessarily have to be the same.  Therefore, we have to think about how to interpret the ``value one setting'' such that it is related to the definition of \textsc{Roman Domination} where exactly one army is put on a certain vertex. We suggest two ways of modelling this effect in hypergraphs:
\begin{enumerate}
    \item If a vertex has the value one under a Roman dominating function, then it hits only its own closed-neighborhood hyperedge.  
    \item If a vertex has the value one under a Roman dominating function, then its closed neighborhood needs \emph{not} be hit by some incident vertex with value~2.
\end{enumerate}
Following the first interpretation, we need a function, called \emph{correspondence}, which maps a vertex  to an incident hyperedge, such that this hyperedge is dominated if the vertex has the value~1. We will call the related decision problem \textsc{Roman Hitting Function}. The second interpretation implies that the value 1 is related to a hyperedge  rather than to its incident vertex. The  related decision problem will be called \textsc{Roman Hitting Set}.
We will use these ideas to define these two different hypergraph problems more formally in the next section. Moreover, we briefly discuss the notion of domination in hypergraphs that has been introduced before in the literature.

To wrap up our main results quickly:  \textsc{Roman Hitting Set} leads to a generalization of \textsc{Roman Domination} whose extension version is polynomial-time solvable and hence, minimal Roman hitting sets can be enumerated with polynomial delay. However, for \textsc{Roman Hitting Function}, it seems to be crucial whether or not the correspondence is surjective. If it is surjective, then we can prove that the extension version of \textsc{Roman Hitting Function} is again polynomial-time solvable and hence, under this condition minimal Roman hitting functions  can be enumerated with polynomial delay. If this surjectivity condition is not met, then the extension version of \textsc{Roman Hitting Function} is \NP-complete, and various attempts to parameterize this problem also fail and do not lead to \FPT-delay enumeration algorithms. This way, we obtain a clear borderline of feasibility for enumerating minimal Roman hitting functions.
We also show how to define notions like Roman vertex cover and transfer polynomial-delay enumeration results.

\section{Definitions and Notation}
\label{sec:defs}

Throughout this paper, we will freely use standard notions from complexity theory without defining them here. This includes notions from parameterized complexity, concerning \FPT and the further lower levels of the \textsf{W}-hierarchy up to \W{3}, as described in textbooks like~\cite{FluGro2006,DowFel2013}.

\subsection{General Notions, Graphs and Hypergraphs}
Let $\mathbb{N}$ denote the set of all nonnegative integers (including 0). For $n\in \mathbb{N}$, we will use the notation $[n]\coloneqq\{1,\ldots,n\}$. For a finite set~$A$ and some $n\in\mathbb{N}$ with $n\leq |A|$, the cardinality of $A$, $\binom{A}{n}$ denotes the set of all subsets of $A$ of cardinality~$n$, while $2^A$ denotes the power set of~$A$.
For two sets $A,B$, $B^A$ denotes the set of all mappings $f:A\rightarrow B$
. If $C\subseteq A$, then $f(C)=\{f(x)\mid x\in C\}\subseteq B$.
We denote by $\chi_{C}\in \{0,1\}^A$ the characteristic function, where $\chi_{C}(x)=1$ holds \iffl $x\in C$. For two functions $f,g\in \mathbb{N}^A$, we write $f\leq g$ \iffl $f(a)\leq g(a)$ holds for all $a\in A$. Further, we define the \emph{weight} of~$f$ by $\weightfunction{f}=\sum_{a\in A}f(a)$. 

We focus on hypergraphs $H=\left(X,\hat{S}=\left(s_i\right)_{i\in I}\right)$ with a finite universe~$X$, also called \emph{vertex} set, and a finite index set $I$, where for each $i\in I$, $s_i\subseteq X$ is a \emph{hyperedge}. With the set $S$ we denote the set  which includes all hyperedges of the sequence $\hat{S}$, i.e., $S=\{s_i\mid {i\in I}\}$. Note that that the same hyperedge may appear multiple times in the sequence $\hat S$. If this is forbidden, we speak of a \emph{simple} hypergraph.
 For all $x\in X$, define $\mathbf{S}(x)=\{s_i\in S\mid  x\in s_i\}$ as the set of hyperedges that is \emph{hit} by the vertex~$x$, hence defining a function $\mathbf{S}:X\to 2^S$. A set $D\subseteq X$ is a \emph{hitting set} \iffl $\mathbf{S}(D)=S$, where $\mathbf{S}(D)=\bigcup_{x\in D}\mathbf{S}(x)$. Similarly, define $\mathbf{I}:X\to 2^I$ by $\mathbf{I}(x)=\{i\in I\mid x\in s_i\}$, exentending to $A\subseteq X$ by $\mathbf{I}(A)=\bigcup_{x\in A}\mathbf{I}(x)$. 
We call \longversion{a function }$\tau: X\rightarrow I$ a \emph{correspondence} if $x\in s_{\tau(x)}$ for all $x\in X$ or if, in other words, 
$\tau(x)\in\mathbf{I}(x)$ for all $x\in X$.

We can consider a (simple undirected) graph $G=(V,E)$ as a hypergraph $G=(V,\hat{E})$, where each hyperedge contains exactly two elements. 
In this case, we call the  hyperedges just edges.
Talking about simple graphs, we can consider $E$ as the index set.
For each vertex $v\in V$, define its neighborhood as $N(v)= \{u\mid \{v,u\}\in E\}$ and the closed neighborhood as $N[v] = \{v\} \cup N(v)$. For vertex sets $U\subseteq V$, we use $N[U]= \bigcup_{v\in U} N[v]$ for the closed neighborhood of $U$.  A \emph{dominating set} of a graph $G=(V,E)$ is a set $D\subseteq V$ such that $N[D]=V$.

\subsection{Roman Dominating Functions}
Let $G=(V,{E})$ be a graph. 
A function $f\in\{0,1,2\}^V$ is a \emph{Roman dominating function} (rdf for short) \iffl, for each vertex $v\in V$ with $f(v)=0$, there exists a $u\in N(v)$ with $f(u)=2$. For $f,g\in \{0,1,2\}^V$,  define $f\leq_{PO} g$ \iffl $f(v)=0$ or $f(v)=g(v)$ for each $v\in V$. A Roman dominating function $f$ is minimal (or PO-minimal, respectively) if for each Roman dominating function $g$ with $g\leq f$ (or $g\leq_{PO} f$, respectively), $f=g$ holds. We will consider the following problems.

\centerline{\fbox{\begin{minipage}{.96\textwidth}
\textbf{Problem name: }\textsc{Roman Domination}, or \textsc{RD} for short\\
\textbf{Given: } A graph $G=(V,{E})$ and $k\in \mathbb{N}$\\
\textbf{Question: } Is there a rdf $f$ with $\weightfunction{f}\leq k$?
\end{minipage}
}}

\noindent
\centerline{\fbox{\begin{minipage}{.96\textwidth}
\textbf{Problem name: }\textsc{(PO-)Extension Roman Domination}, or  \textsc{(PO-)Ext RD} for short.\\
\textbf{Given: } A graph $G=(V,{E})$ and a function  $f: V\to \{0,1,2\}$\\
\textbf{Question: } Is there a minimal rdf $g$ for $G$ with $f\leq_{(PO)} g$?
\end{minipage}
}}
Somewhat surprisingly, the extension problems in the second box were proven to be polynomial-time solvable in \cite{AbuFerMan2022b}. This implies that (PO-)minimal rdf can be enumerated with polynomial delay. In order to explore why these results were possible, we \longversion{are going to }generalize these notions and problems for hypergraphs in three ways.

The first one is probably the most natural one, formed in analogy to the notion of a dominating set in a (simple) hypergraph; see \cite{Ach2007}. Let $H=(V,E)$ be a (simple) hypergraph, i.e., $E\subseteq 2^V$. Then, $f:V\to\{0,1,2\}$ is a Roman dominating function if, for all $v\in V$ with $f(v)=0$, there is a vertex $u\in V$ with $f(v)=2$ that is a neighbor of $v$, i.e., it shares an edge with $v$, which means, more formally, that there exists some $e\in E$ with $\{u,v\}\subseteq e$. The problems can then be defined as for graphs. We will call them \textsc{Hypergraph Roman Domination}, or  \textsc{HRD} for short, and 
\textsc{(PO-)Extension Hypergraph Roman Domination}, or  \textsc{(PO-)Ext HRD} for short.

However, as we show next, we can transfer all interesting properties of Roman domination from the graph case to the hypergraph setting, basically by using the same reduction. More specifically, we can show:

\begin{proposition}
\textsc{HRD} is \NP-complete.
\end{proposition}

\begin{proposition}
\textsc{(PO-)Ext HRD} can be solved in polynomial time.
\end{proposition}

\begin{proposition}
One can enumerate all (PO-)minimal  Roman dominating function of a hypergraph with polynomial delay.
\end{proposition}

Namely, if $G=(V,E)$ is a simple hypergraph, we can construct a graph $G'=(V,E')$ by setting
$\{x,y\}\in E'$ \iffl there is a hyperedge $e\in E$ with $\{x,y\}\subseteq e$. Then, $f:V\to\{0,1,2\}$ is a Roman dominating function of the hypergraph~$G$ \iffl $f$ is a Roman dominating function of the graph~$G'$.
Conversely, we just have to interpret a given graph as a simple hypergraph.

Therefore, we propose two further generalizations of Roman domination towards hypergraphs that allow us to study why Roman domination shows such a peculiar behavior when it comes to its extension version, as well as concerning enumerating minimal Roman dominating functions. These definitions can be found in the next subsections.

The basis of the polynomial-delay enumeration algorithm is a combinatorial characterization of minimal Roman dominating functions.
To be able to formulate the mentioned combinatorial characterization of minimal rdf, we need a further notion.
For $D\subseteq V$ and $v\in D$, define the \emph{private neighborhood} of $v\in V$ with respect to~$D$ as $P_{G,D}\left( v\right)\coloneqq N_G\left[ v\right] \setminus N_G\left[D\setminus \lbrace v\rbrace\right] $. The vertices in $P_{G,D}\left( v\right)$ are called private neighbors of $v$.

\begin{theorem} \cite{AbuFerMan2022b} \label{t_property_min_rdf}
Let $G=\left(V,E\right)$ be a graph and $f: \: V \to \lbrace 0,1,2\rbrace$ be a function. 
Abbreviate
$G'\coloneqq G\left[ f^{-1}(0)\cup f^{-1}(2)\right]$. Then, $f$ is a minimal rdf \iffl \shortversion{we find}\longversion{the following constraints hold}:
\begin{enumerate}
\item$N_G\left[f^{-1}(2)\right]\cap f^{-1}(1)=\emptyset$,\label{con_1_2}
\item $\forall v\in V_2\left(f\right) :\: P_{G',V_2\left(f\right)}\left( v \right) \nsubseteq \lbrace v\rbrace$, also called \emph{privacy condition}, and \label{con_private}
\item $f^{-1}(2)$ is a minimal dominating set of $G'$.\label{con_min_dom}
\end{enumerate}
\end{theorem}
\begin{theorem} \cite{AbuFerMan2022b} \label{t_property_po_min_rdf}
Let $G=\left(V,E\right)$ be a graph and $f: \: V \to \lbrace 0,1,2\rbrace$ be a function. 
Abbreviate
$G'\coloneqq G\left[ f^{-1}(0)\cup f^{-1}(2)\right]$. Then, $f$ is a PO-minimal rdf \iffl \shortversion{we find}\longversion{the following constraints hold}:
\begin{enumerate}
\item$N_G\left[f^{-1}(2)\right]\cap f^{-1}(1)=\emptyset$,\label{con_1_2_po}
\item $f^{-1}(2)$ is a minimal dominating set of $G'$.\label{con_min_dom_po}
\end{enumerate}
\end{theorem}

\subsection{Roman Hitting Functions}\label{sec_def_rhf}
Let $H=(X,\hat{S}=(s_i)_{i\in I})$ be a hypergraph and $\tau: X\rightarrow I$ be a correspondence.
 We call a function $f\in \{0,1,2\}^X$ a \emph{Roman hitting function} (rhf for short) if, for each $i\in I$, there exists an $x\in s_i$ with $f(x)=2$ or if there exists an $x\in X$ with $\tau(x)=i$ and $f(x)=1$. In these scenarios, we say that $x$ hits $i$ or $s_i$. For a function $f\in \{0,1,2\}^X$, we define the partition $P(f)=\lbrace 
 f^{-1}(0),f^{-1}(1),f^{-1}(2)
 \rbrace$.
We now define two decision problems related to rhf.

\centerline{\fbox{\begin{minipage}{.96\textwidth}
\textbf{Problem name: }\textsc{Roman Hitting Function}, or RHF for short\\
\textbf{Given: } A finite set $X$, a hyperedge sequence $\hat{S}=\left(s_i \right)_{i\in I}$, forming the hypergraph $(X,\hat S)$,
a correspondence $\tau: X\to I$, and $k\in \mathbb{N}$\\
\textbf{Question: } Is there a rhf $f$ with $\weightfunction{f}\leq k$?\end{minipage}
}}

\noindent
\centerline{\fbox{\begin{minipage}{.96\textwidth}
\textbf{Problem name: }\textsc{Extension Roman Hitting Function}, or \textsc{Ext RHF} for short\\
\textbf{Given: } A finite set $X$, a hyperedge sequence $\hat{S}=\left(s_i \right)_{i\in I}$, forming the hypergraph $(X,\hat S)$, a correspondence $\tau: X\to I$, and $f: X\to\{0,1,2\}$\\
\textbf{Question: } Is there a minimal rhf $g$ with $f\leq g$?\end{minipage}
}}

To understand in which way this setting generalizes \textsc{RD}, recall that there are alternative ways to specify a graph as a hypergraph. More precisely, the \emph{closed-neighborhood hypergraph} $G_{nb}$ associated to a graph $G=(V,E)$ can be described as $G_{nb}=(V,(N[v])_{v\in V})$. Clearly, $D\subseteq V$ is a dominating set \iffl $D$ is a hitting set of $G_{nb}$. As $v\in N[v]$, the identity can be viewed as a correspondence.
In this interpretation, $f:V\to \{0,1,2\}$ is a Roman domination function of~$G$ \iffl it is a Roman hitting function of~$G_{nb}$.

\subsection{Roman Hitting Sets}

Let $(X,\hat S=(s_i)_{i\in I})$ be a hypergraph. 
 We call $R=\left( R_1,R_2\right)\in 2^I \times 2^X $ a \emph{Roman hitting set}, or rhs for short, \iffl, for all $i\in I$, $s_i\cap R_2\neq\emptyset$ or $i \in R_1$. If $x\in s_i\cap R_2$ we say $x$ \emph{hits} $i$ or $s_i$. In the case that $i\in R_1$ we say $i$ \emph{hits itself}. For two tuples $(P_1,P_2),(R_1,R_2)\in 2^I \times 2^X$ we say $(P_1,P_2)\leq(R_1,R_2)$ if $P_1 \subseteq R_1$ and $P_2\subseteq R_2$. We call a rhs $P$ \emph{minimal} if $R\leq P$ implies $R=P$ for each rhs $R$. Adapting earlier notations, we also write $\weightfun(R_1,R_2)\coloneqq \vert R_1\vert + 2\cdot\vert R_2\vert$, calling it the \emph{weight} of $(R_1,R_2) \leq (I, X)$.

\begin{example}
Consider some $f:V\to \{0,1,2\}$. Then, $f$ is a rhf \iffl $(\tau(f^{-1}(1)),f^{-1}(2))$ is a rhs, 
but for $X=\{a,b,c,d\}$ and $\hat{S}\coloneqq(s_1\coloneqq\{a,b\},s_2\coloneqq\{a\},s_3\coloneqq\{b\},s_4\coloneqq\{a,c\},s_5\coloneqq\{c,d\})$, $I\coloneqq[5]$, $R_1=[3]$ and $R_2=\{c\}$ forms a valid rhs that does not correspond to any rhf for the hypergraph $(X,\hat S)$.
\end{example}

\begin{lemma}
    Let $H=(V,(s_i)_{i\in I})$ be a hypergraph and $\tau: V\to I$ be a correspondence. $\tau$ is surjective \iffl for each rhs $(R_1, R_2)$ of $h$ there exists a rhf $f:V \to \{0, 1, 2\}$ with $f^{-1}(2)=R_2$. 
\end{lemma}

\begin{pf}
    Let $H=(V,(s_i)_{i\in I})$ be a hypergraph and let $\tau: V\to I$ be a correspondence. Furthermore, let $\tau$ be surjective and let $(R_1, R_2)$ be a rhs. Then  $f:V\to \{0,1,2\}$ is a rhf, where $f$ is defined as $f(v)=2$ if $v\in R_2$ and $f(v)=1$ if $v\in\phi(\tau^{-1}(i))$ for some $i\in R_1$, where $\phi$ is Zermelo's choice function, and $f(v)=0$, otherwise.

    Assume $\tau$ is not surjective. Then there is an $i\in I$ with $\tau^{-1}(i)= \emptyset$. $(I,\emptyset)$ is a rhs, but there is no rhf with $f^{-1}(2) = \emptyset$, as for all $j\in I$ there has to be a $v\in f{-1}(1)\cap \tau^{-1}(j)$. This is not the case for $j=i$.
\end{pf}

In particular, if we consider functions $f:V\to \{0,1,2\}$ associated to a graph $G=(V,E)$, then $f$ is a rdf \iffl, in the related hypergraph $G_{nb}$, $f$ is a rhf \iffl $(f^{-1}(1),f^{-1}(2))$ is a rhs in~$G_{nb}$, as we view the identity as our correspondance. In that case,
$\weightfunction{f}=\weightfun(f^{-1}(1),f^{-1}(2))$.
As a further example, consider $X=\{a,b,c,d\}$ and $\hat{S}\coloneqq(s_1\coloneqq\{a,b\},s_2\coloneqq\{b,c\},s_3\coloneqq\{b,e\},s_4\coloneqq\{b,c,d\},s_5\coloneqq\{d,e\})$, $I\coloneqq[5]$.
If the correspondence~$\tau$ that we are considering satisfies $\tau(d)=s_4$ and $\tau(e)=s_3$, then $\tau$ cannot be surjective, as $s_5\notin \tau(X)$.
A rhs with smallest weight~3 is $(\{s_5\},\{b\})$. However, any rhf~$f$ for $(X,\hat S)$ that respects~$\tau$ must have $f(d)=2$ or $f(e)=2$. In either case, for the remaining hypergraph, at least weight~2 is needed, so that $\weightfun(f)\geq 4$ is enforced. 

The concept of rhs gives rise to two decision problems:

\centerline{\fbox{\begin{minipage}{.96\textwidth}
\textbf{Problem name: }\textsc{Roman Hitting Set}, or RHS for short\\
\textbf{Given: } A hypergraph $H=(X,\hat{S}=\left(s_i \right)_{i\in I})$ and $k\in \mathbb{N}$\\
\textbf{Question: } Is there a rhs $(R_1,R_2)$ of~$H$ with  $\weightfun(R_1,R_2)
\leq k$?\end{minipage}
}}

\noindent
\centerline{\fbox{\begin{minipage}{.96\textwidth}
\textbf{Problem name: }\textsc{Extension Roman Hitting Set}, or \textsc{Ext RHS} for short\\
\textbf{Given: } A hypergraph $H=(X,\hat{S}=\left(s_i \right)_{i\in I})$ and some  $U=(U_1,U_2) \leq (I, X)$\\
\textbf{Question: } Is there a minimal rhs $R = ( {R_1}, R_2)$ of~$H$ with $U\leq R$?\end{minipage}
}}

\longversion
{More precisely, if $G=(V,E)$ is a graph, then take $X=V$, $\hat{S}_G=(N [v])_{v\in V}$, defining the closed-neighborhood hypergraph $G_{nb}$. In this case, a function $f$ is a rdf \iffl $(f^{-1}(1),f^{-1}(2))$ is a rhs. Hence, $(G,k)$ is a \yes-instance of \textsc{Roman Domination} \iffl $(X,\hat S_G,k)$ is a \yes-instance of \textsc{RHF}. For the if-part, we have to  show that if there exists a rhs $(R_1,R_2)$ with $ \vert R_1\vert + 2\cdot\vert R_2\vert \leq k$ then there exists a rhs $(P_1,P_2)$ with $ \vert P_1\vert + 2\cdot\vert P_2\vert \leq k$ and $P_1\cap P_2=\emptyset$. In this case, we could construct a rdf $f$ with $f^{-1}(0)\coloneqq V\setminus( P_1 \cup P_2)$, $f^{-1}(1)\coloneqq P_1$ and $f^{-1}(0)\coloneqq P_2$. So if there would exist a rhs $(R_1,R_2)$ with $ \vert R_1\vert + 2\cdot\vert R_2\vert \leq k$ and $R_1\cap R_2\neq \emptyset$, then $(R_1\setminus R_2, R_2)$ is also a rhs, as each $v\in R_1\cap R_2$ is dominated by itself.}  

Slightly abusing notions, we will call  \RHF and \RHS \emph{optimization problems}, as there is an underlying minimization problem in each case, which is not the case for the other problems, summarized as \emph{extension problems}. 
When viewed as parameterized problems, their \emph{standard parameter} is the upper bound~$k$ given on the weight of the rhf or rhs that we are looking for.
Similarly, we address \ExtRHF and \ExtRHS as \emph{extension problems}. Their \emph{standard parameter} is the weight of the given function $f:X\to\{0,1,2\}$ or of the given tuple $(U_1,U_2)\leq (I, X)$.

\subsection{Organization of the Paper and Main Results}
In \autoref{sec:optimization problems}, we will prove that our optimization problems are \NP-complete and, more interestingly, if viewed as standard-parameterized problems, they become \W{2}-complete. Then, we turn our attention to the extension problems for both variations. Recall that the algorithmic results in the case of Roman domination were based on some basic combinatorial insights that we actually repeat at the end of this section for the reader's convenience. Following this logic, first we show in \autoref{sec:rhs} a combinatorial characterization of minimal rhs that we can make use of in  \autoref{sec:Ext-RHS} where we prove that  \ExtRHS can be solved in polynomial time.
After having derived a collection of interesting combinatorial properties of minimal rdf in \autoref{sec:rhf}, we can exhibit a corresponding result for \ExtRHF with surjective correspondences in \autoref{sec:tau_sur}.
In \autoref{sec:bounded-extension-RD}, we return to Roman domination and consider a variant of the extension problem where we give both lower and upper bound conditions to the minimal Roman dominating function that we are looking for. In contrast to the original problem (that only provides an lower bound), this two-sided extension problem turns out to be \NP-complete, as we show.
Also a natural parameterization of this problem by the weight of the lower bound function (motivated by the principle of distance to triviality~\cite{GuoHufNie2004}) does not really help, it rather yields a parameterized problem that is complete in the class \W{3}, as we prove.
These results are then employed in \autoref{sec:Ext RHF complexity I} to prove that in general, the extension problem for rhf is complete in the class \W{3} when parameterized by the weight of the given function. This inspires our studies of \autoref{sec:Ext RHF complexity II}, where we investigate alternative parameterizations for this extension problem, sometimes proving hardness results, but sometimes also arriving at \FPT-results. 
In \autoref{sec:enumerating_rhs}, we show how our results from \autoref{sec:rhs} and from \autoref{sec:Ext-RHS} can be used to develop an algorithm that enumerates all minimal rhs with polynomial space and with polynomial delay. We also analyze this algorithm from the perspective of input-sensitive enumeration and show that its running time cannot be substantially improved in general by providing a rather simple lower-bound example. In \autoref{sec:exact_branch}, we explain a quite simple branching algorithm for computing minimum rhs. 
Motivated by our insights on rhs, we show in \autoref{sec:rvc+rec} how problems like \textsc{Roman Vertex Cover} and \textsc{Roman Edge Cover} can be defined and how our previous results also imply algorithmic results for these problems. In particular, we show an (optimal) \FPT-enumeration algorithm for minimal Roman vertex covers. We conclude with a discussion of our results and further lines of research.
Let us mention already here that with our approach, feasibility in the sense of enumerability of minimal solutions with polynomial delay goes hand-in-hand with feasibility in the sense of polynomial-time solvability of the corresponding extension problem. This need not be the case, even in the area of domination-type problems. As shown in \cite{CasFKMS2019a}, the extension problem for edge dominating sets is \NP-complete, while minimal edge dominating sets can be listed with polynomial delay, as shown in~\cite{KanLMNU2015}. 

\section{Optimization Problems}
\label{sec:optimization problems}

This section will discuss the (parameterized) complexity of the optimization problems. The probably most natural parameterization for both problems is the upper bound~$k$ on the weight of the rhf or rhs. For our results, we will use the $\W{2}$-completeness of \textsc{Roman Domination} (with standard parameterization) shown in~\cite{Fer08}.

\begin{theorem}\label{thm:RHF/RHS-hardness}
\textsc{RHF} and \textsc{RHS} are \NP-complete. From the parameterized perspective, $\textrm{standard-}\textsc{RHF}$ and $\textrm{standard-}\textsc{RHS}$ are \W{2}-complete.
\end{theorem}

\begin{pf}
As discussed in \autoref{sec:defs}, rdf can be interpreted as rhf and hence as rhs, so that all hardness claims translate from \textsc{Roman Domination}. 

We only prove the membership of (standard-)\RHF within $\W{2}$. The reduction for (standard-)\RHS is easier.  

Let $H=(X,\hat{S}=(s_i)_{i\in I})$ be a 
hypergraph, $\tau:X\rightarrow I$ be a correspondence (hence, $x\in \tau(x)$ for each $x\in X$) and $k\in\mathbb{N}$. Define $G=(V,E)$ with
\begin{eqnarray*}
        V &=&\, \{a,b,c\} \cup \overbrace{\{v_x\mid x\in X \}}^{=:V(X)} \cup\, W\cup U, \quad\text{where}\\
        W &=& \{w_{i}\mid  i\in I\}\quad\text{and} \\
        U&=& \{u_{i} \mid  i \in I, \tau^{-1}(i)=\emptyset \},\\
        E &=& \{\,\{a,b\},\{a,c\}\,\} \cup \{\,\{a,v_x\}\mid x\in X\}\cup \{\,\{v_x,w_{i}\}\mid i\in I, x \in s_i\}\\
        &\cup&\{\,\{v_x,u_{i}\}\mid i\in I, x \in s_i, u_{i} \in U\} \cup \{\,\{v_x,v_y\}\mid x,y\in X\}.
\end{eqnarray*}
\begin{claim}
$H=(X,\hat{S})$ with $\tau$ has a rhf of weight at most $k$ \iffl $G$ has a rdf with weight at most $k + 2$.
\end{claim}

\begin{pfclaim}
Let $f\in \{0,1,2\}^X$ be minimal rhf on $H$ with $\tau$. Define $g: V\rightarrow \{ 0,1,2\}$,
$$ v\mapsto \begin{cases} 2,& v\in \{a\}\cup \{v_x \mid  x\in X:\:f(x)=2\} \\ 1,& v\in\{w_{i}\in W \mid  \exists x\in \tau^{-1}(i):\:f(x)=1\} \\ 0,& v\in \{b,c\}\cup \{w_{i}\in W \mid  \forall x\in \tau^{-1}(i):\:f(x) \neq 1\} \cup U\end{cases}.$$

 As $g(a)=2$, $N[a]=\{a,b,c\}\cup V(X)$ is dominated. Let $u_{i}\in U$ for some $ i \in I$. Since $\tau^{-1}(i)=\emptyset$ and as $f$ is a rhf, there exists an $x\in X\cap s_i$ with $f(x)=2$. This implies the existence of $v_x\in V(X) \cap N(u_{i})$ with $g(v_x)=2$. Therefore, each vertex in $U$ is dominated.  Let $w_{i} \in W$ for some $ i \in I$ with $g(w_{i})=0$. Thus, $f(x)\neq 1$ for each $x\in \tau^{-1}(i)$. As $f$ is a rhf, there exists an $x\in X\cap s_i$ with $f(x)=2$. Hence, $w_{s_i}$ is dominated by some $v_x$ and $g$ is a rdf. 
\begin{equation*}
    \begin{split}
        \sum_{v\in V}g(v)&= 2 + \sum_{ x\in X\text{ with }f(x)=2}g(v_x) + \sum_{i\in I, (\exists x\in \tau^{-1}(i):\:f(x)=1)} g(w_{i})\\
        &\leq 2 + \sum_{x\in f^{-1}(2)}f(x) + \sum_{x\in f^{-1}(1)} f(x) \leq k+2\,.
    \end{split}
\end{equation*}
For the other direction, assume $f\in \{0,1,2\}^V$ is a rdf on $G$. 
 \shortversion{W.l.o.g.}\longversion{Without loss of generality}, if there exist $v,w\in V$ with $N[w]\subseteq N[v]$ and $f(v) + f(w) \geq 2$, then $f(v)=2$ and $f(w)=0$. We can assume this, since each other variation would dominate at most the same vertices. Consequently, we have $\{v\in V\mid f(v)=2\} \subseteq \{a\} \cup \{v_x\mid x\in X \}$.

Let $u,v,w\in V$ with $N[u]\cup N[w]\subseteq N[v]$ and $f(u) + f(v) + f(w) \geq 2$. Assume  $f(u)=f(w)=1$. Define $h\in \{ 0,1,2\}^V$ with $$v\mapsto \begin{cases}f(x),& x\notin\{u,v,w\}\\ 0, & x\in\{u,w\}\\ 2,& x=v\end{cases}.$$ $N[u]\cup N[w]$ is dominated (with respect to~$h$) by~$v$. The other vertices are dominated as with~$f$ and $\sum_{v\in V}h(v) \leq \sum_{v\in V}f(v)$. Thus, w.l.o.g., $f(a)=2$ and $f(x)=0$ for $x\in \{b,c\} \cup \{w_{i},u_{i}\mid i\in I: \: \tau^{-1}(i) = \emptyset \}$. Furthermore, $W \cap f^{-1}(2)$ is empty, since we could swap the value of a vertex under $f$ in this set with any of his neighbors. For each $w_{i}\in W$ with $f(w_{i})=1$, there has to exist an $x_{i}\in \tau^{-1}(i)$.

Define $g\in \{0,1,2\}^X$ with $g(x)=2$ for if $f(v_x)=2$, $g(x_{i})=1$ defined for $x_{i}\in \tau^{-1}(i)$ as above, i.e., there is some $w_{i}\in W$ with $f(w_{i})=1$, and $f(x)=0$, otherwise. Consider $i\in I$ with no $x\in X$ such that $g(x)=1$ and $\tau(x)=i$ hold. Thus, $f(w_{i})=0$ and there exists a $v_x\in V(X) \cap N(w_{i})$ with $f(v_x)=2$. Hence, there exists an $x\in s_i$ with $f(x)=2$. Therefore, $g$ is a rhf. Furthermore, 
$$
        \sum_{x\in X}g(x)= \sum_{x\in X:\: f(v_x)=2} g(x) + \sum_{x_{s_i}\in X:\: f(w_{i})=1} g(x_{i}) =    \sum_{v\in V}f(v)-2\leq k.
$$    
\end{pfclaim}
This concludes the description of the working of our polynomial-time reduction and serves to prove both membership in $\NP$ and membership in \W{2}.

To prove \W{2}-membership for \textrm
{standard}-\RHS, we do not need the vertices in  $\{u_{i}\mid i\in I:\:\tau^{-1}(i)=\emptyset\}$. Furthermore, the argument is easier because the elements of $S$ could be part of the solution. We omit further details here. 
\end{pf}

\begin{remark}
     \RHF and \RHS can be reduced at each other easily.

     \smallskip\noindent
     \underline{From \RHF to \RHS.}
     Let $H=(X,(s_i)_{i\in I})$ a hypergraph with correspondence $\tau$ and $k\in \mathbb{N}$. Define $H'=(X,(s_i)_{i\in I\cup I'})$ with $I'=\{i' \mid i\in I, \tau^{-1}(i)=\emptyset\}$ and  $s_{i'}=s_i$ for each $i'\in I'$. We will now show that there is a rhf on $H$ with $\tau$ of weight $k$ \iffl there is a minimal rhs on $H'$ of weight $k$. If there is a rhf $f$ on $H$ of weight at most $k$, then $\left(\tau\left( f^{-1}(1) \right), f^{-1}(2)\right)$ is a rhs on $H'$ with weight at most $k$. Let $(R_1,R_2)$ be a rhs on $H'$ of weight $k$. If there is a $i'\in I'\cap R_1$, then $s_i\cap R_2$ is not empty or $i\in R_1$. In the first case $(R_1\setminus \{i'\}, R_2)$ would also be a rhs. Otherwise, let $x\in s_i$ be an arbitrary element, then $(R_1\setminus \{i,i'\}, R_2\cup \{ x\})$ is also a rhs with the same weight. Therefore, we can assume $I'\cap R_1=\emptyset$. Analogously, $\{i\in I \mid \tau^{-1}(i)=\emptyset\}\cap R_1=\emptyset$ holds. For each $i\in R_1$, $x_i\in \tau^{-1}(i)$ be an arbitrary element. Then $f\in \{0,1,2\}^X$ with $f^{-1}(1)=\{x_i\mid i\in R_1\}$ and $f^{-1}(2)=R_2$ is a rhf with weight at most $k$.

     \smallskip\noindent
     \underline{From \RHS to \RHF.}
     Let $H=(X,(s_i)_{i\in I})$ be a hypergraph and $k\in \mathbb{N}$ with $k<\vert I\vert$, as for $\vert I\vert \leq k$, we have a trivial \yes-instance. Therefore, $(I,\emptyset)$ is not a solution for the \RHS instance. Define the hypergraph $H' \coloneqq (X\cup I,(s'_i)_{i\in I'})$ with $I'\coloneqq I\cup \{a\}$ (for an $a\notin I\cup X$) and $s'_i\coloneqq s_i\cup \{i\}$ for $i\in I$ and $s'_a=X$. The correspondence $\tau$ of $H$ is given by 
     $$\tau: X\cup I \to I', y\mapsto\begin{cases}
         a, & y\in X\\
         y, & y\in I
     \end{cases}.$$
     Let $f$ be a rhf on $H'$ with $\weightfunction{f}\leq k$. As each $i\in I$ is only in one edge, we can assume $f(i)\neq 2$ (otherwise, $f-\chi_{\{i\}}$ is also a rhf with smaller weight). Hence, to hit each edge of $(s'_i)_{i\in I}$ with weight at most $k<\vert I\vert$, $f^{-1}(2)\cap X$ is not empty. Thus, $s'_a$ hit. If there would be an $x\in f^{-1}(1)\cap X$, then $f-\chi_{\{ x \}}$ is also a rhf with smaller weight. Therefore, we have $f^{-1}(1)\subseteq I$ and $f^{-1}(2)\subseteq X$. Furthermore, $(f^{-1}(1),f^{-1}(2))$ is a rhs with weight at most $k$. Now let $(R_1,R_2)$ be a rhs on $H$ with weight $k$, then $f\in \{0,1,2\}^{X\cup I}$ with $f^{-1}(1)=R_1$ and $f^{-1}(2)=R_2$ is a rhf with $\weightfunction{f} = k$.
\end{remark}

The fact that \RD is \NP-complete even on split graphs was mentioned repeatedly in the literature, for instance, in \cite{Cocetal2004,Lieetal2008}, but to the best of our knowledge, no proof of this fact has been published. We will provide a strengthened assertion in the following. Recall that two vertices $u,v$ in a graph are called \emph{true twins} if $N[u]=N[v]$.

\begin{lemma}\label{lem:twin-free}
\textsc{RD} is \NP-complete even on true-twin-free split graphs. Likewise, \textrm{standard}-\RD is \W{2}-complete on true-twin-free split graphs. 
\end{lemma}

\begin{pf}
The construction provided in \autoref{thm:RHF/RHS-hardness} shows how to construct a graph~$G$ as an instance of \RD, given an instance of \RHF. It is not hard to observe that this graph~$G$ is indeed a split graph, as (in the notation of the proof of \autoref{thm:RHF/RHS-hardness}) $V(X)\cup\{a\}$ is a clique, while $\{b,c\}\cup W\cup U$ is an independent set.
Hence, any true twins must belong to $V(X)\cup\{a\}$.
As argued in the proof of \autoref{thm:RHF/RHS-hardness}, $g(a)=2$ is enforced by $b,c\in N(a)$ (ensures $g(v_x)\neq 1$, as \autoref{t_property_min_rdf} would imply the existence of a smaller rhf). As $b,c\notin N(V(X))$, there is no $x\in V(X)$ with $N[x]=N[a]$, i.e., true twins must be in $V(X)$. We now walk through $V(X)$ in an arbitrary but fixed order, possibly deleting  vertices from the graph that are bigger than the currently considered one, and possibly also adding some further vertices to the independent set.
Let $T(x)=\{y\in V(X)\mid N[y]=N[x]\}$ be the set of twins of~$x$. If $|T(x)|>1$, we delete $T(x)\setminus\{x\}$ from the graph.
The resulting graph gives an instance that is a twin-free split graph that has a Roman dominating function of weight at most~$k$ \iffl the original hypergraph has.
\end{pf}

The hardness part of the proof of \autoref{thm:RHF/RHS-hardness} (implicitly) uses the fact that the sequence of hyperedges could include a hyperedge multiple times, as there could be twins (vertices with the same closed neighborhoods) in the original graph. Consider a complete graph $K_n=([n],E_n)$ with $n\geq 2$ vertices. Since the closed neighborhoods are always equal to~$[n]$, if we would use a normal set for the hyperedges instead of a sequence of hyperedges, then the best solution would be $\chi_{\{v\}}$ for any vertex~$v$. This would even not be a rdf. Namely, minimal rdf would be of the form $2\cdot \chi_{\{v\}}$ for any vertex~$v$, or they would be constant~$1$. Nevertheless, the following holds, revisiting \autoref{lem:twin-free}.

\begin{corollary}
\RHF and \RHS are \NP-complete even on simple hypergraphs. Furthermore, standard-\RHF and standard-\RHS are \W{2}-complete.
\end{corollary}

Let us think about \RHF and \RHS as minimization problems for some moments.
We mentioned already that \textsc{Minimum RD} can be approximated up to a logarithmic factor. The idea of the proof of \cite{Pagetal2002} can be easily adapted to \RHF and \RHS and gives us the next result, which we also combine with supplementing hardness results.
\begin{theorem}\label{t_approx_rhs}There is a polynomial-time algorithm that, given an instance $H=(X,(s_i)_{i\in I})$ of 
   \textsc{Minimum RHS}, outputs a rhs $(R_1,R_2)\in 2^I\times 2^X$ such that
   $\omega(R_1,R_2)\leq 2\cdot (\ln(|I|)+1)\cdot \text{opt}(H)$.
Moreover, unless $\PTIME=\NP$, there is no polynomial-time algorithm that achieves an approximation ratio better than $\frac{1}{2}(\ln(|I|)-o(1))\cdot \text{opt}(H)$.
\end{theorem}

\begin{pf}
The claimed algorithm would run the well-known greedy algorithm for \textsc{Hitting Set} (or \textsc{Set Cover} in the dual interpretation; see~\cite{Chv79}) on the underlying simple hypergraph and will get a collection $C\subseteq X$ that covers $S=\left\{ s_i\mid i\in I\right\}$. Compared to the optimum cover $C^*$, Chv{\'{a}}tal~\cite{Chv79} proved that $|C|\leq (\ln(|S|)+1)\cdot |C^*|$.
Clearly, $(\emptyset,C)$ is a valid rhs with $\omega(\emptyset,C)=2\vert C\vert$. Conversely, let $R^*=(R_1^*,R_2^*)$ be an optimum rhs solution, i.e., $\omega(R^*)=\text{opt}(H)$. We can obtain a valid solution to the interpretation of $H$ as a \textsc{Hitting Set} instance by defining $C_{R^*}=\phi(R_1^*)\cup R_2^*$, where $\phi$ applied to an index set like $R_1^*$ will pick, for each index~$i\in R_1^*$, an arbitrary element from $s_i\subseteq X$. Clearly, $|C^*|\leq |C_{R^*}|\leq \omega(R^*)\leq 2\cdot |C_{R^*}|$. Putting the derived inequalities together and finally observing $|S|\leq |I|$ gives the claim on the approximation factor of the described greedy algorithm:
$$\omega(R)=2|C|\leq 2\cdot (\ln(|S|)+1)\cdot |C^*|\leq 2\cdot (\ln(|I|)+1)\cdot \omega(R^*)\,.$$
Now, consider some polynomial-time approximation algorithm~$A$ for \textsc{Minimum RHS} that achieves an approximation ratio of $r(|I|)$, given an instance with index set~$I$. Let $H=(X,S)$ be a simple hypergraph, as an instance of \textsc{Minimum Hitting Set} with an optimum hitting set $C^*\subseteq X$. We interpret this as an instance of \textsc{Minimum RHS} and run~$A$ on it, getting some solution $R=(R_1,R_2)$. Compared to an optimum solution $R^*$ for this \textsc{Minimum RHS} instance, we know that $\omega(R)\leq r(|S|)\cdot \omega(R^*)$. Let $C_R=\phi(R_1)\cup R_2$, similar as above. Again, $|C_{R}|\leq \omega(R)\leq 2\cdot |C_{R}|$. If $C_R$ achieves an approximation ratio better than $(\ln(|S|)-o(1))\cdot |C^*|$, then $\PTIME=\NP$ by the famous results of Dinur and Steuer~\cite{DinSte2014}. But this would be the case if $r(|S|)\leq \frac{1}{2}(\ln(|S|)-o(1))$, because then we find:
$$|C_R|\leq r(|S|)\cdot \omega(R^*)\leq r(|S|)\cdot\omega((\emptyset,C^*))\leq r(|S|)\cdot 2|C^*|\leq (\ln(|S|)-o(1))\cdot |C^*|\,,$$
because we could interpret again the \textsc{Minimum Hitting Set} solution $C^*$ as a solution $(\emptyset,C^*)$ of weight $2|C^*|$ that is clearly not better than an optimum \textsc{Minimum RHS} solution~$R^*$.
\end{pf}

There is clearly a small gap between lower and upper bound in the previous theorem, but there is also a gap in the corresponding (in)approximability result concerning \textsc{Minimum Roman Domination} in \cite{Pagetal2002} that is simply not made explicit. However, our reasoning, together with the reductions given at the beginning of this section, prove that also in that case, we still have a multiplicative gap of four. 
We can get a similar result for \textsc{Minimum RHF}. 
\begin{corollary}
    There is a polynomial-time algorithm that, given an instance $H=(X,(s_i)_{i\in I})$ of \textsc{Minimum RHF}, outputs a rhf $f\in \{0,1,2\}^X$ such that
   $\weightfun(f)\leq 2\cdot (\ln(|I|)+1)\cdot \text{opt}(H)$.
Moreover, unless $\PTIME=\NP$, there is no polynomial-time algorithm that achieves an approximation ratio better than $\frac{1}{2}(\ln(|I|)-o(1))\cdot \text{opt}(H)$.
\end{corollary}

The proof works the same as the proof for \autoref{t_approx_rhs}, but the result of our algorithm is $f\in \{0,1,2\}^X$ with $f^{-1}(1)=\emptyset$ and $f^{-1}(2)=C$. Furthermore, for a rhf $g$ we define the hitting set $C_g\coloneqq f^{-1}(1) \cup f^{-1}(2)$.

Together with the construction given in \autoref{lem:twin-free}, we can deduce the same inapproximability result for \textsc{Minimum RD} on true-twin-free split graphs, hence sharpening previously published results of this type.

The \NP-completeness of both optimization problems also motivates our analysis of the extension problems; it could also help speed up an exact branching algorithm for solving these decision problems.

\section{Combinatorial Properties of Minimal \longversion{Roman Hitting Sets}\shortversion{rhs}}
\label{sec:rhs}

In order to pave the ground for our algorithmic results, we are now going to derive a combinatorial characterization of minimal rhs, tightly linking them to the notion of inclusion-wise minimal hitting sets.

\begin{lemma}\label{lem_rhs_disjoint}
Let $H=(X,\hat S=(s_i)_{i \in I})$ a hypergraph and $(R_1,R_2)$ a minimal rhs. Then for each $i\in I$, $s_i\cap R_2=\emptyset$.
\end{lemma}

\begin{pf}
Assume there is a minimal rhs $R=(R_1,R_2)$ of~$H$ with an $i\in I$  such that $s_i\cap R_2\neq \emptyset$. Then $R'\coloneqq(R_1\setminus \{i\},R_2)\leq (R_1,R_2)$. Since $s_i\cap R_2$ is not empty, $i$ is hit by $R'$. For $j\in I\setminus \{i\}$, $s_j$ is hit by $R'$ in the same way as by $R$.  Therefore, $R$ is not minimal.
\end{pf}

\begin{lemma}\label{lem_rhs_minhs}
Let $H=(X,\hat S=(s_i)_{i \in I})$ a hypergraph and $(R_1,R_2)$ a minimal rhs. Then $R_2$ is a minimal hitting set on $H'=(X, (s_i)_{i\in I\setminus R_1})$.
\end{lemma}

\begin{pf}
Assume there is a minimal rhs $R=(R_1,R_2)$ of $H$ such that $R_2$ is no minimal hitting set. $R_2$ has to be hitting set, as otherwise $R$ would not be a rhs. This implies there exists a $x\in R_2$ such that for each $i\in \mathbf{I}(x)$ there exists a $y_i\in s_i\cap (R_2\setminus \{x\})$. Define $R'\coloneqq(R_1,R_2\setminus \{x\})$. The indices in $I\setminus\mathbf{I}(x)$ are hit by $R'$ in the same way as by $R$. As for each $i\in \mathbf{I}(x)$, $ s_i\cap (R_2\setminus \{x\})$ is not empty, $\mathbf{I}(x)$ is hit by $R'$. Therefore, $R'$ is a rhs and $R$ not minimal. 
\end{pf}
 
\begin{theorem}\label{t_property_min_rhs}
Let $X$ be a vertex set, $\hat{S}=\left(s_i\right)_{i\in I}$  be a hyperedge sequence. Then, a tuple $(R_1,R_2)\leq (I, X)$ is a minimal rhs \iffl the following constraints hold:
\begin{enumerate}
\item $\forall i\in R_1:\: s_i \cap R_2 = \emptyset$, \label{con_rhs_disjoint}
\item $R_2$ is a minimal hitting set on $\{s_i\in S\mid i\in I \setminus R_1\}$. \label{con_rhs_minHS}
\end{enumerate}
\end{theorem}

\begin{pf}
    The ``only if''-part of this proof follows by \autoref{lem_rhs_disjoint} and \ref{lem_rhs_minhs}.
    For the other part, we assume  that the tuple $R=(R_1,R_2)$ fulfills the two constraints. Since $R_2$ is a minimal hitting set on $\{s_i\in S\mid i\in I \setminus R_1\}$, for each $i\in I$, $i$ in $R_1$ or  $s_i \cap R_2$ is not empty. Therefore, $R$ is a rhs. Now assume there is a minimal rhs $R'=(R'_1,R'_2)$ with $R'\leq R$. 
    Let $x\in R_2\setminus R_2'$. Then for each $i\in \mathbf{I}(x)$ either $i \in R'_1\subseteq R_1$ (this contradicts the first constraint) or $s_i\cap (R'_2)\subseteq s_i\cap (R_2 \setminus \{ x\})$ is not empty (this contradicts the second constraint). This implies $R_2=R'_2$. Let $i \in R_1 \setminus R'_1$. Then there has to be a $x\in s_i \cap R'_2=s_i \cap R_2$. This is a contradiction to Constraint \ref{con_rhs_disjoint}. Therefore, $R'=R$. 
\end{pf}

\begin{remark}\label{rem:rhs_vs_po_rdf}
Now we will compare this characterization theorem with \autoref{t_property_po_min_rdf}. Therefore, let $G=(V,E)$ be a graph and $G_{nb}=(V,(N[v])_{v\in V})$ be the corresponding closed-neighborhood hypergraph. We want to show that for a function $f:V \rightarrow \{0,1,2\}$ the constraints from \autoref{t_property_po_min_rdf} with respect to~$G$ are the same as the constraints of \autoref{t_property_min_rhs} for the tuple $R_f=(f^{-1}(1),f^{-1}(2))$ with respect to~$G_{nb}$.

First, we show that the first constraint  of \autoref{t_property_po_min_rdf} implies the first constraint of \autoref{t_property_min_rhs} by contraposition. 
If there would exist $v\in f^{-1}(1)$ and $w\in N[v]\cap f^{-1}(2)$, then $N[f^{-1}(2)]\cap f^{-1}(1)$ would not be empty. Conversely, the existence of an element $v\in N[f^{-1}(2)]\cap f^{-1}(1)$ would imply that there is a $w\in f^{-1}(2)\cap N[v]$. This would contradict Constraint~\ref{con_rhs_disjoint} in \autoref{t_property_min_rhs}. Thus, this constraint is equivalent to the Constraint~\ref{con_1_2_po} of \autoref{t_property_po_min_rdf} in this setting. 

The equivalence of the other two constraints follows by the fact that the closed-neighborhood hypergraph $G_{nb}'$ of $G'=G[V\setminus f^{-1}(1)]$ from \autoref{t_property_po_min_rdf} can be expressed as $(V\setminus f^{-1}(1),\linebreak[3](N[v]\setminus f^{-1}(1))_{v\in V\setminus f^{-1}(1)})$. In \autoref{sec_def_rhf}, we mentioned that a set is a dominating set in a graph \iffl it is a hitting set on its closed-neighborhood hypergraph. Clearly, also the minimality adapts. 
\end{remark}

\section{\textsc{Extension Roman Hitting Set} and \textsc{PO-Extension Roman domination}} \label{sec:Ext-RHS}

In this section, we will consider the complexity of \textsc{Extension Roman Hitting Set}. We do this by using the fact that PO-\textsc{Extension Roman Domination} is polynomial-time solvable (see~\cite{AbuFerMan2022}).  Furthermore, we introduce an explicit polynomial-time algorithm.

\begin{theorem}
\textsc{Extension Roman Hitting Set} is polynomial-time solvable. 
\end{theorem}

\begin{pf}
Let $H=(X,\hat{S}=(s_i)_{i\in I} )$ be a \longversion{(not necessarily simple) }hypergraph and $U=(U_1, U_2)\leq (I , X)$ with $X=\{x_1,\ldots,x_n\}$, i.e., $I=[m]$. Define $G=(V,E)$ and  $f\in \{0,1,2\}^V$ with 
\begin{equation*}
    \begin{split}
        V\coloneqq{}& \{a, b\} \cup \{v_1,\ldots,v_n\}\cup\{w_i \mid i\in I\}\\
        E\coloneqq{}& \binom{\{a, v_1,\ldots,v_n\}}{2}\cup \{\{v_j,w_i\} \mid x_j\in s_i\}\cup \{\{a,b\}\}\\
        & \text{and, for each $v\in V$,}\\
        f(v)\coloneqq{}&\begin{cases}
        0,& v\in \{ b \} \cup \{w_i\mid i\notin U_1\} \cup \{v_j\mid x_j \notin U_2\}\\
        1,& v\in \{w_i\mid i\in U_1\}\\
        2,& v\in \{ a \} \cup \{v_j\mid x_j\in U_2\}
        \end{cases}.
    \end{split}
\end{equation*}
Notice that $G$ is a split graph, where the clique basically consists of the original hypergraph vertices, with $v_j$ modeling $x_j$, plus~$a$, and the independent set is formed by vertices~$w_i$ representing the hyperedges $s_i$, plus $b$, and adjacency between $w_i$ and $v_j$ models incidence of $s_i$ with $x_j$. 

 Assume there exists a PO-minimal rdf~$g$ on $G$ with $f\leq g$. Define $R=(R_1,R_2)$ with $R_1 \coloneqq\{i \mid f(w_i)\neq 0\} $ and $R_2\coloneqq \{ x_i\mid f(v_i)=2\}$. Let $i\notin R_1$. Then there exists a $v_j\in N(w_i)\cap g^{-1}(2)$. This implies the existence of an $x_j \in R_2\cap s_i$. Hence, $R$ is a rhs. Assume there is an $i\in R_1$ such that $(R_1\setminus \{i\},R_2)$ is also a rhs. Thus, there exists an $x_j\in s_i \cap R_2$. This would contradict the PO-minimality of $g$, as $w_i\in f^{-1}(\{1,2\})$, $f(v_j)=2$ and $N[w_i]\subseteq N[v_j]$ (i.e., $g-\chi_{\{w_i\}}$ is also a rdf). Assume there is a $x_j\in R_2$ such that $(R_1,R_2\setminus\{x_j\})$ is also a rhs. Therefore, each $i\in I$ with $x_j\in s_i$ is either in $R_1$ or there is another $x_{t_i}\in R_2 \cap (s_i \setminus\{x_j\})$. Thus, vertex~$v_j$ has no private neighbor, which means that $g$ is not PO-minimal. Hence, $(R_1,R_2)$ is a minimal rhs with $(U_1,U_2)\leq ({R_1},R_2)$. 
 
 Conversely, let $(R_1,R_2)$ be a minimal rhs with $(U_1,U_2)\leq ({R_1},R_2)$. Define 
 $$g:\:V\rightarrow\{0,1,2\}, \, v\mapsto \begin{cases}
 0, & v\in \{ b \} \cup \{w_i\mid i\notin R_1\} \cup \{v_j\mid x_j\notin R_2\}\\
        1,& v\in \{w_i\mid i\in R_1\}\\
        2,& v\in \{ a \} \cup \{v_j\mid x_j\in R_2\}
 \end{cases}.$$
 
 As $g(a)=2$, the vertices $a,b,v_1\ldots,v_n$ are dominated. Since $(R_1,R_2)$ is a rhs, for each $w_i\in g^{-1}(0)$ there is a $v_j\in N(w_i)\cap g^{-1}(2)$. Thus, $g$ is a rdf. Assume there exists a $w_i\in g^{-1}(1)$ which has a neighbor in $v_j\in g^{-1}(2)$. This implies there exists an $i\in R_1$ with $x_j\in s_i \cap R_2$. But then $(R_1 \setminus \{i\}, R_2)$ is also a rhs. This leaves us to show that each $v\in g^{-1}(2)$ has a private neighbor. By construction, $b$ is a private neighbor of~$a$. Assume there exists a $v_j\in V$ such that each $u\in N[v_j]$ has a neighbor $t\in (N[u]\cap g^{-1}(2))\setminus\{v_j\}$. Therefore, each $s_i\in \mathbf{S}(x_j)$ has an $x_z\in (R_2\cap s_i)\setminus\{x_j\}$. This contradicts \autoref{t_property_min_rhs}. Therefore, $g$ is PO-minimal. 
\end{pf}

Let $G=(V,E)$ be a graph and $f:V\rightarrow \{0,1,2\}$ be a mapping. Theorem~6 of~\cite{AbuFerMan2022} implies that, if we solve the \textsc{Ext-RHS} on $H=(V,(N[v])_{v\in V})$ with the tuple $(f^{-1}(1),f^{-1}(2))$, then we solve the \textsc{PO-Ext-RDF} instance $(G,f)$. Therefore, we can view \textsc{Ext-RHS} as a generalization of \textsc{PO-Ext-RDF}. 

\begin{algorithm}[ht]
\caption{ExtRHS Algorithm}\label{alg:ext_rhs}
\begin{algorithmic}[1]
\Procedure{ExtRHS Solver}{$X,\hat{S},(R_1,R_2)$}\newline
 \textbf{Input:} set $X$, $\hat{S}\coloneqq(s_i)_{i\in I}$, $(R_1,R_2)\leq (I,X)$.\newline
 \textbf{Output:} Is there a minimal Roman hitting  set $M$ with $R\leq M$?

\For{$x\in R_2$}\label{alg_for_no_rhs}
\For {$i\in I(x)$}
\If{$i\in R_1$}
\State \textbf{Return \no}\label{alg_disjunct_rhs}
\EndIf
\EndFor
\If{$\mathbf{I}\left( x\right) \subseteq \mathbf{I}\left( R_2\setminus \{ x \}\right)$}
\State \textbf{Return \no} \label{alg_private_rhs}
\EndIf
\EndFor
\State $M_1\coloneqq R_1$  \label{alg_invariant_rhs}
\State $M_2 \coloneqq R_2$
\For{$i\in I\setminus \left( \mathbf{I} \left( M_2\right) \cup M_1\right)$} \label{alg_fil_for_rhs}
\State Add $i$ to $M_1 $. 
\EndFor 
\State\textbf{Return \yes}  $(M_1,M_2)$ 
\EndProcedure
\end{algorithmic}
\end{algorithm}

\begin{theorem}
    Algorithm \ref{alg:ext_rhs} solves \textsc{Ext-RHS} for instances $(X,\hat{S},(R_1,R_2))$ in polynomial time. 
\end{theorem}

\begin{pf}
    Let $H=\left( X,(s_i)_{i\in I}\right)$ be a hypergraph and let $(R_1,R_2)\leq  (I, X)$ be a tuple. Assume that the algorithm returns \yes. Then $R_1\subseteq M_1$ and $R_2=M_2$, since we define $M_1,M_2$ as $R_1,R_2$ in the beginning and never delete any vertices from these sets. Further, we put each index from $I\setminus \left( \mathbf{I} \left( M_2\right) \cup M_1\right)$ into $M_1$. Therefore $M=(M_1,M_2)$ is a rhs at the end of this algorithm. Now we want to check if $M$ is minimal. Assume that there is an $i\in R_1$ such that $s_i\cap M_2 \neq \emptyset$. Since the for-loop in Line~\ref{alg_fil_for_rhs} does not put such an index into $M_1$, $i$ has to be in~$R_1$. Because $M_2=R_2$, Line~\ref{alg_disjunct_rhs} would have returned \no. Assume $R_2$ is not a minimal hitting set on $(s_i)_{i \in I\setminus M_1}$. Since $(M_1,M_2)$ is a rhs, $M_2$ is a hitting set on $(s_i)_{i \in I\setminus M_1}$. If $M_2=R_2$ is not minimal, then Line~\ref{alg_private_rhs} will return \no.

    Assume there is a minimal rhs $(P_1,P_2)$ on~$H$ with $(R_1,R_2)\leq (P_1,P_2)$, but the algorithm returns \no. If the algorithm returns \no in Line~\ref{alg_disjunct_rhs}, then there exists a $i\in R_1\subseteq P_1$ with an $x\in s_i\cap R_2\subseteq s_i \cap P_2$. This contradicts \autoref{t_property_min_rhs}. Assume the algorithm returns \no in Line~\ref{alg_private_rhs}. Then there is an $x\in R_2\subseteq P_2$ such that, for each $i\in \mathbf{I}(x)$, there exists a $y\in s_i\cap (R_2\setminus \{x\})$. Then $P_2$ can not be a minimal hitting set on $(s_i)_{i \in I\setminus P_1}$. As these are the only cases in which the algorithm returns \no, if the input is a \yes-instance, then the algorithm will return \yes.
\end{pf}

We will explain in \autoref{sec:enumerating_rhs} how we can make use of this extension algorithm for enumerating all minimal rhs with polynomial delay.

\section{Combinatorial Properties of Minimal \longversion{Roman Hitting Functions}\shortversion{rhf}}
\label{sec:rhf}

In this section, we will prove combinatorial properties of minimal rhf. This will help us analyze the complexity of \textsc{Extension Roman hitting functions}. 

\begin{lemma}\label{t_samepic}
Let $X$ be a vertex set, $\hat{S}=\left(s_i\right)_{i\in I}$  be a hyperedge sequence and $\tau:X\rightarrow I$ be a correspondence. Then for each minimal rhf $f\in \{0,1,2\}^X$ of $X$, $\hat S$, $\tau$ and for all $x,y\in f^{-1}(1)$ with $x\neq y$, $\tau(x)\neq \tau(y)$ holds, i.e., $\tau\vert_{f^{-1}(1)}$ is injective.
\end{lemma}
\begin{pf}
Assume $f$ is a minimal rhf of $X$, $\hat{S}$ and $\tau$ such that there exist $x,y\in f^{-1}(1)$ with $x\neq y$ but $\tau(x)=\tau(y)= i$. Define $\tilde{f}=f-\chi_{\{ y \}}$.  Trivially, $\tilde{f}\leq f$ and $\tilde{f} \neq f$. Since $f^{-1}(2)=\tilde{f}^{-1}(2)$ and $\tau(\tilde{f}^{-1}(1))=\tau(f^{-1}(1)\setminus \lbrace y\rbrace) = \tau(f^{-1}(1))$ hold, $\tilde{f}$ is a rhf. This implies that $f$ is not minimal, which is a contradiction.
\end{pf}
\begin{lemma}\label{t_Hmdisjoint}
Let $X$ be a vertex set, $\hat{S}=\left(s_i\right)_{i\in I}$ be a hyperedge sequence and $\tau:X\rightarrow I$ be a correspondence. Then for each minimal rhf $f\in \{0,1,2\}^X$ of $X$, $\hat S$, $\tau$ and each $x\in f^{-1}(1)$, the sets $s_{\tau(x)}$ and $f^{-1}(2)$ are disjoint.
\end{lemma}
\begin{pf}
Assume $f$ is a minimal rhf of $X$, $\hat{S}$ and $\tau$ such that there exists an $x\in f^{-1}(1)$ with $s_{\tau(x)}\cap f^{-1}(2)\neq\emptyset$. Define $\tilde{f}=f - \chi_{\{x\}}$.
Trivially, $ \tilde{f}\leq f$ and $f\neq \tilde{f}$. By definition of $\tilde{f}$, we have $f^{-1}(2)=\tilde{f}^{-1}(2)$. Hence, $s_{\tau(x)}\in \mathbf{S}(\tilde{f}^{-1}(2))$. Since $s_i$ ,for $i \in I\setminus \{\tau(x)\}$, is hit by $\tilde{f}$ in the same way as by $f$, $\tilde{f}$ is a rhf. This contradicts the minimality of $f$. 
\end{pf}

\begin{lemma}\label{t_minHS}
Let $X$ be a vertex set, $\hat{S}=\left(s_i\right)_{i\in I}$  be a hyperedge sequence and $\tau:X\rightarrow I$ be a correspondence. Then for each minimal rhf $f\in \{0,1,2\}^X$ of $X$, $\hat S$, $\tau$, $f^{-1}(2)$ is minimal hitting set on $S'\coloneqq\{s_i\in S\mid i\in I, \tau^{-1}(i)\cap f^{-1}(1) = \emptyset\}$, i.e., each $s_i\in S'$ contains a vertex~$x$ with $f(x)=2$.
\end{lemma}
\begin{pf}
Assume $f^{-1}(2)$ is not a minimal hitting set on $S'$. If there is an $s\in S'$ that is not hit by $f^{-1}(2)$, then $f$ is no rhf, contradicting our assumption. 
Therefore, we assume that $f^{-1}(2)$ is not minimal. Then there exists an $x\in f^{-1}(2)$ such that, for each $s\in S'$ with $x\in s$, there exists a $y\in \left(f^{-1}(2) \setminus \lbrace x\rbrace\right) \cap s.$ Define $\tilde{f}=f-2\cdot\chi_{\{ x \}}$. Let $i\in I$.
 As $\tau(f^{-1}(1))= \tau(\tilde{f}^{-1}(1))$ holds by definition, we only need to consider $i\in I\setminus \tau(f^{-1}(1))$. For $i$ with  $s_i\in \mathbf{S}(f^{-1}(2))\setminus \mathbf{S}(x)$, trivially $s_i\cap \tilde{f}^{-1}(2)\neq\emptyset$ . If $s_i\in \mathbf{S}(x)$, as we mentioned before, there exists a $y\in \left(f^{-1}(2) \setminus \lbrace x\rbrace\right) \cap s_i=\tilde{f}^{-1}(2)\cap s_i$. This implies that $\tilde{f}$ is a rhf and $f$ is not minimal, contradicting our assumption. 
\end{pf}

\noindent
Altogether, these three lemmas imply the following observation: 

\begin{corollary}
Let $f$ be a minimal rhf. For all $x\in f^{-1}(1)$ and for all $i\in I$ that satisfy $s_i = s_{\tau(x)}$, we have $ f^{-1}(1)\cap \tau^{-1}(i)\neq\emptyset$.
\end{corollary}

\begin{lemma}\label{t_12Switch}
Let $X$ be a vertex set, $\hat{S}=\left(s_i\right)_{i\in I}$  be a hyperedge sequence and $\tau:X\rightarrow I$ be a correspondence. Then for each minimal rhf $f\in \{0,1,2\}^X$ of $X$, $\hat S$, $\tau$,  and for all $x\in f^{-1}(2)$, there exists $i\in I\setminus \{\tau(x)\}$ with $s_i\cap f^{-1}(2)=\{x\}$. 
\end{lemma}
\begin{pf}
Let $f$ be a minimal rhf. Assume there exists an $x\in f^{-1}(2)$ such that for each $i\in I\setminus \{\tau(x)\}$, $s_i \cap f^{-1}(2)\neq \{x\}$. We want to show that $\tilde{f}=f-\chi_{\{ x \}}$ is also a rhf. 
By \autoref{t_minHS}, there exists an $i\in I\setminus \tau(f^{-1}(1))$ such that $s_i\cap f^{-1}(2)=\{x\}$. The assumption on $x$ implies that $i$ has to be $\tau(x)$.
Let $j\in I$. If $x\notin s_j\cap f^{-1}(2)\neq \emptyset$, $s_j$ is hit by $\tilde{f}$ in the same way as by~$f$. The index $j=\tau(x)$ is hit by $\tilde{f}(x)=1$. For $x\in s_j\cap f^{-1}(2)$  with $j\neq \tau(x)$, there has to exist another $y\in s_i\cap f^{-1}(2)$ besides~$x$. This leaves the indices which are hit by $f^{-1}(1)\subseteq \tilde{f}^{-1}(1)$. Therefore, $\tilde{f}$ is a rhf.
\end{pf}

\begin{theorem}\label{t_property_min_rhf}
Let $X$ be a vertex set, $\hat{S}=\left(s_i\right)_{i\in I}$  be a hyperedge sequence and $\tau:X\rightarrow I$ be a correspondence. Then, a function $f\in \{ 0,1,2\}^X$ is a minimal rhf \iffl the following constraint items hold:
\begin{enumerate}\setcounter{enumi}{-1}
\item $\forall x,y\in f^{-1}(1) :\: x\neq y \Rightarrow \tau(x) \neq \tau(y)$, \label{con_difvalue}
\item $\forall x\in f^{-1}(1):\: s_{\tau(x)} \cap f^{-1}(2) = \emptyset$, \label{con_disjoint}
\item $ \forall x\in f^{-1}(2) \,\exists i\in I\setminus \{\tau(x)\} :\: s_i\cap f^{-1}(2)=\{x\}$, and \label{con_12switch}
\item $f^{-1}(2)$ is a minimal hitting set on $\{s_i\in S\mid i\in I, \tau^{-1}(i)\cap f^{-1}(1) = \emptyset\}$. \label{con_minHS}
\end{enumerate}
\end{theorem}
\begin{pf}
The only-if-part follows from \autoref{t_Hmdisjoint}, \autoref{t_samepic}, \autoref{t_minHS} and \autoref{t_12Switch}.

Assume $f$ fulfills the constraints. By Constraint 
\ref{con_minHS}, we know for all $i \in I$, either $s_i\cap f^{-1}(2)\neq \emptyset$, or there exists an $x\in f^{-1}(1)$ with $\tau(x)=i$. Therefore, $f$ is a rhf. Let $g\in\{0,1,2\}^X$ be a minimal rhf with $g\leq f$. Thus, $g^{-1}(2)\subseteq f^{-1}(2)$ and $ \mathbf{S}(g^{-1}(2))\subseteq \mathbf{S}(f^{-1}(2))$ hold. Furthermore, $g^{-1}(1)\subseteq f^{-1}(1)\cup f^{-1}(2)$.  Since for each $x\in X$, $\lbrace s_{\tau(x)}\rbrace \subseteq \mathbf{S}(x)$, for each $i\in \tau(g^{-1}(1))$,  $ i\in \tau(f^{-1}(1))$ or $s_i\in \mathbf{S}(f^{-1}(2))$. Let $x\in X$ be a element with $g(x)< f(x)$. 

\smallskip\noindent
\textbf{Case 1:} $g(x)=0<2=f(x)$.
This implies that, for each $i \in \mathbf{I}(x)$, there exists a $y\in s_i$ with $2=g(y)\leq f(y)=2$ or $y\in \tau^{-1}(i)\cap g^{-1}(1) \subset \tau^{-1}(i)\cap (f^{-1}(1)\cup f^{-1}(2))$. This either contradicts Constraint 
\ref{con_disjoint} or Constraint 
\ref{con_12switch}.

\smallskip\noindent
\textbf{Case 2:} $g(x)=1<2=f(x)$. This case works analogously, somehow simpler. We only need to exclude $i=\tau(x)$.

\smallskip\noindent
\textbf{Case 2:} $f(x)=1$.
This implies $g(x)=0$. Since $g$ is a rhf, either $s_{\tau(x)}\cap g^{-1}(2)$ is not empty or there exists a $y\in g^{-1}(1)$ with $\tau(x)=\tau(y)$.

\smallskip\noindent
\textbf{Case 2.1:} $\tau(x)\cap g^{-1}(2)\neq \emptyset$. As $g^{-1}(2)\subseteq f^{-1}(2)$, this contradicts Constraint 
\ref{con_disjoint}.

\smallskip\noindent
\textbf{Case 2.2:} There exists a $y\in g^{-1}(1)$ with $\tau(x)=\tau(y)$. Therefore, either there exists a $y\in f^{-1}(1)\setminus \lbrace x\rbrace$ such that $\tau(x)= \tau(y)$ (this contradicts Constraint 
\ref{con_difvalue}) or $y\in f^{-1}(2)$ (this contradicts Constraint 
\ref{con_disjoint}).

Thus, $g=f$ holds. Therefore, $f$ is minimal.
\end{pf}

\begin{remark}\label{rem:rhf_vs_rdf}
    As in \autoref{rem:rhs_vs_po_rdf}, we will compare this theorem with \autoref{t_property_min_rdf}. Let $G=(V,E)$ be a graph and let $G_{nb}=(V,(N[v])_{v\in V})$ be the closed-neighborhood hypergraph. The function $\tau: V\to V$ is the identity function. Let $f:V \to \{0,1,2\}$.
    
    As $\tau$ is bijective, Constraint~\ref{con_difvalue} holds for each~$f$. We will show next that $f$ fulfills Constraint~\ref{con_disjoint} of \autoref{t_property_min_rhf} \iffl $f$ fulfills Constraint \ref{con_1_2} of \autoref{t_property_min_rdf}. Assume there exists a $v\in f^{-1}(1)$ such that $N[v]\cap f^{-1}(2)\neq \emptyset$. This implies $N[f^{-1}(2)]\cap \{v\} \subseteq N[f^{-1}(2)]\cap f^{-1}(1)$ is not empty. This contradicts Constraint~\ref{con_1_2} of \autoref{t_property_min_rdf}. Conversely, the existence of an element $v\in N[f^{-1}(2)]\cap f^{-1}(1)$ would imply that there is a $w\in f^{-1}(2)\cap N[v]$. This would contradict Constraint~\ref{con_disjoint} in \autoref{t_property_min_rhf}.

    For each $v\in f^{-1}(2)$, there is a $u \in V\setminus \{v\}$ such that $N[u] \cap f^{-1}(2)=\{v\}$ \iffl each vertex in $f^{-1}(2)$ has a private neighbor that is not itself.  We can reformulate the set occurring in the last Constraint of \autoref{t_property_min_rhf} to $\{N[v]\mid v\in V, \tau^{-1}(v)\cap f^{-1}(1) = \emptyset\} = \{N[v]\mid v\in V,\,  v\notin f^{-1}(1)\}$. This is the set of hyperedges of $(f^{-1}(\{0,2\}), (N[v])_{v\in V\setminus f^{-1}(1) })$. $f^{-1}(2)$ is a minimal hitting set of this hypergraph \iffl it is a minimal hitting set of ($(f^{-1}(\{0,2\}), (N[v]\setminus f^{-1}(1))_{v\in V\setminus f^{-1}(1) })$), which is the closed-neighborhood hypergraph of $G'=G[f^{-1}(\{0,2\})]$. Therefore, the last Constraints of \autoref{t_property_min_rdf} and \autoref{t_property_min_rhf} are equivalent in this case.
\end{remark}

We call a $f\in \{0,1,2\}^X$ \emph{extensible} on the hypergraph $H=(X,\hat{S})$  with correspondence  $\tau$ if there exists a minimal rhf $g$ with $f\leq g$.

\begin{corollary}\label{cor_extensible12}
Let $H=(X,\hat{S}=(s_i)_{i\in I})$ be a hypergraph with correspondence~$\tau$ and $f\in \{0,1,2\}^X$ be a function with $x\in f^{-1}(2)$, $y\in f^{-1}(1)$ and $x\in s_{\tau(y)}$. Then, $f$ is extensible \iffl $f+\chi_{\{y\}}$ is extensible.
\end{corollary}

\begin{pf}
Since $\leq$ on $\mathbb{N}^X$ is transitive, if $f+\chi_{\{y\}}$ is extensible, then so is~$f$. For the other direction, we assume that there exists a minimal rhf $g$ with $f\leq g$. Thus, $g(x)=f(x)=2$ and $g(y)\in\{1,2\}$  for $x\in f^{-1}(2)$, $y\in f^{-1}(1)$. By Constraint~\ref{con_disjoint} of \autoref{t_property_min_rhf},  $y\in g^{-1}(2)$, as $x\in s_{\tau(y)}$. For the remaining $z\in X\setminus \{ y \}$, $(f+\chi_{\{y\}})(z)=f(z)\leq g(z)$ holds. This implies $f+\chi_{\{y\}}\leq g$, so that $f+\chi_{\{y\}}$ is extensible. 
\end{pf}

\begin{corollary}\label{cor_extensible11}
Let $H=(X,\hat{S}=(s_i)_{i\in I})$ be a hypergraph with correspondence~$\tau$ and $f\in \{0,1,2\}^X$ be a function with $x, y \in f^{-1}(1)$, $x\neq y$ and $\tau(x) = \tau(y)$. Then, $f$ is extensible \iffl $f+\chi_{\{x,y\}}$ is extensible.
\end{corollary}

\begin{pf}
Since $\leq$ on $\mathbb{N}^X$ is transitive, if $f+\chi_{\{x,y\}}$ is extensible, then so is~$f$. For the other direction, we assume that there exists a minimal rhf $g$ with $f\leq g$. Thus, $g(x),g(y)\in\{1,2\}$. By Constraint~\ref{con_difvalue} of \autoref{t_property_min_rhf}, $g(x)=2$ or $g(y)=2$. This implies that if $f$ is extensible, then $f+\chi_{\{x\}}$ or $f+\chi_{\{y\}}$ is extensible. In both cases, we can use \autoref{cor_extensible12} and conclude that $f+\chi_{\{x,y\}}$ is extensible if function~$f$ is extensible.  
\end{pf}

\begin{theorem}\label{t_extrhf_equiv}
Let $H=(X,\hat{S}=(s_i)_{i\in I})$ be a hypergraph with correspondence~$\tau$, $\tau:X\rightarrow I$. Let $f \in \{0,1,2\}^X$  be a function such that $x\neq y$ implies $\tau(x)\neq \tau(y)$ for each $x,y \in f^{-1}(1)$. Then, $f$ is extensible \iffl there exist a set $R_2$ with $f^{-1}(2)\subseteq R_2\subseteq f^{-1}(1)\cup f^{-1}(2)$ and a mapping $\rho:R_2\to I$,
satisfying the following constraints.
\begin{enumerate}
    \item $\forall x\in R_2:\:\rho(x)\neq \tau(x)$.\label{con_tauVSPrivate}
    \item $\forall x\in R_2:\:s_{\rho(x)}\cap R_2= \{x\}$.\label{con_privateEdge}
    \item $\forall x\in f^{-1}(1)\setminus R_2:\:\tau(x)\cap R_2=\emptyset$.\label{con_1VS2}
    \item $\forall i\in  I \text{ such that } \tau^{-1}(i)=\emptyset:\\ s_i\subseteq\left(\bigcup_{x\in f^{-1}(1)\setminus R_2}s_{\tau(x)}\right) \cup \left( \bigcup_{x\in R_2} s_{\rho(x)}\right)\implies s_i \cap R_2\neq \emptyset$. \label{con_hitting}
\end{enumerate}  
\end{theorem}

\begin{pf}
Define $I'\coloneqq\{i\in I \mid \tau^{-1}(i)=\emptyset\}$. First, we assume that~$f$ is extensible. 
Let $g\in \{0,1,2\}^X$ be a minimal rhf with $f\leq g$. By \autoref{t_minHS}, $g^{-1}(2)$ is a minimal hitting set on $\{s_i\in S\mid i\in I, \tau^{-1}(i)\cap f^{-1}(1) = \emptyset\}$ $(*)$. With \autoref{t_property_min_rhf}, Constraint \ref{con_12switch}, this implies that, for each $x\in g^{-1}(2)$, there exists an $\rho(x) \in I\setminus \{\tau(x)\}$ such that $s_{\rho(x)}\cap g^{-1}(2)=\{x\}$. Define $R_2=(f^{-1}(1)\cup f^{-1}(2))\cap g^{-1}(2)$. Clearly, $f^{-1}(2)\subseteq R_2\subseteq f^{-1}(1)\cup f^{-1}(2)$. We have to check the four constraints claimed for~$R_2$. The first two are even true in a slightly more general fashion by $(*)$. If there would exist a $y\in f^{-1}(1)\setminus R_2 \subseteq g^{-1}(1)$ with $\emptyset \neq s_{\tau(y)}\cap R_2 \subseteq s_{\tau(y)}\cap g^{-1}(2)$, then this would contradict \autoref{t_property_min_rhf}, Constraint~\ref{con_12switch}, showing the third constraint of this theorem. We now turn to the fourth and last constraint. Let $i\in I'$. Since $\tau^{-1}(i)=\emptyset$, there has to be a $y\in g^{-1}(2)\cap s_i$. If $y\in R_2$, then the constraint is satisfied. Hence, we can assume that $y\in g^{-1}(2)\setminus R_2=g^{-1}(2)\cap f^{-1}(0)$.
Consider $x\in f^{-1}(1)\setminus R_2$. As $f\leq g$ and $x\notin R_2$, we have $g(x)=1$. By Constraint~\ref{con_disjoint} of \autoref{t_property_min_rhf}, $y\notin s_{\tau(x)}$.
If $s_i\subseteq\left(\bigcup_{x\in f^{-1}(1)\setminus R_2}s_{\tau(x)}\right) \cup \left( \bigcup_{x\in R_2} s_{\rho(x)}\right)$ and $s_i \cap R_2= \emptyset$ hold, then this would contradict $s_{\rho(x)}\cap R_2 =\{x\}$. Therefore, all the constraints hold.

Let now $f$, $R_2$ and $\rho: R_2\to I$ fulfill the constraints of this theorem. For this part of the proof we will define a hypergraph $H'$ that includes each edge where $\tau(X)$ does not include its index and the edge is not hit, yet. We will show that there is a minimal hitting set $D$ on~$H'$ which does not include any vertex of $s_{\rho(x)}$ for $x \in R_2$ or $s_{\tau(x)}$ for $f^{-1}(1)\setminus R_2$. $R_2\cup D$ will describe the set of vertices with value~2. We will hit the remaining vertices by assigning the value 1~to some vertices.

Therefore, we define the hypergraph $H'=(X', (s'_i)_{i\in I''})$ with
\begin{align*}
I''\coloneqq\, &I' \cap \{i\in I \mid s_i\cap R_2 = \emptyset \}\,,\\
X' \coloneqq\, &\left( \bigcup_{i\in I''} s_i \right) \setminus \left( \left(\bigcup_{x\in f^{-1}(1)\setminus R_2}s_{\tau(x)}\right) \cup \left( \bigcup_{x\in R_2} s_{\rho(x)}\right)\right)
\end{align*}
and $s'_i\coloneqq s_i\cap X'$. If $s'_i$ is empty for an $i\in I''$, then there would not exist any hitting set on~$H'$. Therefore, we need to ensure that such an index does not exist. Let $i\in I''$. Hence, $\tau^{-1}(i)=\emptyset$ and $s_i\cap R_2 =\emptyset$. The contraposition of the implication of Constraint~\ref{con_hitting} implies
$$s_i\nsubseteq\left(\bigcup_{x\in f^{-1}(1)\setminus R_2}s_{\tau(x)}\right) \cup \left( \bigcup_{x\in R_2} s_{\rho(x)}\right).$$ 
Hence, $s'_i\neq \emptyset$ for each $i\in I''$. Thus, there exists a minimal hitting set~$D$ on~$H'$.  Furthermore, the construction of $H'$ and $D$ implies that $\tau^{-1}(i)$ is not empty for each $i\in I\setminus \mathbf{I}(R_2 \cup D)$. For each $i \in I\setminus \mathbf{I}(R_2 \cup D)$, $x_i$ will describe an arbitrary vertex in $\tau^{-1}(i)$, unless there exists an $x_i\in \left(f^{-1}(1) \setminus R_2\right) \cap \tau^{-1}(i)$ (by assumption on~$f$, there is at most one such element). In this case, we choose this~$x_i$. 
 
 Define $g\in \{0,1,2\}^X$ with $g^{-1}(1)= \{x_i \mid i \in I\setminus \mathbf{I}(R_2 \cup D)\}$ and $g^{-1}(2)=D\cup R_2$. We will now use \autoref{t_property_min_rhf} to show that $g$ is a minimal rhf. By the construction of $g^{-1}(1)$, the first two constraints of \autoref{t_property_min_rhf}, i.e., Constraints~\ref{con_difvalue} and~\ref{con_disjoint}, are fulfilled. Since $g^{-1}(1)$ hits each edge in $I\setminus \mathbf{I}(D \cup R_2)$, each hyperedge in $\{s_i \mid i\in I,\tau^{-1}(i)\cap g^{-1}(1)= \emptyset \}$ is hit by $g^{-1}(2)$. As D is minimal and $D\cap \left(\cup_{x\in R_2}s_{\rho(x)}\right)=\emptyset$, Constraint \ref{con_privateEdge} implies that is $D\cup R_2$ also a minimal hitting set on $\mathbf{I}(D\cup R_2)$. The remaining constraint of \autoref{t_property_min_rhf} follows by the first two constraints of~$f$ together with definition of $H'$ and $D$, as $H'$ only contains hyperedges $s'_i$ where $\tau^{-1}(i)=\emptyset$.\end{pf}

 \begin{remark}
     Let $H=(X, (s_i)_{i\in I})$ be hypergraph with a correspondence $\tau$. With a combination of \autoref{cor_extensible11} and \autoref{t_extrhf_equiv} we can check if a function $f\in \{ 0, 1, 2\}^X$ is extensible (and therefore if $(H,\tau,f)$ is a yes instance of \ExtRHF). First, we use \autoref{cor_extensible11} exhaustively, and then we can use \autoref{t_extrhf_equiv}.
 \end{remark}

\begin{remark}
    This Theorem already gives an idea why \ExtRHF with surjective correspondence (and therefore also \textsc{Ext-RDF}) runs in polynomial time. Let $H=(X,(s_i)_{i\in I})$ be a hypergraph and $f\in \{0,1,2\}^X$ be a function with the surjective correspondence $\tau$. First of all we can disregard Constraint \ref{con_hitting}, as $\tau(i)\neq \emptyset$ for each $i\in I$. 
    Then, we could use the corollaries \ref{cor_extensible12} and \ref{cor_extensible11}. We set $R_2 \coloneqq f^{-1}(2)$.
    By Constraint \ref{con_1VS2}, we have to add each element in $x\in f^{-1}(1)$ with $s_{\tau(x)} \cap R_2 \neq \emptyset$ to $R_2$. This can be done in polynomial time (we will show how in Algorithm \ref{alg:tau_sur}). Now we can check in polynomial time if each for each $x \in R_2$ there exists a $i_x\in I$ that fulfills the constraints \ref{con_tauVSPrivate} and \ref{con_privateEdge}. 
\end{remark}

\noindent
This will also be our strategy in the next section.

\section{\textsc{Ext-RHF} with Surjective Correspondence and \textsc{Ext-RDF}}\label{sec:tau_sur}

This section is presented polynomial time algorithm for \textsc{Ext-RHF} instances with a surjective correspondence function~$\tau$. At the end of this section, we explain how this algorithm can be viewed as a natural generalization of \textsc{Ext-RDF} that was studied before~\cite{AbuFerMan2022b}.

\begin{algorithm}[ht]
\caption{ExtRHF Algorithm}\label{alg:tau_sur}
\begin{algorithmic}[1]
\Procedure{ExtRHF Solver}{$X,\hat{S},\tau,g$}\newline
 \textbf{Input:} set $X$, $\hat{S}\coloneqq(s_i)_{i\in I}$,  $\tau$ correspondence function, $g\in \{0,1,2\}^X$ with $\{i\in I \mid \tau^{-1}(i) =  \emptyset \} \subseteq \mathbf{I} \left( g^{-1}(2) \right) $.\newline
 \textbf{Output:} Is there a minimal Roman hitting  function $f$ with $g\leq f$?
\State $M_2\coloneqq g^{-1}(2)$  \label{alg_invariant}
\State $M_1 \coloneqq g^{-1}(1)$
\label{alg_before_while}
\For{$x\in M_1$}\label{alg_tau_not_injective}
\For{$y\in M_1\setminus\lbrace x\rbrace $}
\If{$\tau(x)=\tau(y)$}
\State Add $x,y$ to $M_2$ and delete them in $M_1$.
\State Continue with the next $x$.
\EndIf
\EndFor
\EndFor
\State $ M \coloneqq M_2$ \{ All $x\in f^{-1}(2)$ are considered in the following. \}
\While{$M\neq \emptyset$}\label{alg_whileloop}
\State Choose $x\in M$.
\For {$y\in \tau^{-1}( \mathbf{I}\left( x\right))$}
\If{$y\in M_1$}
\State Add $y$ to $M_2$ and $M$. Delete $y$ in $M_1$.
\EndIf
\EndFor
\State Delete $x$ from $M$. 
\EndWhile
\For{$x\in M_2$}\label{alg_for_no}
\If{$\mathbf{I}\left( x\right)\setminus \mathbf{I}\left( M_2\setminus \{ x \}\right)\subseteq \{\tau(x)\}$}
\State \textbf{Return \no} \label{alg_no}
\EndIf
\EndFor
\For{$i\in I\setminus \left( \mathbf{I} \left( M_2\right) \cup \tau(M_1)\right)$} \label{alg_fil_for}
\State Add one element $x\in \tau^{-1}(i)$ to $M_1 $.
\EndFor 
\State\textbf{Return \yes}  $\{\,f^{-1}(0)=X\setminus \left( M_1\cup M_2\right), f^{-1}(1)=M_1, f^{-1}(2)=M_2\, \}$ 
\EndProcedure
\end{algorithmic}
\end{algorithm}

\begin{theorem}
Algorithm \ref{alg:tau_sur} solves \textsc{Ext-RHF} for instances $(X,\hat{S},\tau,g)$ satisfying $\{i\in I \mid \tau^{-1}(i)= \emptyset \} \subseteq \mathbf{I}\left( g^{-1}(2) \right)$ in polynomial time.
\end{theorem}

\begin{pf}
Assume the algorithm returns \yes. Define $f\in\{0,1,2\}^X$ with $f^{-1}(0)=X\setminus \left( M_1\cup M_2\right)$, $f^{-1}(1)=M_1$ and $ f^{-1}(2)=M_2$. As we  put vertices from~$M_1$ into~$M_2$ and from $X\setminus (M_1\cup M_2)$ into~$M_1$ only, $g\leq f$ holds. Since $\tau^{-1}(i)$ is not empty for each $i\in I\setminus \mathbf{I} \left( M_2\right)\subseteq I\setminus \mathbf{I} \left( g^{-1}(2) \right)$, $f$ is a rhf. $\tau(M_1)\cap \mathbf{I}(M_2)$ does not contain any elements after the while-loop. As the for-loop starting in line \ref{alg_fil_for} puts elements $x\in X$ in~$M_2$ with $s_{\tau(x)}\cap M_2=\emptyset$ only, $\tau(M_1)\cap \mathbf{I}(M_2)$ is empty at the end of the algorithm. The first for-loop ensures, that two different elements in~$M_1$ have different values under~$\tau$. Only in the last for-loop, we put an element~$x$ into~$M_1$, which has a unique $\tau(x)$ and where $\tau(x)$ is not hit so far. If there is a vertex $x\in M_2$ with $\mathbf{I}(x)\setminus \mathbf{I}\left(M_2\setminus \{ x \}\right)\subseteq \{ \tau(x) \}$, then the algorithm would return \no. \autoref{t_property_min_rhf} implies that $f$ is a minimal rhf.

Assume the algorithm return \no, but there is a minimal rhf with $g \leq h$. Therefore, the constraints of \autoref{t_property_min_rhf} hold for $h$. This implies that $\mathbf{I}\left(h^{-1}(2)\right) \cap \tau(h^{-1}(1))=\emptyset$ and  $\tau(x) \neq \tau(y)$ hold for each  $x,y\in h^{-1}(1)$ with $x\neq y$. Therefore, $M_1\subseteq h^{-1}(1)$ and $M_2\subseteq h^{-1}(2)$ is fulfilled before line \ref{alg_for_no}. Therefore, $\mathbf{I}(x)\setminus \mathbf{I}\left( h^{-1}(2)\setminus\{x\}\right) \subseteq \mathbf{I}(x)\setminus \mathbf{I}(M_2\setminus\{x\}) =\{\tau(x)\}$ holds for each $x\in M_2$. Thus, $h$ is not a minimal rdf. Hence, Algorithm~\ref{alg:tau_sur} solves \textsc{Ext-RHF} with $I\setminus \tau(X)\subseteq \mathbf{I}\left(g^{-1}(2)\right)$.  

Trivially, the for-loops run in polynomial time. As in each iteration of the while-loop starting in Line \ref{alg_whileloop} we delete a vertex in $M$ and never put this vertex in to it again, this while-loop needs only polynomial time. Thus, Algorithm~\autoref{alg:tau_sur} runs in polynomial time. 
\end{pf}

\noindent
This implies the following corollary. 

\begin{corollary}
\textsc{Ext-RHF} with surjective $\tau$ is polynomial-time solvable.
\end{corollary}

Let $G=(V,E)$ be a graph. Recall the closed-neighborhood hypergraph $G_{nb}$  associated to~$G$, i.e., $G_{nb}=(V,(N[v])_{v\in V})$. If we compare this algorithm on $G_{nb}$ with Algorithm~1 in~\cite{AbuFerMan2022a} on $G$, then these algorithms would work the same. Since the correspondence~$\tau$ is bijective, $\tau(x)\neq \tau(y)$ holds for each $x,y\in X$ with $x\neq y$. Therefore, the for-loop starting in Line \ref{alg_tau_not_injective} would not change anything. In the while loop we go through all $x\in M_2$ and look for $y\in \tau^{-1}(\mathbf{I}(x))\cap M_1$. The very definition of $\hat{S}$ implies that we update the value of such a $y\in N[x]$ to $f(y)=2$. This also happens in the ExtRDF-algorithm of~\cite{AbuFerMan2022a}. In the next step, both algorithms check if each vertex with value 2 has private edge which is not the $\tau$-hyperedge (private neighbor that is not the itself). In the end Algorithm~\ref{alg:tau_sur} takes an index $i$ not hit and sets $f(x)=1$ for one vertex $x\in \tau^{-1}(i)$. As $\tau$ is injective, the vertex~$x$ is unique. Therefore, both algorithm would work the same.   
Our presented algorithm hence generalizes the polynomial-time solvability of \ERD.  



\section{Bounded \textsc{Extension Roman Domination}}
\label{sec:bounded-extension-RD}
 
In this section, we will discuss a two-sided bounded version of \textsc{Extension Roman Domination} which was also suggested by a colleague of ours.
 
\noindent
\centerline{\fbox{\begin{minipage}{.96\textwidth}
\textbf{Problem name: } Bounded \textsc{Extension Roman Domination}, or \textsc{bounded-ExtRD} for short\\
\textbf{Given: } A graph $G=\left( V,E\right)$ and functions $f,h\in \lbrace 0,1,2 \rbrace^V.$\\
\textbf{Question: } Is there a minimal rdf $g \in \lbrace 0,1,2\rbrace^V$ with $f\leq g \leq h$?
\end{minipage}
}}

\smallskip
\noindent
Namely, as we will show in \autoref{cor:bounded-ExtRD-NP}, \textsc{bounded-ExtRD} is \NP-complete. A standard way to act in such circumstances is to look for \FPT-algorithms.  
One natural parameterization for this problem could be $\weightfunction{2-h}$, because for $2 - h = 0$, i.e., $\omega(2 - h)=0$, we are back to \textsc{ExtRD} as a special case, which is known to be solvable in polynomial time. Hence, this parameterization can be viewed as a `distance-from-triviality' parameter~\cite{GuoHufNie2004}. 
However, as we will prove in \autoref{thm:Bounded ExtRD W1}, this parameterization strategy fails.
We employ the well-known \W{1}-completeness of \textsc{Independent Set}, parameterized by the size of the solution, in the reduction presented in the proof of \autoref{thm:Bounded ExtRD W1}.
In that theorem, we actually discuss a slightly different parameterization, namely $\kappa_{h^{-1}(0)}(G,f,h)\coloneqq\vert h^{-1}(0)\vert$.

\begin{theorem}\label{thm:Bounded ExtRD W1}
$\kappa_{h^{-1}(0)}$-bounded-\textsc{ExtRD} is \W{1}-hard even on bipartite graphs.
\end{theorem}

\begin{pf}
Let $G=(V,E)$ be a graph and $k\in \mathbb{N}$. Define $G'=(V',E')$ with 
\begin{equation*}
    \begin{split}
        V' \coloneqq&\, \{a,b\} \cup \{ w_e, u_{e, x}, u_{e, y} \mid e = \{x, y\} \in E\} \cup \{r_i \mid i \in \left[ k \right]\} \cup\{s_v \mid v \in V\} \\
        &\cup \{v_i,t_{v, i} \mid v\in V, i\in \left[ k \right]\},\\
        E' \coloneqq&\, \{\{u_{e,x},w_e\}, \{ u_{e,y},w_e\}, \{u_{e,x}, x_i\}, \{u_{e,y}, y_i\} \mid e=\{x,y\}\in E, i\in \left[ k \right]\}\\
        &\cup \{\{t_{v,i},s_v\},\{t_{v,j}, v_i\}, \{a,v_i\},\{r_i, v_i\}\mid v\in V,\, i,j\in \left[k\right],\, i\neq j\}\cup \{\{a,b\}\}.
    \end{split}
\end{equation*}

\begin{figure}
    \centering
    	
\begin{subfigure}[b]{.45\textwidth}
    \centering
    	
	\begin{tikzpicture}[transform shape,fill lower half/.style={path picture={\fill[#1] (path picture bounding box.south west) rectangle (path picture bounding box.east);}}]
			\tikzset{every node/.style={ fill = black,circle,minimum size=0.1cm}}
			\node[draw,fill=white,label={left:$t_{v,1}$}] (t1) at (2,-2.) {}; 
			\node[draw,fill=white,label={left:$t_{v,i-1}$}] (tim) at (2,-0.75) {}; 
			\node[draw,fill=white,label={left:$t_{v,i+1}$}] (tip) at (2,0.75) {}; 
			\node[draw,fill=white,label={left:$t_{i}$}] (ti) at (2,0) {}; 
			\node[draw,fill=white,label={left:$t_{v,k}$}] (tk) at (2,2) {}; 
			\node[draw,fill=white,label={above:$b$}] (b) at (-1.5,0) {}; 
			\node[draw,fill=white,label={above:$a$}] (a) at (-0.75,0) {}; 
			\node[draw,fill=white,label={above:$v_i$}] (vi) at (0,0) {}; 
			\node[draw,fill=white,label={below:$r_i$}] (ri) at (0,-0.75) {}; 
			\node[draw,fill=white,label={right:$s_v$}] (sv) at (3.5,0) {}; 
			\node[fill=none] at (2,1.45) {\vdots}; 
			\node[fill=none] at (2,-1.25) {\vdots};
			
			\path (sv) edge (t1);
			\path (sv) edge (ti);
			\path (sv) edge (tip);
			\path (sv) edge (tim);
			\path (sv) edge (tk);
			\path (vi) edge (t1);
			\path (vi) edge (tip);
			\path (vi) edge (tim);
			\path (vi) edge (tk);
			\path (b) edge (a);
			\path (a) edge (vi);
			\path (ri) edge (vi);
        \end{tikzpicture}

    \subcaption{Construction for $v\in V$ and $i\in [k]$.}
    \label{fig:boundedExtrdf_vertex_gadget}
\end{subfigure}
\quad 
\begin{subfigure}[b]{.45\textwidth}
\centering
	\begin{tikzpicture}[transform shape]

			\tikzset{every node/.style={ fill = white,circle,minimum size=0.1cm}}
			\node[draw,fill=white, label={left:$y_1$}] (y1) at (-2,0.5) {}; 
			\node[draw,fill=white, label={left:$y_k$}] (yk) at (-2,-0.5) {}; 
			\node[draw,fill=white, label={right:$x_1$}] (x1) at (2,0.5) {}; 
			\node[draw,fill=white, label={right:$x_k$}] (xk) at (2,-0.5) {}; 
			\node[draw,fill=white, label={above:$u_{e,x}$}] (uex) at (1,0) {}; 
			\node[draw,fill=white, label={above:$u_{e,y}$}] (uey) at (-1,0) {};
			\node[draw,fill=white, label={above:$w_e$}] (we) at (0,0){};
			\node[fill=none] at (-2,0.1) {\vdots};
   		\node[fill=none] at (2,0.1) {\vdots};
   		\node[fill=none] at (0,2) {};
   		\node[fill=none] at (0,-2) {};
			\path (y1) edge (uey);
        	\path (yk) edge (uey);
			\path (x1) edge (uex);
        	\path (xk) edge (uex);
			\path (we) edge (uey);
			\path (we) edge (uex);
        \end{tikzpicture}

    \subcaption{Construction for $e=\{x,y\}\in E$.}
    \label{fig:boundedExtrdf_edge_gadget}
\end{subfigure}
    \caption{Gadgets for \autoref{thm:Bounded ExtRD W1}}
\end{figure}
\autoref{fig:boundedExtrdf_vertex_gadget} and \autoref{fig:boundedExtrdf_edge_gadget} gives an idea of how parts of the graph could look like.
Furthermore, we need the maps $f,h\in \{0,1,2\}^{V'}$ with $f(z)=2$ for $z\in \{a\}\cup \{ w_e \mid e \in E\} \cup\{s_v\mid v\in V\}$, $f(z)=0$ otherwise, and $h(z) = 0$ for $z\in \{r_i \mid i\in [k]\}$, $h(z)=2$ otherwise.

Since $f(a)=2$, all vertices in $\{v_i\mid i\in [k],\, v\in V\}$ are dominated, as well as $a,b$. For each $e=\{x,y\}\in E$, $w_e$ dominates $w_e,u_{e,x},u_{e,y}$. Furthermore,  $s_v,t_{v,1},\ldots,t_{v,k}$ are dominated by~$s_v$ for $v\in V$. This leaves to dominate the set $\{r_i\mid i\in [k]\}$. As the neighborhoods are pairwise disjoint, for each $i\in [k]$, exactly one vertex $v_i\in N(r_i)=\{v_i\mid v\in V\}$ must have the value~$2$ in a rdf $g$ with $f\leq g$. This also implies that $f(z)=g(z)$ for all $z\in V'\setminus\{v_i\mid i\in [k],\, v\in V\}$, as $z$ would not have a private neighbor (or $h(z)=0$).  
\begin{claim}
There is an independent set $I$ of size $k$ \iffl there exists a minimal rdf $g$ on $G'$ with $f\leq g\leq h$.
\end{claim}

\begin{pfclaim}
Let $I=\{p^1,\ldots, p^k\}$ be an independent set with $p^i\neq p^j$ for $ i,j\in [k]$ with $i \neq j$. Define $g:\: V'\rightarrow\{0,1,2\}$ with 
$$z\mapsto 
\begin{cases}
2, & z\in \{a\}\cup \{ w_e \mid e \in E\} \cup\{s_v\mid v\in V\}\cup \{p^i_i\mid i\in [k] \} \\
0, & \text{otherwise}
\end{cases}. $$

With the argument above, $g$ is a rdf. Furthermore, $f \leq g \leq h$ holds. Assume $g$ is not minimal. As $g$ is a rdf and there exists no $z\in V'$ with $g(z)=1$, there has to be a $z\in V'$ such that $N(z)\setminus N[g^{-1}(2)\setminus \{z\}]$ empty. Since $p^i\neq p^j$ for $i,j\in[k]$ with $i \neq j$, $t_{p^i,i}$ is a private neighbor of $s_{p^i_i}$. For each $i\in [k]$, $r_i$ is a private neighbor of $p^i_i$. Furthermore, $b$ is only dominated by $a$. Assume there exists an $e=\{x,y\}\in E$ such that $w_e$ has no private neighbor but itself. This implies $u_{e,x}$ and $u_{e,y}$ are dominated by some $p^j_j\in\{x_i\mid i\in [k]\}\cap f^{-1}(2)$ and $p^l_l\in \{x_i\mid i\in [k]\}\cap f^{-1}(2)$. This contradicts the independence of~$I$. Therefore, $g$ is a minimal.

To show the converse, let $g$ be a minimal rdf on $G'$ with $f\leq g \leq h$. As mentioned before, $g^{-1}(2)\setminus f^{-1}(2)=\{d^1_1,\ldots,d^k_k\}$ where $d^i \in V$ for $i\in [k]$. Define $I=\{d^1,\ldots,d^k\}$. Assume there exist $i,j\in [k]$ with $i\neq j$ and $d^i=d^j$. $N(s_{d^i})\subseteq N(\{d^i_i, d^j_i\})$ implies that $g$ was not minimal in the first place. To show that $I$ is independent, we assume there exist $d^i,d^j\in I$ such that $\{d^i,d^j\}=e\in E$. This contradicts minimality, as $w_e$ has no private neighbor but itself ($u_{e,d^i}\in N(d^i)$ and $u_{e,d^j}\in N(d^j)$). Therefore, $I$ is an independent set of size~$k$.
\end{pfclaim}
This concludes the proof of the validity of the reduction. Clearly, the reduction can be computed in polynomial time. $G'$ is also bipartite with the classes $A=\{a\} \cup \{u_{e,x}\mid x\in e\in E\} \cup \{t_{v,i},r_i \mid v\in V,\, i\in [k]\}$ and $B=\{b\}\cup \{w_e\mid e\in E\}\cup \{s_v, v_i \mid v\in V,\, i\in [k]\}$. These classes can be checked in the definition of $E$, as in each edge the vertex of $A$ is mentioned first and the vertex of $B$ is mentioned second.
\end{pf} 

Since this reduction is also a polynomial-time reduction and since membership in~$\NP$ is easily seen using guess-and-check, we can conclude:

\begin{corollary}\label{cor:bounded-ExtRD-NP}
\textsc{bounded-ExtRD} is \NP-complete, even on bipartite graphs.
\end{corollary}

Since $h$ maps no vertex to~$1$ in the reduction presented in the proof of \autoref{thm:Bounded ExtRD W1}, bounded-\textsc{Ext-RDF} is \W{1}-hard, parameterized by $\kappa_{2-h}(G,f,h)\coloneqq\sum_{v\in V}(2-h(v))$. Namely, in the construction  of \autoref{thm:Bounded ExtRD W1}, $\kappa_{2-h}(G,f,h) = 2 \cdot \kappa_{\vert h^{-1}(0) \vert} (G,f,h)$.  

Another parameterization could be $\weightfunction{f}$: If $\weightfunction{f} = 0$, there is a minimal rdf $g\in \{0,1,2\}^V$ with $f \leq g \leq h$ \iffl $h$ is a rdf (this can be checked in polnomial time). If there exists such a $g$, then $h$ is also a rdf. If $h$ is a rdf, then we can decrease the value of the vertices until we can no longer decrease the value of any vertices without losing the rdf property.    

To understand the complexity of this parameter, we need the following extension version of \textsc{Hitting Set}.

\noindent
\centerline{\fbox{\begin{minipage}{.96\textwidth}
\textbf{Problem name: }\textsc{Extension Hitting Set}, or \textsc{ExtHS} for short\\
\textbf{Given: } A simple hypergraph $H=\left( X,S\right)$, $S\subseteq 2^X$, and a set $U\subseteq X$.\\
\textbf{Question: } Is there a minimal hitting set $T\subseteq X$ with $U \subseteq T$?
\end{minipage}
}} 

\smallskip
\noindent
In~\cite{BlaFLMS2019}, it was proven that \textsc{ExtHS} is \W{3}-complete when parameterized by $|U|$.

\begin{theorem}\label{thm:Bounded ExtRD W3}
\textsc{bounded-ExtRD} is \W{3}-hard on split graphs when parameterized by the weight $\weightfun(f)$ of the lower-bound function~$f$. 
\end{theorem}

\begin{pf}
Let $(X,S,U)$ with $X=\{x_1,\ldots,x_n\}$, $S=\{S_1,\ldots, S_m\}\subseteq 2^X$ and $U\neq \emptyset$ be an instance of \textsc{ExtHS}. Define $G=(V,E)$ with 
\begin{equation*}
    \begin{split}
        V &\coloneqq \{c_1,\ldots,c_n\}\cup\{t_1,\ldots,t_m\},\\
        E &\coloneqq \{\{c_i,t_j\} \mid x_i\in S_j\} \cup \binom{\{c_1,\ldots,c_n\}}{2}.
    \end{split}
\end{equation*}
Obviously, $G$ is a split graph. Let  $T\coloneqq\{t_1,\ldots, t_m\}$, $C\coloneqq \{c_1,\ldots,c_n\}$  and $\varphi:X\to C, x_i\mapsto c_i$. Define the functions $f,h\in V^{\{0,1,2\}}$ with $f(t) = f(c) = h(t) = 0$, $f(u) = h(c) = h(u) = 2$, for $u\in \varphi(U), \, t\in T,\, c\in C\setminus \varphi(U)$. This implies $\sum_{v\in V} f(v)= 2\cdot \vert U\vert.$ Notice that $f(v)=2$ \iffl $\varphi^{-1}(v)\in U$.

Let $H$ be a minimal hitting set with $U\subseteq H$. Define
$$ g_H: V\rightarrow \{0,1,2\}, x\rightarrow \begin{cases}0,& x\in T \cup \{c_i\in C\mid x_i\notin H\}\\
2, & x\in \{c_i\in C\mid x_i\in H\}=\varphi(H)
\end{cases}.$$
$U\subseteq H$ and $g_H(T)=\{0\}$ imply $f(v)\leq g_H(v) \leq h(v)$ for each $v\in V$.
Since $U$ is not empty, $g_H$ already dominates $C$. As $H$ is a hitting set, for each $s_j\in S$, there exists an $x_i\in S_j\cap H$. Therefore, for each $t_j\in T$, there exists a $c_i\in N(t_j)$ with $g_H(c_i)=2$. Hence, $g_h$ is a rdf. Assume there exists a $c_i\in C$ with $g_H(c_i)=2$ such that $g_H- \chi_{\{c_i\}}$ is also a rdf. This implies that for each $t_j\in N(c_i)$ there exists a $c_q\in N(t_j)\cap g_H^{-1}(2)$. Hence, for each $s_j\in S$ with $x_i\in s_j$, there exists an $x_q\in (s_j\cap H)\setminus \{x_i\}$. This contradicts the minimality of $H$. Therefore, $g_H$ is a minimal rdf.  Thus,  $(G,f,h)$ is a \yes-instance of \textsc{bounded-ExtRD} if $(X,S,U)$ is a \yes-instance of \textsc{ExtHS}.

Let $g$ be a minimal rdf on $G$ with $f\leq g \leq h$. Define $H_g =\varphi^{-1}(C\cap g^{-1}(2)) = \{x_i \mid c_i\in C \land  g(c_i)=2\}$. $U$ is a subset of $H_g$, as $f(v)\leq g(v)$ holds for each $v\in V$. Since $g(t_j)=0$ for each $t_j\in T$, there has to exist a $c_i\in N(t_j)$ such that $g(c_i)=2$. Therefore, $H_g$ is a hitting set. Assume there exists an $x_i \in H_g$ such that for each $ S_j\in S$ with $x_i\in S_j$ there exists an $x_q\in H_g\subseteq S_j$. Therefore, for each $t_j\in N(x_i) \cap T$, there is a $c_q\in N(t_i)\setminus\{ x_i \}$ with $g(c_q)=2$. The existence of such a~$c_q$ implies that the private neighbor of $c_i$ cannot be in~$C$. This is a contradiction to the minimality of~$g$.  Thus,  $(X,S,U)$ is a \yes-instance of \textsc{ExtHS}  if $(G,f,h)$ is a \yes-instance of \textsc{bounded-ExtRD}.
\end{pf}

Since this reduction is also a polynomial-time reduction and since membership in~$\NP$ is easily seen using guess-and-check, we can conclude:

\begin{corollary}\label{cor:bounded-ExtRD-NP-split}
\textsc{bounded-ExtRD} is \NP-complete on split graphs.
\end{corollary}

\begin{remark}
This result is interesting, as we can show that \textsc{Extension Dominating Set} is polynomial-time solvable on split graphs. As mentioned above, \textsc{Ext RD} can be solved in polynomial time but \textsc{Ext DS} is \NP-complete. Now we have a graph class for which bounded-\textsc{Ext RD} is \NP-complete but there is a polynomial-time algorithm for \textsc{Ext DS}.
\end{remark}

Our last proof would also work if $T$ would be a clique. This change does not affect the proof, as $h(t)=0$ holds for each $t\in T$. The resulting graph is then co-bipartite. Hence, we can state the following results.

\begin{corollary}
    \textsc{bounded-ExtRD} is \NP-complete on co-bipartite graphs. \textsc{bounded-ExtRD} is \W{3}-hard on co-bipartite graphs when parameterized by the weight $\weightfun(f)$ of the lower-bound function~$f$. 
\end{corollary}

We will make use of the parameterized reduction of \autoref{thm:Bounded ExtRD W3} in the next section, when we turn to discuss the (parameterized) complexity of \textsc{Ext RHF}. 
The reductions provided there will also show that \textsc{bounded-ExtRD}, parameterized by the weight of the lower-bound function, belongs to~\W{3}. Therefore, the $\W{3}$-hardness results proved in this section turn into $\W{3}$-completeness results.

Finally, let us mention that, in any given \textsc{bounded-ExtRD} instance $(G,f,h)$, we can always assume (1) $f\leq h$ and, moreover, (2) $h(v)=0$ implies $h(u)=2$ for some $u\in N(v)$. Otherwise, there cannot exist a rdf $g$ with $f\leq g\leq h$. In particular, if $h(v)=0$, as $g\leq h$, then also $g(v)=0$, and as $g$ should be a rdf, $g(u)=2$ for some $u\in N(v)$, so that $h(u)=2$ as $g\leq h$. Both conditions are easy to check.

\section{Complexity of \textsc{Extension-\longversion{Roman Hitting Function}\shortversion{RHF}} I}
\label{sec:Ext RHF complexity I}
 

In this section, we will show that  there are also instances of \textsc{Roman Hitting Function} which are \NP-complete and \W{3}-complete, considering their standard parameterizations. For the \W{3}-membership, we make use of the problem \textsc{Multicolored Independent Family (MultIndFam)} that is defined next.

\centerline{\fbox{\begin{minipage}{.96\textwidth}
\textbf{Problem name: }\textsc{Multicolored Independent Family}\\
\textbf{Given: } A $(k+1)$-tuple $(S_1,\ldots,S_k,T)$ of subsets of $2^U$
on the common universe $U$, i.e., $(U,S_1),\dots,(U,S_k),(U,T)$ are $k+1$ many simple hypergraphs.\\
\textbf{Question: } Are there hyperedges $s_1\in S_1,\ldots,s_k\in S_k$ such that no $t\in T$ is a subset of $\bigcup_{i=1}^k s_i\subseteq U$?\end{minipage}
}}

This problem is known to be \W{3}-complete when parameterized by~$k\in\mathbb{N}$ (see \cite{BlaFLMS2019,BlaFLMS2022}). As there are no good canonical  \W{3}-complete \longversion{parameterized }problems, we use a reduction to \textsc{MultIndFam} to prove membership in~\W{3}.
Unfortunately, this reduction is quite technical.

\begin{theorem}
standard-\textsc{Extension Roman Hitting Function} is in \W{3}.
\end{theorem}
\begin{pf}
Let  $H=(X,\hat{S}=(s_i)_{i\in I})$ be a (not necessarily simple) hypergraph with correspondence $\tau:X\to I$ and let $f: X\to \{0,1,2\}$ be function, comprising an instance of standard-\ExtRHF. $(*)$ We can assume that there are not two elements $x,y\in X$ such that $f(x)=f(y)=1$ with $\tau(x)=\tau(y)$ or $f(x)=2, f(y)=1$ with $x\in s_{\tau(y)}$. Otherwise, we could use Corollaries~\ref{cor_extensible12} and~\ref{cor_extensible11}. 

We are going to construct an equivalent instance of \textsc{MultIndFam} next. To this end, we define its universe as $$U\coloneqq X\cup\{r_{i,x}, x'\mid x\in f^{-1}(\{1,2\}), i\in I\} \cup \{\tau_x\mid x\in f^{-1}(1)\}\,.$$ For the construction of the hypergraphs, we need to define some additional (auxiliary) sets:
\begin{itemize} \item 
For $x \in f^{-1}(\{1,2\})$, $i\in \mathbf{I}(x)$ abbreviate $\tilde{s}_{x,i}\coloneqq s_i\cup\{r_{i,x},x'\}$. 
 \item 
 Define $t_i\coloneqq s_i\cup \{\tau_x\mid x\in f^{-1}(1)\cap s_i\}$ for $i\in I $ \longversion{such that $\tau^{-1}(i)$ and $s_i\cap f^{-1}(2)$ are empty.}\shortversion{with $\emptyset=\tau^{-1}(i)=s_i\cap f^{-1}(2)$.} 
 \item 
For each $x\in f^{-1}(2)$, let $S_x\coloneqq\{\tilde{s}_{x,i}\mid \tau(x) \neq i \}$ and\\ for each $x\in f^{-1}(1)$, let $S_x\coloneqq\{s_{\tau(x)}\cup \{\tau_x\}\}\cup \{\tilde{s}_{x,i}\mid \tau(x) \neq i\}$. 
 \item Furthermore, we need the target set $T=T'\cup T''$, where
\begin{equation*}
    \begin{split}
        T'\coloneqq\,\,&\{\,t_i\mid i\in I\land \tau^{-1}(i)=\emptyset \land s_i \cap f^{-1}(2)=\emptyset\,\}\quad\text{and}\\
     T'\coloneqq\,\,&\left\{\, \{ r_{i,x},y'\}\mid i\in I\land \{x,y\}\subseteq (s_i \cap f^{-1}(\{1,2\}))\land x\neq y \,\right\}\\ 
        &\cup \left\{\,\{\tau_x,y'\} \mid x\in g^{-1}(1)\land y\in  s_{\tau(x)}\land x\neq y \,\right\}\,.
    \end{split}
\end{equation*}
\end{itemize}

Now we will explain the idea of each element. It is important to keep in mind that we want to use \autoref{t_extrhf_equiv}: 
If $x'$ is in a chosen hyperedge, then we assign the value~2 to~$x$ in the minimal rhf. The element $r_{i,x}$ gives us information about the mapping $\rho$. $r_{i,x}$ is in one of the chosen edges \iffl $\rho(x)=i$ holds. Therefore, the sets $\{r_{i,x},y'\}$ verify the Constraint~\ref{con_privateEdge} of \autoref{t_extrhf_equiv}. $\tau_x$ will only be in a set we chose if we assign the value~1 to~$x$. Hence, Constraint~\ref{con_1VS2} will be checked by the sets $\{\tau_x ,y\}$. The sets in $T'$ correspond to the sets which we consider in Constraint~\ref{con_hitting}. This is also the reason why $\tau_x$ is in included in $t_i$. Since there exists a $\tau_x$, $f(x)=1$. If $x\in R_2$, $s_i\cap R_2\neq \emptyset$. In the \textsc{MultIndFam} instance, this corresponds to: $\tau_x$ will not be in our sets, which implies that $t_i$ will not be covered completely.

\begin{claim}
$(H,\tau,f)$ is a \yes-instance of  \textsc{Ext RHF} \iffl $(U,(S_x)_{x\in f^{-1}(\{1,2\})},T)$ is a \yes-instance of \longversion{the \textsc{Multicolored Independent Family} problem.}\shortversion{\textsc{MultIndFam}}.
\end{claim}

\begin{pfclaim}
First assume that $(H,\tau,f)$ is a \yes-instance of  \textsc{Ext RHF}, i.e., we have to show that $(U,(S_x)_{x\in f^{-1}(\{1,2\})},T)$ is a \yes-instance of \textsc{MultIndFam}. 
To prove this, we will use \autoref{t_extrhf_equiv}, keeping the observation $(*)$ in mind. As $(H,\tau,f)$ is a \yes-instance of  \textsc{Ext RHF}, there exist a set $f^{-1}(2)\subseteq R_2 \subseteq f^{-1}(\{ 1,2 \})$ and a mapping $\rho: R_2\to I$ that fulfill all four constraints of \autoref{t_extrhf_equiv}.  For each $x\in R_2$, choose some $\tilde{s}_{x,\rho(x)}\in S_x$, and for each $x \in f^{-1}(1)\setminus R_2$, choose $s_{\tau(x)}\cup \{\tau_x\}\in S_x$. We will show that this choice of hyperedges from $(S_x)_{x\in f^{-1}(\{1,2\})}$ proves that $(U,(S_x)_{x\in f^{-1}(\{1,2\})},T)$ is a \yes-instance of\shortversion{ the problem} \textsc{MultIndFam}. 
For the sake of contradiction, assume that there exists a hyperedge in~$T$ which is a subset of 
$$\mathcal{S}\coloneqq\left(\bigcup_{x\in R_2}\Tilde{s}_{x,\rho(x)}\right)\cup\left(\bigcup_{x\in f^{-1}(1)\setminus R_2}s_{\tau(x)}\cup \{\tau_x\}\right)\,.$$ 
Observe that, for each $x\in f^{-1}(\{1,2\})$, $x'\in \mathcal{S}$ \iffl there is an $i\in \mathbf{I}(x)$ such that $\tilde{s}_{x,i}$ was chosen. Therefore, $x$ has to be in $R_2$. 

If there would exist a hyperedge in $T''$ which is in included in $\mathcal{S}$, this would contradict  Constraints~\ref{con_tauVSPrivate} or~\ref{con_privateEdge}. Now, consider a hyperedge $t_i$ in $T'$. Hence, we have some $i\in I$ with $t_{i}\subseteq \mathcal{S}$, $s_i\cap f^{-1}(2) = \emptyset$ and $\tau^{-1}(i)=\emptyset$. This implies that $$s_i\subseteq\left(\bigcup_{x\in f^{-1}(1)\setminus R_2}s_{\tau(x)}\right) \cup \left(\bigcup_{x\in R_2} s_{\rho(x)}\right)\,.$$ Furthermore, $ f^{-1}(1) \cap R_2  \cap s_i = \emptyset$, as otherwise, there exists a $\tau_x\in t_i$, which is not in $\mathcal{S}$. This would contradict the Constraint~\ref{con_hitting}. Hence, $(U,(S_x)_{x\in f^{-1}(\{1,2\})}, T)$ is a \yes-instance.

For the if-part, let $q_x\in S_x$ for each $x\in f^{-1}(\{1,2\})$ be a solution for the \textsc{MultIndFam} instance, i.e., their union does not contain any hyperedge from~$T$. Define 
$$R_2  \coloneqq f^{-1}(2)\cup \{ x \in f^{-1}(1) \mid \tau_x\notin q_x\}$$
and, for each $x\in R_2$, determine $\rho(x)$ such that $\tilde{s}_{x,\rho(x)}=q_x$. We again use \autoref{t_extrhf_equiv} and the four constraints mentioned in this characterization theorem. Constraint~\ref{con_tauVSPrivate} is fulfilled by the construction of~$R_2$. The next two constraints are met, as otherwise a set of $T''$ would be included in $\bigcup_{x\in f^{-1}(\{ 1,2 \})} q_x$. Assume there exists a $j\in I$ with $\tau^{-1}(j)=\emptyset$ and 
$$s_j\subseteq \left(\bigcup_{x\in f^{-1}(1)\setminus R_2}s_{\tau(x)}\right) \cup \left( \bigcup_{x\in R_2} s_{\rho(x)}\right)\,.$$
As for each $i\in I$, $t_i$ is no subset of $\bigcup_{x\in f^{-1}(\{1,2\})} q_x$, there must be some $\tau_x\in t_i$, with $ x\in f^{-1}(1)\cap s_j$, which is not in $\bigcup_{x\in f^{-1}(\{1,2\})} q_x$. Thus, $R_2\cap s_i\neq \emptyset.$ Hence, also Constraint~\ref{con_hitting} is met. Therefore, $(H,f,\tau)$ is a \yes-instance  of  \textsc{Ext RHF}.
\end{pfclaim}

\noindent
The following two claims are easy to prove but needed to conclude the proof.
\begin{claim}
Given an instance of $\textsc{Ext RHS}$, the equivalent instance of $\textsc{MultIndFam}$ as described above can be constructed in polynomial time.
\end{claim}

\begin{claim}
The parameter~$k$ of the constructed instance of $\textsc{MultIndFam}$ is bounded by the cardinality of $f^{-1}(\{1,2\})$ and hence by the weight of~$f$, which is the standard parameter of the original instance of $\textsc{Ext RHF}$.
\end{claim}
\noindent
Hence,  \textsc{Ext RHF} belongs to $\W{3}$.
\end{pf}

As mentioned in the previous proof, the described reduction is also a poly\-nomial-time reduction. Therefore, \textsc{Extension Roman Hitting Function} is a member of \NP, but this is also (easily) observed by the guess-and-check characterization of \NP.
For the hardness results, we will use bounded-\textsc{Extension Roman Domination} as described in the previous section.

\begin{theorem}\label{thm:ExtRHF W3}
standard-\textsc{Extension Roman Hitting Function} is \W{3}-hard 
even if the correspondence function is injective.
\end{theorem}

\begin{pf} We will make use of \autoref{thm:Bounded ExtRD W3}, reducing from bounded-\textsc{ExtRD}.
Let $(G,f,h)$ be a instance of the bounded-\textsc{ExtRD}, with $G=(V,E)$. We can assume (1) $f\leq h$ and, moreover, (2) $h(v)=0$ implies $h(u)=2$ for some $u\in N(v)$. We parameterize by $\omega(f)$.
For $v\in X \coloneqq V\setminus h^{-1}(0)$, define $T_v\coloneqq\left(N(v) \setminus h^{-1}(\{0,1\})\right)\cup\{v\}$, and  for $v\in h^{-1}(0)$, define $T_v\coloneqq \left(N(v) \setminus h^{-1}(\{0,1\})\right)$. Further, we set
$\hat{S} \coloneqq (T_v)_{v\in V}$ and we define~$\tau$ as the correspondence satisfying $\tau(v)= v$  and we let $\overline{f} :X \rightarrow \{0,1,2\},\, v \mapsto f(v)$, i.e., $\overline{f}=f|_X$. Let $H=(X,\hat{S})$. Then, $(H,\tau,\overline{f})$ describes an instance of \textsc{Ext RHF}. The standard parameter is the weight of $\overline{f}$ for this instance.  As $f\leq h$, $h(v)=0$ implies $f(v)=0$. Therefore, $\omega(f)=\omega(\overline{f})$, so that the parameter value does not change when moving from the bounded-\textsc{ExtRD} instance to the standard-\ExtRHF instance. Clearly, the described construction can be carried out in polynomial time. Trivially,
the correspondence~$\tau$ is injective. What remains to be shown is that the construction enjoys the reduction property, mapping \yes-instances to \yes-instances and \no-instances to \no-instances.

Let $g$ be a minimal rdf on $G$ with $f \leq g \leq h$. Define $\overline{g} :X \rightarrow \{0,1,2\},\, v \mapsto g(v)$, i.e., $\overline{g}=g|_X$. Thus, $\overline{f}\leq \overline{g}$. Let $v\in V$. For $g(v)\in\{1,2\}$, $T_v$ is hit by~$v$. If $g(v)=0$ holds, there has to exist a $w\in N(v)$ with $g(w)=2$. This $w$ hits~$T_v$. Therefore, $\overline{g}$ is a rhf. 
Assume $\overline{g}$ is not minimal. Since $\tau$ is injective, Constraint~\ref{con_difvalue} of \autoref{t_property_min_rhf} is satisfied. Hence, either Constraint~\ref{con_disjoint}, Constraint~\ref{con_12switch}, or the minimality of Constraint~\ref{con_minHS} of \autoref{t_property_min_rhf} has to be contradicted. If Constraint \ref{con_disjoint} is contradicted, this implies the existence of $v,w\in V$ with $g(v)=\overline{g}(v)=2$, $g(w)=\overline{g}(w)=1$ and $v\in T_w =\left(N(w) \setminus h^{-1}(\{0,1\})\right)\cup\{w\} \subseteq N[w]$. This contradicts the minimality of~$g$, as we can set $g(w)=0$ and still have a rdf.

Consider Constraint~\ref{con_12switch}. Assume there exists a $v\in \overline{g}^{-1}(2)$ such that for each $w\in V\setminus \{v\}$, $T_w\cap \overline{g}^{-1}(2)\neq \{v\}$ holds. Thus, for each $w\in N(v)$ there exists a $u\in (T_w\cap N[w] \cap \overline{g}^{-1}(2))\setminus \{ v \} \subseteq (N[w] \cap g^{-1}(2))\setminus \{v\}$. This would imply that $g$ is not minimal. 

Concerning Constraint~\ref{con_minHS}, consider $\mathcal{T}\coloneqq\{T_v\mid v\in V, \tau^{-1}(v)\cap \overline{g}^{-1}(1)=\emptyset\}=\{(N(v)\cap h^{-1}(2))\cup\{v\}\mid v\in V,\overline{g}(v)\neq 1\}$. If $\overline{g}(v)=2$, then~$v$ trivially hits $(N(v)\cap h^{-1}(2))\cup\{v\}$. If $\overline{g}(v)=0$, then for some $w\in N(v)$, $g(w)=2$, as $g$ is a rdf of~$G$. If $g(w)=2$, then $h(w)=2$, so that then $v$ hits $(N(v)\cap h^{-1}(2))\cup\{v\}$. Hence, $\overline{g}^{-1}(2)$ is a hitting set of~$\mathcal{T}$. Assume that there exists some $z\in \overline{g}^{-1}(2)$ such that $\overline{g}^{-1}(2)\setminus\{z\}$ is also a hitting set of~$\mathcal{T}$. Hence, for each $T_v\in\cal T$ with $z\in T_v$,
there is some~$y\in \overline{g}^{-1}(2)\setminus\{z\}$ with $y\in T_v$. If $v=z$, then $y\in (N(v)\cap h^{-1}(2))$. Hence, $h(y)=2$ and hence $\overline{g}(y)=2$. As $y\in N(z)$ in this case, we could set $g(z)=0$, possibly leading to a smaller rdf. Now assume $v\neq z$. Then, $z\in (N(v)\cap h^{-1}(2))$. Consider $y\in T_v$. If $y=v$, we can again argue that we may set $g(z)=0$, possibly leading to a smaller rdf, because we found some $y\in N(z)$ with $g(y)=2$. If $y\neq v$, then $\{y,z\}\subseteq (N(v)\cap h^{-1}(2))$. As $h(v)=2$ implies $g(v)=2$, also in this third case, we can set $g(z)=0$. In summary, in each case, we can find a  rdf for $G$ that is smaller than~$g$ by setting $g(z)=0$. This contradicts the minimality of~$g$. 
By \autoref{t_property_min_rhf}, $\overline{g}$ is minimal.   

Let $\overline{g}$ be a minimal rhf, a solution to the instance $(H,\tau,\overline{f})$. Define 
$$ g:\: V\rightarrow \{0,1,2\}, v\mapsto \begin{cases} \overline{g}(v), & v\in X\text{, i.e., }h(v)\neq 0 \\ 0,& v\notin X\text{, i.e., }h(v)=0\end{cases}\,.$$ 
As $\overline{g}$ extends $\overline{f}$, we have 
$\overline{f}\leq \overline{g}$. By construction, $f\leq h$, so that $f(v)=0$ for $v\notin X$. Hence, $f\leq g$. Consider some $y\in h^{-1}(1)$.
We show that then, $g(y)\leq 1$, which proves $g\leq h$. Since $y\in h^{-1}(1)$, $y$ is only contained in the set $T_y=T_{\tau(y)}$ among all sets in~$\hat S$. If $g(y)=2$, i.e., $\overline{g}(y)=2$,  this would contradict \autoref{t_property_min_rhf}, Constraint~\ref{con_12switch}.
We now prove that $g$ is a rdf. Consider some  $v\in g^{-1}(0)=\overline{g}^{-1}(0)\cup h^{-1}(0)$.
If $v\in h^{-1}(0)$, there has to exist a $u\in \overline{g}^{-1}(2) \cap T_v \subseteq g^{-1}(2) \cap N[v]$, as there is no $u\in X$ with $\tau(u)=v$.  If $v\in \overline{g}^{-1}(0)$, there is no vertex in $\tau^{-1}(v) \cap \overline{g}^{-1}(1)$, as $\tau(v)=v$. Therefore, there exists a  $u\in \overline{g}^{-1}(2) \cap T_v \subseteq g^{-1}(2)\cap N[v]$.

Next, we argue why $g$ is minimal. 
Assume there exist $v,w\in V$ with $\overline{g}(v)=g(v)=2$, $\overline{g}(w)=g(w)=1$ and $\{v,w\}\in E$, violating the first property of \autoref{t_property_min_rdf}. This would contradict the minimality of $\overline{g}$, as $v\in \left(N(w) \setminus h^{-1}(\{0,1\})\right)\cup\{w\} =T_w$, because  $g(v)=2$ implies $h(v)=2$ by $g\leq h$. Assume that there exists a $v\in g^{-1}(2) = \overline{g}^{-1}(2)$ with $N[v]\setminus N[g^{-1}(2) \setminus \{v\}]\subseteq \{v\}$, violating the privacy property of \autoref{t_property_min_rdf}. This implies that, for each $w\in N(v)$, there is a $t\in (N[w]\cap \overline{g}^{-1}(2))\subseteq (N(w)\setminus h^{-1}(\{0,1\}))\cup \{ w \}$ with $t\neq v$. 
This contradicts Constraint~\ref{con_12switch} of \autoref{t_property_min_rhf}.  Consider
$G'\coloneqq G\left[ g^{-1}(0)\cup g^{-1}(2)\right]$. As~$g$ is a rdf,  $D\coloneqq g^{-1}(2)$ is a dominating set of $G'$. If $D$ is not a minimal dominating set (hence violating the last constraint of \autoref{t_property_min_rdf}), then there must be some $v$ with $g(v)=2$ such that $D\setminus \{v\}$ is also a dominating set. Hence, for each $u\in N[v]$, there is some $v_u\in N[u]\cap (D\setminus \{v\})$. By construction, for each such $v_u$, we have $g(v_u)=2$ and hence $\overline{g}(v_u)=h(v_u)=2$ and $T_{v_u}\neq \{v_u\}$ (*). On the other hand, as $\overline{g}$ is a minimal rhf, by the last constraint of \autoref{t_property_min_rhf}, $D=g^{-1}(2)=\overline{g}^{-1}(2)$ is a minimal hitting set of $\mathcal{T}\coloneqq\{T_x\mid x\in V\setminus \overline{g}^{-1}(1)\}$. As $T_x\subseteq N[x]$ for all $x\in V$,  $(D\setminus \{v\})$ is not only a  dominating set of $G'$ but also a hitting set of  $\mathcal{T}$ because of  (*), contradicting mininality as stated in \autoref{t_property_min_rhf}.
As all possible violations of the characterization for minimal rdf as formulated in of \autoref{t_property_min_rdf} have been shown to contradict our assumptions, we must conclude that $g$ is indeed a minimal rdf. 
\end{pf}

\noindent
As this is also a  polynomial-time reduction, it implies following corollary.

\begin{corollary}
\textsc{Extension Roman Hitting Function} is \NP-complete.
\end{corollary}

\section{Complexity of \textsc{Extension-\longversion{Roman Hitting Function}\shortversion{RHF}} II}
\label{sec:Ext RHF complexity II}

We know already  that \textsc{Extension Roman Hitting Function} is polynomial-time solvable if the correspondence function is surjective. Since \textsc{Extension Roman Hitting Function} with injective correspondence function is as hard as the general \textsc{Hitting Set} problem, this leads to the question if  $\kappa_1(\mathcal{I})=\vert \{i\in I \vert  \tau^{-1}(i)=\emptyset\}\vert$ could be a good parameter for this problem for each instance $\mathcal{I}=(H,\tau,f)$ with $H=(V,(s_i)_{i\in I})$, $\tau:V\to I$, $f:V\to\{0,1,2\}$, somehow measuring the \emph{distance from triviality} again. In other words, we try to use parameterized complexity as a way to study the phenomenon that classical function properties as surjectivity seem to be crucial for finding polynomial-time algorithms for \textsc{Extension Roman Hitting Function}.

\begin{theorem}
$\kappa_1$-\ExtRHF is \W{1}-hard.
\end{theorem}

For the proof we use the fact that the reduction in the proof of \autoref{thm:ExtRHF W3} is also a \FPT-reduction from $\kappa_1$-\ExtRHF to $\kappa_{h^{-1}(0)}$-bounded-\textsc{ExtRDF}, with $\kappa_{h^{-1}(0)}(G,f,h)\coloneqq\vert h^{-1}(0)\vert$. 

\begin{lemma}
$\kappa_1$-\ExtRHF, $\kappa_{ h^{-1}(0) }$-bounded-\textsc{Ext-RDF}, $\kappa_{2-h}$-bounded-\textsc{Ext-RDF}${}\in\XP$.
\end{lemma}
\begin{pf}
For $\kappa_1$-\ExtRHF and $\kappa_{h^{-1}(0) }$-bounded-\textsc{Ext-RDF}, it is enough to show that $\kappa_1$-$\textsc{Ext-RHF}\in\XP$. Let $(H,\tau,f)$ be an instance with $H=(X,\hat{S})$, $\hat{S}=(s_i)_{i\in I}$.  This \XP-algorithm is not difficult, as we branch over each index in $\{i\in I \mid \tau^{-1}(i)=\emptyset \:\}$ and try to hit the edge $s_i$ with an element from~$X$. If each edge is hit, we check with Algorithm \ref{alg:tau_sur} if there exists a minimal rhf which is bigger than this function. If Algorithm \ref{alg:tau_sur} returns \yes, we also can return \yes. Otherwise, we try another way to hit all the edges in $S\setminus\tau(X)$.
This algorithm does run in time $\mathcal{O}(n^ {\kappa_1(H,\tau,f)})$.

Now we consider $\kappa_{2-h}$-bounded-\textsc{Ext-RDF}. Let $(G,f,h)$ be an instance with $G=(V,E)$ with $f\leq h$. First we want to set $g(x)\coloneqq2$ for each vertex  $x\in N(f^{-1}(2)) \cap f^{-1}(1)$. However, if the set $N(f^{-1}(2)) \cap f^{-1}(1)\cap h^{-1}(\{0,1\})$ is non-empty, we face a trivial \no-instance, as for the solution $g$ we are looking for, any vertex~$x$ with $f(x)=2$ satisfies $g(x)=2$ and moreover, a vertex $y\in N(x)$ with $f(y)=1$ must be set to~$2$ by~$g$, which is ruled out if $h(y)\in\{0,1\}$. As $f\leq g\leq h$, we can set $g(x)\coloneqq 1$ for $x\in f^{-1}(1)\cap h^{-1}(1)$. After these settings, the \XP-algorithm branches on each vertex in $h^{-1}(0)$ and tries to dominate it by a neighbor in $h^{-1}(2)\setminus N(f^{-1}(1)\cap h^{-1}(1))$. As $2 \cdot \vert h^{-1}(0) \vert \leq \weightfunction{2-h}$, in total we have at most $n^{\frac{\weightfunction{2-h}}{2}}$ possibilities to do this. After each of these possible settings for~$g$, where we define $g(x)\coloneqq f(x)$ for any hitherto undefined vertex~$x$, we run the polynomial-time algorithm for \textsf{Ext-RDF} from \cite{AbuFerMan2022b}, with $g$ as the lower-bound function. This algorithm will work, as each vertex in $h^{-1}(0)$ is dominated (the undominated vertices are the ones which are set to~$1$) and there is no vertex in $N(f^{-1}(2)) \cap f^{-1}(1)\cap h^{-1}(1)$.
\end{pf}

We are discussing several other parameterizations in the following that either lead to  \paraNP-hardness results (in \autoref{t_paraNP_EXTRHF}) or to \FPT-results (in \autoref{t_FPT_EXTRHF}).
The parameterization function is denoted by $\kappa_Y$, where $Y$ specifies a subset of the universe~$X$ of the instance $(X,\hat S,f)$, such that $\kappa_Y(X,\hat S,f)=\vert Y\vert$.

\begin{theorem}\label{t_paraNP_EXTRHF}
$\kappa_\zeta$-\ExtRHF is \paraNP-hard for each parameterization described by 
$\zeta\in\{ f^{-1}(0),  f^{-1}(1),  f^{-1}(2),  f^{-1}(\{0,2\}) \}$.
\end{theorem}

\begin{pf}
For all the reductions that we show in this proof, we use the \NP-com\-plete\-ness of \textsc{Extension Hitting Set}. Let $H=(X,\hat S=(s_i)_{i\in I})$ be a hypergraph with $U\subseteq X$, $I\cap X = \emptyset$ and an element $a,i_b,i_c\notin X\cup I$. We define $X'=X \cup \{a\}$, $I'\coloneqq I \cup X$ and $I''= I\cup X\cup \{i_b, i_c\}$. For each parameterization, we prove that even if the parameter takes values zero or one and is hence fixed, each of the obtained problems can be used to solve arbitrary instances of \textsc{Extension Hitting Set}.

First, we discuss $\zeta= f^{-1}(0)$. Define the hypergraph $H'= (X,\hat{S'}=(s_i')_{i\in I'})$. Furthermore, we need $\tau: X\to I', x\mapsto  x$, and for $i\in I'$, we set
$$  s_i'\coloneqq\begin{cases}
    s_i, & i\in I\\
    \{ i \}, & i\in X
\end{cases} \quad\text{and}\quad f: X\to \{0,1,2\}, x\mapsto  \begin{cases}
    2, & x\in U\\
    1, & x\in X \setminus U 
\end{cases}.$$ 
This implies $\vert f^{-1}(0)\vert =0$, i.e., the parameter value is constant.
\begin{claim}
    $(H,U)$ is a \yes-instance of \textsc{Extension Hitting Set} \iffl $(H',\tau,f)$ is a \yes-instance of \textsc{Extension Roman Hitting Function}.
\end{claim}

\begin{pfclaim}
First, we assume there is a minimal rhf $g$ on $H', \tau$ with $f \leq g$. Let $D= g^{-1}(2)$. As $\tau(i)=\emptyset$ for each $i\in I$, $s_i\cap D$ is not empty and $D$ a hitting set. If there exists an $x\in D$  such that for each $i\in \mathbf{I}(x)$, there is a $y\in s_i\cap (D\setminus \{x\})$, then this would contradict Constraint~\ref{con_12switch} of \autoref{t_property_min_rhf}. Hence, $D$ is a minimal hitting set. 

Assume there is a minimal hitting set $D\subseteq X$ on $H$ with $U\subseteq D$. Define $g: X\to \{0,1,2\}$ with $g^{-1}(2)=D$ and $g^{-1}(1) = X \setminus D$. Trivially, $f\leq g$ holds. We will use \autoref{t_property_min_rhf} to show that $g$ is a minimal rhf. The definition of $\tau$ implies the correctness of the first two constraints, i.e., Constraints~\ref{con_difvalue} and~\ref{con_disjoint}. As $D$ is a minimal hitting set on $H$ and each hyperedge of $H$ is also a hyperedge of $H'$, Constraint \ref{con_12switch} also holds. The remaining constraint holds as $X=f^{-1}(\{1,2\})= \tau(f^{-1}(\{1,2\}))$ and $D$ is a minimal hitting set on $H$. By \autoref{t_property_min_rhf}, $g$ is a minimal rhf.
\end{pfclaim}

Next, we take a look at the parameter described by $\vert f^{-1}(1)\vert$. Define the hypergraph $H''= (X',(s_i'')_{i\in I''})$. Furthermore,  we need 
$$\tau': X'\to I'', x\mapsto \begin{cases}
    x, & x\in X\\
    i_b, & x=a
\end{cases}, \quad s_i''\coloneqq\begin{cases}
    s_i, & i\in I\\
    \{ a, i \}, & i\in X\\
    X', & i=i_b\\
    \{ a \}, & i=i_c\\
\end{cases}$$ 
for $i\in I''$. Define $f': X'\to \{0,1,2\}$ with $f'^{-1}(2)=U\cup \{a\}$ and $f^{-1}(0)=X\setminus U$. Thus $\vert f'^{-1}(1)\vert =0$, i.e., the parameter value is constant.
\begin{claim}
    $(H,U)$ is a \yes-instance of \textsc{Extension Hitting Set} \iffl $(H'',\tau',f')$ is a \yes-instance of \textsc{Extension Roman Hitting Function}.
\end{claim}
\begin{pfclaim}
First, we assume that there is a minimal rhf $g$ on $H'', \tau$ with $f' \leq g$. Define $D\coloneqq g^{-1}(2)\setminus \{a\}$. As $\tau'^{-1}(i)=\emptyset$ for each $i\in I$, $s_i\cap D$ is not empty and $D$ a hitting set on $H$. If there exists a $x\in D$  such that for each $i\in \mathbf{I}(x)$ there is a $y\in s_i\cap (D\setminus \{x\})$, then it would contradict Constraint \ref{con_12switch} of \autoref{t_property_min_rhf}. Hence $D$ is a minimal hitting set. 

Assume there is a minimal hitting set $D\subseteq X$ on $H$ with $U\subseteq D$. Define $g: \{0,1,2\}^X$ with $g^{-1}(2) = D \cup \{a\}$ and $g^{-1}(0) = X \setminus D$. Trivially, $f'\leq g$ holds. We will use \autoref{t_property_min_rhf} to show that $g$ is a minimal rhf. Since $g^{-1}(1)$ is empty, the first two constraints, i.e., Constraints~\ref{con_difvalue} and~\ref{con_disjoint}, are fulfilled. As $D$ is a minimal hitting set on $H$ and each hyperedge of $H$ is also a hyperedge of $H'$, Constraint \ref{con_12switch} also holds. For the last constraint we have to consider $i\in I$, only, since $g(a)=2$. From the fact that $D$ is a minimal hitting set on $H$ and $s''_{i_c}$ is only hit by $a$, the last constraint of \autoref{t_property_min_rhf} follows. Thus, $g$ is a minimal rhf.
\end{pfclaim}


The cardinality $\vert f^{-1}(2)\vert$ describes the next parameter that we will consider. We will reuse $H''$ and $ \tau'$. Furthermore, we need 
$$f'': X'\to \{0,1,2\}, x\mapsto \begin{cases}
    2, & x=a\\
    1, & x\in U\\
    0, & x\in X\setminus U
\end{cases}.$$ 
Thus, $\vert f''^{-1}(2)\vert =1$, i.e., the parameter value is constant. 
\begin{claim}
    $(H,U)$ is a \yes-instance of \textsc{Extension Hitting Set} \iffl $(H'',\tau',f'')$ is a \yes-instance of \textsc{Extension Roman Hitting Function}.
\end{claim}

\begin{pfclaim}
We use \autoref{cor_extensible12} exhaustively. This implies $(H'',\tau',f'')$ is a \yes-instance  of \textsc{Extension Roman Hitting Function} \iffl  $(H'',\tau',\tilde{f})$ with $\tilde{f}(x)=2$ for $x\in U\cup \{a\}$ and $\tilde{f}(x)=0$ for $x\in X \setminus U$ is a \yes-instance. Since $\tilde{f}=f'$, this case follows by the last one.
\end{pfclaim}


The last case is the parameter $\kappa_\zeta$ with $\zeta= f^{-1}(\{0,2\})$. Define the hypergraph $H'''= (X',(s_i''')_{i\in I''})$. Again, we reuse $\tau'$. Furthermore, we need 
$$ f''': X'\to \{0,1,2\}, x \mapsto \begin{cases}
    2, & x=a\\
    1, & x \in X
\end{cases}, \quad s_i'''\coloneqq\begin{cases}
    s_i, & i\in I\\
    \{a,i\}, & i\in U\\
    \{ i \}, & i\in X\setminus U\\
    X', & i=i_b\\
    \{ a \}, & i=i_c\\
\end{cases}$$ 
for $i\in I'$. This implies $\vert f'''^{-1}(\{0,2\})\vert =1$, i.e., the parameter value is constant.
\begin{claim}
    $(H,U)$ is a \yes-instance of \textsc{Extension Hitting Set} \iffl $(H''',\tau',f''')$ is a \yes-instance of \textsc{Extension Roman Hitting Function}.
\end{claim}
\begin{pfclaim}
First, we use \autoref{cor_extensible12} exhaustively. This implies that $(H''',\tau',f''')$ is a \yes-instance of \ExtRHF \iffl  $(H''',\tau',\tilde{f})$ with $\tilde{f}(x)=2$ for $x\in U\cup \{a\}$ and $\tilde{f}(x)=1$ for $x\in X \setminus U$ is a \yes-instance of \ExtRHF. 
Assume there is a minimal rhf $g$ on $H'''$, $\tau'$ with $\tilde{f} \leq g$. Define $D= g^{-1}(2)\setminus \{a\}$. As $a\notin s_i'=s_i$ for each $i\in I$ (i.e., $\tau'^{-1}(i)=\emptyset$), $s_i\cap D$ is not empty for such an~$i$. If there existed a $x\in D$  such that for each $i\in \mathbf{I}(x)$ there is a $y\in s_i\cap (D\setminus \{x\})$, then this would contradict Constraint \ref{con_12switch} of \autoref{t_property_min_rhf}. Hence, $D$ is a minimal hitting set.  

Assume that there is a minimal hitting set $D\subseteq X$ on $H$ with $U\subseteq D$. Define $g\in \{0,1,2\}^{X'}$ with $g(x)=2$ for $x\in D\cup \{ a \}$ and $g(x)=1$ for $x\in X\setminus D$. As before, we use \autoref{t_property_min_rhf} to show that $g$ is a minimal rhf. Since $s_{\tau(x)}=\{ x \}$ for each $x\in g^{-1}(1)= X\setminus D\subseteq X\setminus U$, the two first constraints hold. The fact that $D$ is a minimal hitting set on~$H$ and $s_{i_c}\cap g^{-1}(2)= \{a\}$ imply Constraint \ref{con_12switch}. Constraint~\ref{con_minHS} follows, because $D$ is a hitting set on $H$ and the hyperedges $(s_i''')_{i \in I''\setminus I}$ are hit by $\{a\}\cup (X\setminus D)$. Therefore, $g$ is a minimal rhf of~$H'''$. 
\end{pfclaim}
This finishes the discussion of all parameters.
\end{pf}

\begin{theorem}\label{t_FPT_EXTRHF}
$\kappa_\zeta$-$\ExtRHF\in\FPT $ for 
$\zeta=\weightfunction{2-f}$ and $\zeta=  f^{-1}(\{0,1\}) \}$.
\end{theorem}

\begin{pf}
To show this result, we will construct a simple \FPT-algorithm. Let $(H=(X,\hat{S}),\tau,f)$ be an instance. The idea is to walk through all functions $g\in \{0,1,2\}^X$ with $f\leq g$ and test if $g$ is a minimal rhf. This runs in \FPT-time, as for vertices $x\in X$ with $f(x)=0$ there are 3 choices, for $f(x)=1$ there are 2 choices and for $f(x)=2$ there is only 1 choice. Furthermore, we can check in polynomial time  if a function is a minimal rhf (such an algorithm can be constructed by simply modifying \autoref{alg:tau_sur}). Therefore, there are  
\begin{eqnarray*}\left( \prod_{x\in  f^{-1}(0)}3\right) \cdot \left( \prod_{x\in  f^{-1}(1)}2 \right)&\leq& \left( \prod_{x\in  f^{-1}(0)}2^{2-f(x)}\right) \cdot \left( \prod_{x\in  f^{-1}(1)}2^{2-f(x)} \right)\\& =& 2^{\weightfunction{2-f(x)}}\end{eqnarray*}
or
\begin{eqnarray*}\left( \prod_{x\in  f^{-1}(0)}3\right) \cdot \left( \prod_{x\in  f^{-1}(1)}2 \right) &\leq& \left( \prod_{x\in  f^{-1}(0)}3\right) \cdot \left( \prod_{x\in  f^{-1}(1)}3 \right)\\&=& 3^{\vert f^{-1}(\{0,1\})\vert } \end{eqnarray*}
many possibilities for $g$.
\end{pf}

\section{Enumerating Minimal Roman Hitting Sets}
\label{sec:enumerating_rhs}


Now we want to enumerate all minimal Roman hitting sets. We do this by constructing a branching algorithm. Let $H=(X,\hat{S}=(s_i)_{i \in I})$ be a hypergraph. We construct a set $X'\subseteq X$ which will include all vertices that are already in $R_2$, i.e., $R_2\subseteq X'$, or at least can be in $R_2$. To do this, we need the tuple $( R_1, R_2)$, where $R_1\subseteq I$ includes all indices and $R_2\subseteq X'$ includes all elements which are a part of the rhs that we are going construct. In the branching process, we will either put an $x\in X'$ 
into $R_2$ 
or \emph{delete~$x$}, which describes the following operation: remove~$x$ from~$X'$ and then also remove $x$ from all the hyperedges that it belonged to. Therefore, we can consider the hypergraph $$H':=(X',(s_i \cap X')_{i\in I})$$ as describing the `current hypergraph' during the computation process. However, we will modify the hyperedges $s_i$ during the computation, so that always $s_i\subseteq X'$ is satisfied, but in terms of the original hyperedge $s_i$, we can always think of $s_i \cap X'$ as the current counterpart.

We define $I'=I \setminus (R_1 \cup \mathbf{I}(R_2) )$ as the set of indices\longversion{ from $I$} that describe edges that are not yet hit. This also defines a new function $\mathbf{I'}$. If we \emph{add an $i\in I'$ to $R_1$} (as a further operation), we will also delete all $x\in s_i$  from~$X'$, as these vertices cannot be in a minimal rhs if $i$ is also.

Our measure is given by 
$$\mu\coloneqq\mu(X',\hat S', R_1,R_2)\coloneqq\vert X'\setminus R_2\vert + \vert I'\vert\,.$$ 
Initially, we have $X'=X$, $I'=I$ and $R_2=\emptyset$, so that $\mu=|X|+|I|$ at the beginning. As our main result of this section, we are going to show that all minimal rhs can be enumerated in $\mathcal{O}^*\left(\sqrt[3]{3}^{|X|+|I|}\right)$ time with polynomial delay. We also provide an example hypergraph family showing that there cannot be any significantly better enumeration algorithm, as
a hypergraph with $n$ vertices and $m$ indices from this family has at least $\sqrt[3]{3}^{n+m}$ many 
minimal rhs.

Before each branching step, we check if there is a minimal rhs $(P_1,P_2)$ on $H'$ with $(R_1,R_2) \leq (P_1,P_2)$ and try to use a reduction rule. Among all applicable branching rules, we will use the rule  with the smallest number. 
First, we define some reduction rules that are always tried before the branching rules.

\begin{redrule}\label{rr:empty_I'x}
If there is an $x\in X'\setminus R_2$ with $\mathbf{I}(x)\subseteq \mathbf{I}(R_2)$, then delete~$x$.
\end{redrule}

This reduction rule is sound, as if we would put such a vertex into a rhs that includes $R_2$, then it cannot be minimal by \autoref{t_property_min_rhs}.

\begin{redrule}\label{rr:empty_hyperedge}
If there is an $i\in I'$ with $s_i=\emptyset$, then put $i$ into $R_1$.
\end{redrule}

\noindent
The soundness of this rule follows easily, since $i$ has to be hit.

\begin{brarule}\label{br:single_elemente_hyperedge}
Let $i\in I'$ with $s_i=\{x\}\subseteq X'\setminus R_2$. Then we branch as follows:
\shortversion{ 
(1)~Delete $x$. (2) Add $x$ to $R_2$. 
}
\longversion{
\begin{enumerate}
    \item Add $i$ to $R_1$ and delete $x$.
    \item Add $x$ to $R_2$.
\end{enumerate}
}
\end{brarule}

\begin{lemma}The case distinction of  \autoref{br:single_elemente_hyperedge} is complete. 
The worst-case branching vector is $(2,2)$.
\end{lemma}

\begin{pf}
Since $s_i$ only includes $x$, $i\in R_1$ or $x\in R_2$ must hold. Both $i\in R_1$ and $x\in R_2$ is impossible by Constraint~\ref{con_rhs_disjoint} of \autoref{t_property_min_rhs}. Therefore, we delete $x$ and then add $i$ to $R_1$ by \autoref{rr:empty_hyperedge}, or we put~$x$ into $R_2$. In both cases, the measure decreases by two, as $i$ is no longer in $I'$ and $x$ is deleted from~$X'$ or put into $R_2$. 
\end{pf}

\begin{brarule}\label{br:element_in_3_sets}
Let $x\in X'$ with $\vert \mathbf{I'}(x)\vert \geq 3$. Then we branch as follows:
\shortversion{ 
(1) Add $x$ to $R_2$. (2) Delete $x$. 
}
\longversion{
\begin{enumerate}
    \item Add $x$ to $R_2$.
    \item Delete $x$.
\end{enumerate}
}
\end{brarule}

\begin{lemma}\label{lem:}The case distinction of  \autoref{br:element_in_3_sets} is complete. 
The worst-case branching vector is $(4,1)$.
\end{lemma}

\begin{pf}
In this branching scenario, we branch if $x$ is part of the solution or not. Trivially, this is a complete case distinction. By adding $x$ to $R_2$, the measure decreases by at least~4, as the resulting $I'$ does not contain any hyperedge which includes $x$. The decrease in the other case is trivially one.  
\end{pf}

\begin{brarule}\label{br:element_in_1_set_including_2}
Let $x,y\in X'$ with $ \mathbf{I'}(x)=\{i\}$ and $s_i=\{x,y\}$. Then we branch as follows:
\shortversion{ 
(1) Add $x$ to $R_2$ and delete $y$. (2) Add $y$ to $R_2$ and delete $x$. (3) Add $i$ to $R_1$ and delete $x$ and $y$. 
}
\longversion{
\begin{enumerate}
    \item Add $x$ to $R_2$ and delete $y$.
    \item Add $y$ to $R_2$ and delete $x$.
    \item Add $i$ to $R_1$ and delete $x$ and $y$.
\end{enumerate}
}
\end{brarule}

\begin{lemma}\label{lem:element_in_1_set_including_2}
The case distinction of  \autoref{br:element_in_1_set_including_2} is complete. 
The worst-case branching vector is $(3,3,3)$.
\end{lemma}

\begin{pf}
This is a complete case distinction, as $x$ and $y$ cannot be in the same minimal rhs $P=(P_1, P_2)$ with $R_1\subseteq P_1\subseteq I$ and $R_2\subseteq P_2\subseteq X'$. If there exists such a minimal rhs, then $R_2\setminus \{x\}$ is also a hitting set on $\left(s_i\right)_{i\in I\setminus R_2}$, which contradicts Constraint \ref{con_rhs_minHS} of \autoref{t_property_min_rhs}. As in each case, $i$ is no longer in $I'$ and $x$ and $y$ are no longer in $X'\setminus R_2$, the measure reduces by 3.
\end{pf}

\begin{brarule}\label{br:element_in_1_set_including_3}
Let $x\in X'$ with $ \mathbf{I'}(x)=\{i\}$ with $\vert s_i\vert \geq 3$. Then we branch as follows:
\shortversion{ 
(1) Add $x$ to $R_2$ and delete the remaining vertices in $s_i$. (2) Delete~$x$. 
}
\longversion{
\begin{enumerate}
    \item Add $x$ to $R_2$ and delete the remaining vertices in $s_i$.
    \item Delete $x$.
\end{enumerate}
}
\end{brarule}

\begin{lemma}\label{lem:element_in_1_set_including_3} 
The case distinction of  \autoref{br:element_in_1_set_including_3} is complete. 
The worst-case branching vector is $(4,1)$.
\end{lemma}

\begin{pf}
    Analogously to \autoref{lem:element_in_1_set_including_2}, $x$ cannot be in $R_2$ if there is another vertex in $R_2\cap s_i$. Therefore, we can delete the other vertices in $s_i$, $i$ is not in $I'$ anymore and the measure is dreasing by 4. The other case holds trivially.
\end{pf}

\begin{remark}\label{rem:size_I'x}
From now on, we can assume that $\vert \mathbf{I'}(x)\vert = 2$. This is seen as follows.\\
If $\vert \mathbf{I'}(x)\vert > 2$, \autoref{br:element_in_3_sets}  would trigger.\\ For $\vert \mathbf{I'}(x)\vert = 0$ we would use \autoref{rr:empty_I'x}.\\
The branching rules \ref{br:single_elemente_hyperedge}, \ref{br:element_in_1_set_including_2} and \ref{br:element_in_1_set_including_3} handle the case $\vert \mathbf{I'}(x) \vert =1$, distinguishing cases for $\mathbf{I'}(x)=\{i\}$ concerning $\vert s_i\vert$.
\end{remark}

\noindent
 We will use this fact for the analysis of the following branching rules.  

\begin{brarule}\label{br:elements_in_the_same_edges}
Let $x,y\in X'$ with $x \neq y$ and $\mathbf{I'}(x) = \mathbf{I'}(y)$. Then we branch as follows:
\shortversion{ 
 (1) Add $x$ to $R_2$ and delete~$y$. (2) Add $y$ to $R_2$ and delete~$x$. (3) Delete~$x$ and~$y$.
}
\longversion{
\begin{enumerate}
    \item Add $x$ to $R_2$ and delete $y$.
    \item Add $y$ to $R_2$ and delete $x$ .
    \item Delete $x$ and $y$.
\end{enumerate}
}
\end{brarule}

\begin{lemma}The case distinction of  \autoref{br:elements_in_the_same_edges} is complete. 
The worst-case branching vector is $(4,4,2)$.
\end{lemma}

\begin{pf}
    There cannot exist a minimal rhs $(P_1,P_2)$ with $R_1\subseteq P_1\subseteq I$ and $R_2\cup \{x,y\}\subseteq P_2 \subseteq X'$, as $(P_1,P_2\setminus \{x\})$ would also be a rhs. Therefore, this branching gives a complete case distinction. In all three cases, $x$ and $y$ leave $X'\setminus R_2$. In the first two cases, the indices in $\mathbf{I'}(x)$ are no longer in $I'$.
\end{pf}

\begin{brarule}\label{br:hyperedge_with_2_elements}
Let $i\in I'$ with $s_i=\{x,y\}$. Then we branch as follows:
\shortversion{ 
 (1) Add $x$ to $R_2$. (2) Add $y$ to $R_2$ and delete~$x$. (3) Add $i$ to $R_1$ and delete $x$ and $y$.
}
\longversion{
\begin{enumerate}
    \item Add $x$ to $R_2$.
    \item Add $y$ to $R_2$ and delete $x$ .
    \item Add $i$ to $R_1$ and delete $x$ and $y$.
\end{enumerate}
}
\end{brarule}

\begin{lemma}The case distinction of  \autoref{br:hyperedge_with_2_elements} is complete. 
The worst-case branching vector is $(3,4,3)$.
\end{lemma}

\begin{pf}
    This is an asymmetric branch, where we first consider if $x$ is in the solution. If this is not the case, then we consider~$y$. If both vertices are not in the solution, we have to add~$i$ to~$R_1$. Thus, this is a complete case distinction. Due to \autoref{rem:size_I'x}, $\vert\mathbf{I'}(x)\vert = \vert\mathbf{I'}(y)\vert=2$. Hence, putting $x$ or $y$ into $R_2$ would imply that two indices will leave $I'$. Therefore, we get the branching vector $(3,4,3)$.
\end{pf}

\begin{brarule}\label{br:hyperedge_with_3_elements}
Let $i\in I'$ with $s_i=\{x,y,z\}$. Then we branch as follows:
\shortversion{ 
 (1) Add $x$ to $R_2$. (2) Add $y$ to $R_2$ and delete $x$. (3) Add $z$ to $R_2$ and delete $x,y$. (4) Add $i$ to $R_1$ and delete $x,y$ and $z$.
}
\longversion{
\begin{enumerate}
    \item Add $x$ to $R_2$.
    \item Add $y$ to $R_2$ and delete $x$.
    \item Add $z$ to $R_2$ and delete $x,y$.
    \item Add $i$ to $R_1$ and delete $x,y$ and $z$.
\end{enumerate}
}
\end{brarule}

\begin{lemma}The case distinction of  \autoref{br:hyperedge_with_3_elements} is complete. 
The worst-case branching vector is $(3,4,5,4)$.
\end{lemma}

\begin{pf}
       This is an asymmetrical branching as in \autoref{br:hyperedge_with_2_elements} (where we have one more case: if $x,y$ are not in $R_2$, we branch on $z$). We also get the branching vector in an analogous fashion.   
\end{pf}

\begin{remark}\label{rem:hyperedges_at_least_4_elements}
    From now on we can assume $\vert s_i\vert \geq 4$ for each $i\in I'$. Also, $\mathbf{I'}(x)\neq \mathbf{I'}(y)$ for $x,y\in X'$ with $x\neq y$. We can see these properties as follows.\\ The case $\vert s_i \vert = 0$ is handled by \autoref{rr:empty_hyperedge}. \\\autoref{br:single_elemente_hyperedge} considers the case $\vert s_i \vert = 1$.\\ From this point on, we can use \autoref{rem:size_I'x}. \autoref{br:hyperedge_with_2_elements} handles $\vert s_i\vert =2$ and \autoref{br:hyperedge_with_3_elements} handles $\vert s_i\vert =3$.\\
    Furthermore, we can assume $\mathbf{I'}(x)\neq \mathbf{I'}(y)$ for $x,y\in X'$ with $x\neq y$ by \autoref{br:elements_in_the_same_edges}.   
\end{remark}

\begin{brarule}\label{br:element_in_two_sets_size_4}
Let $x\in X'$ with $\mathbf{I'}(x)=\{i,j\}$ and $y\in s_i\setminus \{ x\}$ with $\mathbf{I'}(y)=\{i,k\}$. Then we branch as follows:
\shortversion{ 
(1) Delete $x$. (2) Add $x$ to $R_2$ and delete $y$. (3) Add $x$ and $y$ to $R_2$ and  delete all Elements in $s_k \setminus \{y\}$.
}
\longversion{
\begin{enumerate}
    \item Delete $x$.
    \item Add $x$ to $R_2$ and delete $y$.
    \item Add $x,y$ to $R_2$ and delete all Elements in $s_k\setminus \{y\}$.
\end{enumerate}
}
\end{brarule}

\begin{lemma}The case distinction of  \autoref{br:element_in_two_sets_size_4} is complete. 
The worst-case branching vector is $(1,4,8)$.
\end{lemma}

\begin{pf}
This is an asymmetric branch. To analyze this branch, we split this branching rule into two part. At first, we branch on $x$. If $x$ is not in $R_2$, the measure decreases by 1. The measure decreases by 3 if $x$ is in $R_2$, as $i,j$ are no longer in $I'$. In this case, we now use \autoref{br:element_in_1_set_including_3}, as $ \vert \mathbf{I'}(y) \vert =1$ after the first part. In this use of \autoref{br:element_in_1_set_including_3} the branching vector is $(1,5)$, as $\vert s_k \vert \geq 4$ by \autoref{rem:hyperedges_at_least_4_elements}. This results in the branching vector $(1,4,8)$.   
\end{pf}

\begin{remark}
    It can be shown that this branching rule has even the braching vector $(1,4,10)$, as $R_2 \cap (s_j \setminus \{x\})$ has to be empty in the last case. This tighter analysis for this branching vector is difficult, since $s_j$ and $s_k$ do not have to be disjoint. Furthermore, this branching vector is not the worst one and we already can use the branching vector $(1,4,8)$ to show the tight bound, which is given by $(3,3,3)$, see \autoref{thm:rhs-tight-example}.
\end{remark}

\begin{table}\centering
\begin{tabular}{c|c|c}
   Rule  & branching vector & branching number\\ \hline
   \ref{br:single_elemente_hyperedge}  & $(2,2)$ & $\sqrt{2}\leq 1.4143$\\
   \ref{br:element_in_3_sets} \& \ref{br:element_in_1_set_including_3}  & $(4,1)$ & $1.3803$\\
   \ref{br:element_in_1_set_including_2} & $(3,3,3)$ & $\sqrt[3]{3}\leq 1.4423$\\
   \ref{br:elements_in_the_same_edges} & $(4,4,2)$ & $\sqrt{2}\leq 1.4143$\\
   \ref{br:hyperedge_with_2_elements} & $(3,4,3)$ & $1.3954$\\
   \ref{br:hyperedge_with_3_elements} & $(3,4,5,4)$ & $ 1.4253$\\
   \ref{br:element_in_two_sets_size_4} & $(1,4,8)$ & $1.4271$

\end{tabular}
\caption{\label{tab:rhs_branches} Branching vectors and numbers for enumerating minimal rhs.}
\end{table}

\begin{corollary}\label{cor_rhs_enum_complete_case}
The branching algorithm is a complete case distinction.
\end{corollary}
\begin{pf}
    \autoref{rem:size_I'x} and \autoref{rem:hyperedges_at_least_4_elements} imply that $\vert I'(x) \vert =2$ for each $x\in X'\setminus R_2$ and for each $i\in I'$, $\vert s_i\vert \geq 4$. As \autoref{br:elements_in_the_same_edges} ensures that two elements cannot be in the same hyperedges, the remaining cases are handled by \autoref{br:element_in_two_sets_size_4}.
\end{pf}

\begin{theorem}\label{thm:rhs-enumeration}
For a hypergraph $H=(X,(s_i)_{i\in I})$ with $\vert X\vert =n$ and $\vert I\vert =m$, all minimal rhs can be enumerated in $\mathcal{O}^*(\sqrt[3]{3}^{n+m})$ time with polynomial delay and polynomial space. 
\end{theorem}

\begin{pf}
    By \autoref{cor_rhs_enum_complete_case}, the branching algorithm has a branching rule for each possibility. If we calculate the branching numbers for the branching vectors mentioned in \autoref{tab:rhs_branches}, we get $\mathcal{O}^*(\sqrt[3]{3}^{n+m})$ as the running time for the algorithm in the worst case.

    This leaves to show that the algorithm runs with polynomial delay and polynomial space.

As branching algorithms per se use polynomial space, the only potential problem consists in enumerating solutions multiple times, because this would enforce us to store previous results. Therefore, we show next that there is no minimal rhs that we count multiple times. The Branching Rules \ref{br:single_elemente_hyperedge}, \ref{br:element_in_3_sets} and  \ref{br:element_in_1_set_including_3} branch by putting $x$ into $R_2$ or deleting it. Since we never change such a decision afterwards, these branching rules prevent us from enumerating any minimal rhs twice. The same happens (with some side effects) in the remaining branchings, as these are just asymmetrical branches where we branch again in some cases.  

Concerning polynomial delay, knowing that our branching algorithm never outputs solutions again is also helpful. Now, the only potential problem is that the algorithm spends exponential time when diving into sub-branches where no minimal rhs solution exists. We prevent this by evoking our polynomial-time extension check procedure. In order to be able to do so, we only need to show that working on restricted instances, as described by $H'$, suffices. This is shown in our last claim that hence concludes our argument.

    \begin{claim}
        Let $H=(X,\left( s_i \right)_{i\in I})$ be a hypergraph, $R_1 \subseteq I$ and $R_2 \subseteq X' \subseteq X$. Then there is a solution $(P_1, P_2)$ for \ExtRHS with given $H,R_1,R_2$ such that  $P_2\subseteq X'$ \iffl there is a solution for \ExtRHS with given $H'=(X',\left( s_i\cap X' \right)_{i\in I}),R_1,R_2$.
    \end{claim}
\begin{pfclaim}
Let $P_2 \subseteq X'$. Then $ P_2\cap (X' \cap  s_i) = P_2 \cap s_i$ for each $i\in I$. Hence, $(P_1,P_2)$ hits $H$ \iffl it hits $H'$. With this observation, it is easy to verify that $(P_1,P_2)$ is a solution for the \ExtRHS problem with $H,R_1,R_2$ such that $P_2 \subseteq X'$ \iffl it is a solution for the \ExtRHS problem with $H',R_1,R_2$.
\end{pfclaim}
     Therefore, if we use \autoref{alg:ext_rhs} with the parameters $X',(s_i\cap X')_{i\in I}, R=(R_1,R_2)$, we check if there is a minimal rhs $P=(P_1,P_2)$ with $R\leq P$ and $P_2\subseteq X'$. Since we do this after each branch and each rule increases $R$, we check if this branch will enumerate a minimal rhs. If get a no-answer, we can go back and need not dive into the recursion of the branching algorithm. As we branch at most $\vert  X\vert + \vert I\vert $ many times, this only needs polynomial time on each path of the branching tree, hence ensuring polynomial delay.  
\end{pf}

Next, we prove that our (relatively simple)  enumeration algorithm is optimal in the sense that, ignoring polynomial factors of the running time, any algorithm has to spend the claimed amount of time on some examples.

\begin{theorem}\label{thm:rhs-tight-example}
    There is a hypergraph $H=(X, (s_i)_{i \in I})$ such that there are at least $\sqrt[3]{3}^{\vert X \vert + \vert  I\vert}$ many minimal rhs. 
\end{theorem}

\begin{pf}
    Define $X=\{x_{1},\ldots,x_{2n}\}$ and $I=\{1,\ldots,n\}$ with $s_i=\{x_{2i-1}, x_{2i}\}$ for each $i\in I$. For each $i\in I$, there are three ways to hit $i$ by a minimal rhs $(R_1,R_2)$: $x_{2i-1}\in R_2$ or $x_{2i}\in R_2$ or $i \in I$.
    As the sets $s_1,\ldots, s_n$ are pairwise disjoint, for each $i\in I$, each $x_{2i-1}$ and $x_{2i}$ cannot be in $R_2$ at the same time and the way one hyperedge is hit does not effect the way another hyperedge is hit. Therefore, there are $3^n=\sqrt[3]{3}^{3n}=\sqrt[3]{3}^{\vert X\vert + \vert I\vert}$  many minimal rhs. 
\end{pf}

Since we constructed an enumeration algorithm for minimal rhs, the next question could be if there is an enumeration algorithm for minimal rhf with polynomial delay. We do not know if this is the case, but we can show following theorem. 

\begin{theorem}
    If there is an algorithm which enumerates all minimal rhf of an instance $(H,\tau)$ with polynomial delay, then there is an algorithm enumerates all minimal hitting sets of a hypergraph $H'$ with polynomial delay. 
\end{theorem}

\begin{pf}
    Let $H=(X,(s_i)_{i\in I})$ be a hypergraph. We will use the same construction for a hypergraph as $H''$ in the prove of \autoref{t_paraNP_EXTRHF}. As recall we define $X'=X \cup \{a\}$ and $I''= I\cup \{i_b,i_c\}$ as well as the hypergraph $H''= (X',(s_i'')_{i\in I''})$. Furthermore, we need 
$$\tau': X'\to I'', x\mapsto \begin{cases}
    x, & x\in X\\
    i_b, & x=a
\end{cases}, \quad s_i''\coloneqq\begin{cases}
    s_i, & i\in I\\
    \{ a, i \}, & i\in X\\
    X', & i=i_b\\
    \{ a \}, & i=i_c\\
\end{cases}$$ 
for $i\in I''$. Since $\tau^{-1}(i_c) = \emptyset$ and $s''_{i_c}=\{a\}$, $f(a)=2$ for each rhf $f:X'\to \{ 0, 1, 2\}$. Therefore, for each minimal rhf $f:X'\to \{ 0, 1, 2\}$ and $x\in X$, $f(x)\neq 1$, otherwise it contradicts Constraint \ref{con_1_2} of \autoref{t_property_min_rhf}. Hence, $f^{-1}(1)$ is empty for each minimal rhf $f$ of~$H''$.

We will show that there is a bijective function~$T$ between the minimal hitting sets of~$H$ and the minimal rhf of~$H''$. This function maps each minimal hitting set~$D$ to $g:X'\to \{0,1,2\} $ with $g^{-1}(0)= X\setminus D$, $g^{-1}(2)=\{a\}\cup D$. In the proof of \autoref{t_paraNP_EXTRHF}, we already have shown that $g$ is a minimal rhf of~$H''$. The inverse function maps a minimal rhf~$g$ to $g^{-1}(2)\setminus \{a\}$ which is a minimal hitting set by the proof of \autoref{t_paraNP_EXTRHF}. Since we can calculate $T(D)$ and $T^{-1}(g)$ for a minimal hitting set and for a minimal rhf~$g$ in polynomial time, the theorem holds.   
\end{pf}

This reduction links the question if rhf enumeration is possible with polynomial delay with the question if minimal hitting sets can be enumerated  with polynomial delay.  Hence, we can say by the previous theorem that, assuming HSTT,\footnote{a hypothesis introduced in \autoref{sec:intro}} there is no polynomial-delay enumeration algorithm for minimal rhf.
Notice that the argument that we give is via a parsimonious reduction. The idea to use this type of reductions in the world of enumeration complexity is clearly not new; we only refer to \cite{CapStr2019,CreKPSV2019,KanLMN2014,Mar2013a,Str2019} to get a first impression of this area.

\section{Branching algorithm for optimizing \textsc{Roman Hitting Sets}}
\label{sec:exact_branch}

In this section, we will present a simple branching algorithm that returns minimum rhs.
Let $H=(X,(s_i)_{i\in I})$ be a hypergraph. We will reuse the idea of the branching algorithm of \autoref{sec:enumerating_rhs} (including the sets $X',I',R_1,R_2$ as well as the measure $\mu$ and \autoref{rr:empty_hyperedge}). To the algorithm we add the idea of finding a minimum rhs with minimal $\vert R_2\vert$.  
This idea also motivates our next reduction rule.

\begin{redrule}\label{rr:element_in_2_edges_optirhs}
    If there is an $x \in X'\setminus R_2$ with $\vert \mathbf{I'}(x)\vert \leq 2$, then delete $x$.
\end{redrule}

\begin{lemma}\label{lem:element_in_2_edges_optirhs}
    \autoref{rr:element_in_2_edges_optirhs} is sound.
\end{lemma}

\begin{pf}
    Let $H=(X,(s_i)_{i\in I})$ be a hypergraph and $(R_1, R_2)$ be a minimum rhs for which there exists an $x\in R_2$ with $\mathbf{I}(x)=\{i,j\}$ ($i=j$ could hold). Then, $(R_1\cup \{ i, j \}, R_2\setminus \{x\})$ is also a rhs with $\weightfunction{R_1\cup \{ i, j \},R_2\setminus \{x\}} \leq \weightfunction{R_1,R_2}$. 
\end{pf}

\begin{redrule}\label{rr:3edges_with_same_element_optirhs}
    If there is a $x \in X'\setminus R_2$ with $\vert \{i\in I' \mid s_i=\{x\} \}\vert \leq 3$, put $x$ into $R_2$.
\end{redrule}

The soundness of this reduction rule follows from the fact that putting all the hyperedges of the set  $\{i\in I' \mid s_i=\{x\} \}$ into $R_1$ costs more than putting~$x$ into $R_2$ (in this case, we could also hit more edges).

\begin{brarule}\label{br:element_in_3edges_optirhs}
Let $x\in X'$ with $\vert \mathbf{I'}(x) \vert =3$.  Then we branch as follows:
\shortversion{ 
(1) Delete~$x$. (2) Add~$x$ to $R_2$ and delete all vertices from $\bigcup_{i\in \mathbf{I'}(x)}s_i$. 
}
\longversion{
\begin{enumerate}
    \item Delete $x$.
    \item  Add $x$ to $R_2$ and delete all vertices from $\bigcup_{i\in \mathbf{I'}(x)}s_i$. 
\end{enumerate}
}
\end{brarule}

\begin{lemma}The case distinction of  \autoref{br:element_in_3edges_optirhs} is complete. 
The worst-case branching vector is $(1,5)$.
\end{lemma}

\begin{pf}
    Let $H=(X,(s_i)_{i\in I})$ be a hypergraph and $(R_1, R_2)$ be a minimum rhs. To see that this is a complete case distinction, we assume there is an $x \in R_2$ with $\mathbf{I}(x)=\{i_1, i_2, i_3\}$ with $\vert \mathbf{I}(x) \vert = 3$ and $\vert s_{i_1} \cap R_2 \vert = 2$.
    Then, $(R_1',R_2')=(R_1 \cup \{i_2, i_3\}, R_2 \setminus \{x\})$ is also a rhs ($i_1$ is hit by the remaining element in $s_{i_1} \cap (R_2\setminus \{x\})$ and $i_2, i_3$ hit themselves). Furthermore. $\weightfunction{R_1',R_2'}\leq \weightfunction{R_1,R_2}$. Therefore, this is a complete case distinction.

    For the first branch,  it is trivial that the measure decreases by~1. Consider the second branch. By \autoref{rr:3edges_with_same_element_optirhs}, we know that $\left( \bigcup_{i\in \mathbf{I}(x)} s_i \right)\setminus\{x\} $ is not empty. Therefore, two elements get out of $X'\setminus R_2$ and three hyperedges move out of~$I'$. Hence, the measure  decreases by~$5$.
\end{pf}

\begin{brarule}\label{br:element_in_4_sets_optirhs}
Let $x\in X'$ with $\vert \mathbf{I'}(x)\vert \geq 4$. Then we branch as follows:
\shortversion{ 
(1) Add $x$ to $R_2$. (2) Delete $x$. 
}
\longversion{
\begin{enumerate}
    \item Add $x$ to $R_2$.
    \item Delete $x$.
\end{enumerate}
}
\end{brarule}

\noindent
The completeness and the branching vector follows analogously to \autoref{br:element_in_3_sets}.

\begin{corollary}The case distinction of  \autoref{br:element_in_4_sets_optirhs} is complete. 
The worst-case branching vector is $(5,1)$.
\end{corollary}

\begin{theorem}
    For a hypergraph $H=(X,(s_i)_{i\in I})$ with $\vert X\vert =n$ and $\vert I\vert =m$, a minimum rhs can be computed in  time $\mathcal{O}\left(1.3248^{n+m}\right)$, using polynomial space.
\end{theorem}
\begin{pf}
We first check if this branching algorithm is a complete case distinction. By \autoref{rr:empty_hyperedge}, we know that there is no empty hyperedge in $I'$. This implies that there has to be at least one element $x\in X'$. $\vert \mathbf{I'}(x)\vert \leq 2$ triggers \autoref{rr:element_in_2_edges_optirhs}. \autoref{rr:3edges_with_same_element_optirhs} and \autoref{br:element_in_3edges_optirhs} handle the case $\vert \mathbf{I'}(x)\vert= 3$. The remaining cases are covered by \autoref{br:element_in_4_sets_optirhs}.

As Branching Rules \ref{br:element_in_3edges_optirhs} and \ref{br:element_in_4_sets_optirhs} are the only branching rules and both have a branching vector $(1,5)$, we get a running time of $\mathcal{O}\left(1.3248^{n+m}\right)$. \autoref{rr:3edges_with_same_element_optirhs} and  \autoref{br:element_in_3edges_optirhs} deal with $x\in X'$ such that $\vert \mathbf{I'}(x)\vert =3$. The remaing cases are handled  by \autoref{br:element_in_4_sets_optirhs}.
\end{pf}

\begin{remark}
By making use of the closed-neighborhood hypergraph, this implies a running time of $\Oh(1.76^n)$ for \textsc{Minimum Roman Domination} on $n$-vertex graphs. This is clearly better than the trivial $\Oh^*(2^n)$-algorithm, or also better than cycling through all $\Oh(1.94^n)$ many minimal Roman dominating functions, see \cite{AbuFerMan2022b}, but worse than algorithms based on making use of the \textsc{Partial Dominating Set} problem as suggested by \cite{Lie2007,Roo2011,ShiKoh2014}, leading to $\Oh(1.51^n)$ time (and space). 
But our presented algorithm, as well as its analysis, are very simple and clearly leaves room for improvement. For instance, we are not applying any Measure-and-Conquer techniques in the analysis. Yet, we consider this approach quite promising. Also recall that a breakthrough concerning exact algorithms for \textsc{Minimum Dominating Set} was obtained after studying \textsc{Minimum Set Cover} with Measure-and-Conquer techniques; see \cite{FomGraKra2009,FomKra2010}. Here, we are leaving this venue for future research.
\end{remark}

\section{Applications}

In this section, we will consider applications of \RHS. First we will use the fact that we can solve \ExtRHS in polynomial time to show that \textsc{Extension Dominating Set} is polynomial time solvable on split graphs. After this, we will define further Roman versions for problems for which \HS is a generalization.  

\subsection{\textsc{Extension Dominating Set} on split graphs}

\begin{theorem}
    \EDS can be solved in polynomial time on split graphs.
\end{theorem}
\begin{pf}
    We will present a reduction to \ExtRHS. To this end, let $G=(V,E)$ be a split graph, with $V$ partitioned into a clique~$C$ and an independent set~$I$. We can assume that each vertex in $C$ has at least one neighbor in $I$. Otherwise, we put this vertex from~$C$ into~$I$. Let $G$ and  $U\subseteq V$ be the input of \EDS.  Define the hypergraph $H=(C,(N(i))_{i\in I})$ and $U'=(U \cap I, U\cap V)$. 

    \begin{claim}
    $(G,U)$ is a \yes-instance of \EDS \iffl $(H,U')$ is a \yes-instance of \ExtRHS.
    \end{claim}
    \begin{pfclaim}
        Let $D\subseteq V$ be an inclusion-wise minimal dominating set of~$G$ with $U\subseteq D$. Then, define $R=(D\cap I, D \cap C)$. Clearly, $U' \leq R$. As $D$ is a dominating set and $I$ is independent, for each $i\in I\setminus D= I \setminus (D \cap I)$ there exists a $c\in N(i)\cap D\subseteq C\cap D$. Thus, $R$ is a rhs. Since $N[i]\subseteq N[c]$ for each $i\in I$ and $c\in N(i)$, the minimality of $D$ implies that for each $i\in D\cap I$, $N[i] \cap (D\cap C)=N(i)\cap D=\emptyset$. If $D\cap C$ is not a minimal hitting set on $\{N(i)\mid i\in I\setminus D\}$, then there exists an $x\in D\cap C$ such that, for each $i\in N(x)\cap I$, there is a $y\in  N(i)\cap ((D\cap C)\setminus\{x\})$. Therefore, $x$ has no private neighbor in $I$. By $N(x) \cap I\neq \emptyset$, $D \cap C$ contains at least two vertices. Hence, $x$ has no private neighbor in $C$, contradicting the minimality of~$D$. Therefore, \autoref{t_property_min_rhs} implies the minimality of~$R$.

        Now assume there is a minimal rhs $R=(R_1,R_2)$ of $H$ with $U' \leq R$. Define $D=R_1\cup R_2$. If $R_2=D\cap C$ is empty, then  $I=R_1$. Since $N(x) \cap I\neq \emptyset$ for all $x\in C$, this is a dominating set. Furthermore, $D$ is minimal as each $i\in I=R_1$ is its own private neighbor. 
        Now, we assume that $R_2= D\cap C$ is not empty. Therefore, all of $C$ is dominated. As for each $i\in I$, either $i\in R_1$ or there exists a $c\in N(i)\cap R_2 = N(i)\cap D$, $I$ is also dominated. Therefore, $D$ is a dominating set. By Constraint \ref{con_rhs_disjoint}, we already know that each $i\in R_1$ has no neighbor in $N(i)=N(i)\cap R_2=N(i)\cap D$ and is its own private neighbor. Furthermore,  $R_2$ is a minimal hitting set on $\{N(i)\mid i\in I \setminus R_1\}$. Therefore, each $x\in R_2$ has a private neighbor in $I\setminus R_1$. Thus, $D$ is a minimal dominating set.
    \end{pfclaim}
    This reduction is also a polynomial-time  reduction. Therefore, we can use  \autoref{alg:ext_rhs} to solve \EDS on split graphs.
\end{pf}

This is an interesting result, as it is known by \cite{CasFGMS2022} that \EDS is \NP-complete on the related class of bipartite graphs. To our knowledge, there is no such result mentioned for co-bipartite graphs, yet.

But by \cite{KanLMN2014}, it is known that HSTT holds \iffl there is no enumeration algorithm for minimal dominating sets on co-bipartite graphs with polynomial delay.  
The construction of this proof can also be used to show the  \NP-completeness (and the \W{3}-completeness if parameterized by the size of the pre-solution) of \EDS on co-bipartite graphs. For the ease of reference, we make this result explicit.

\begin{corollary}
    \EDS is \NP-complete and \W{3}-complete when parameterized by the size of the pre-solution even on co-bipartite graphs.
\end{corollary}

\subsection{Roman versions for other problems}
\label{sec:rvc+rec}

There are more problems than \DS for which \HS is a generalization. Some other examples are \textsc{Vertex Cover} and \textsc{Edge Cover}.
Let $G=(V,{E})$ be a graph. We call $C\subseteq V$ a \emph{vertex cover} if $e\cap C$ is not empty for each $e\in E$.  In other words, $C\subseteq V$ is a vertex cover \iffl $C$ is a hitting set of the hypergraph $(V,(e)_{e\in E})$.
A set $C\subseteq E$ is an \emph{edge cover} if for each $v\in V$, $\mathbf{E}(v) \cap C\neq \emptyset$ with $\mathbf{E}(v) \coloneqq \{ e\in E \mid v\in e\}$. Hence, $C\subseteq E$ is a edge cover \iffl it is a hitting set on $(E, (\mathbf{E}(v))_{v\in V})$. In the context of these problems we speak of covering instead of hitting.

\centerline{\fbox{\begin{minipage}{.96\textwidth}
\textbf{Problem name: }\textsc{Vertex Cover}, or \textsc{VC} for short\\
\textbf{Given: } A graph $G=(V,{E})$ and $k\in \mathbb{N}$\\
\textbf{Question: } Is there a vertex cover $C\subseteq V$ with $\vert C\vert\leq k$?
\end{minipage}
}}

\smallskip
\noindent
\centerline{\fbox{\begin{minipage}{.96\textwidth}
\textbf{Problem name: }\textsc{Edge Cover}, or \textsc{EC} for short\\
\textbf{Given: } A graph $G=(V,{E})$ and $k\in \mathbb{N}$\\
\textbf{Question: } Is there a edge cover $C\subseteq E$ with $\vert C\vert \leq k$?
\end{minipage}
}}

Now that we generalized \RD for \HS, we could use this to define problems like \textsc{Roman Vertex Cover} and \textsc{Roman Edge Cover}. We will use rhs for the new problems but we could also use rhf for the definition. 

Let $G=(V,E)$ be a graph. We call a tuple $(R_1,R_2) \leq (E,V)$ a \emph{Roman vertex cover} if for each $e\in E$, $e\in R_1$ or $R_2 \cap e$ is not empty. We call a tuple $(R_1,R_2) \leq (V, E)$ a \emph{Roman edge cover} if for each $v\in V$, $v \in R_1$ or $R_2 \cap \mathbf{E}(v)\neq \emptyset$. 

\centerline{\fbox{\begin{minipage}{.96\textwidth}
\textbf{Problem name: }\textsc{Roman Vertex Cover}, or \textsc{RVC} for short\\
\textbf{Given: } A graph $G=(V,{E})$ and $k\in \mathbb{N}$\\
\textbf{Question: } Is there a Roman vertex cover $(R_1,R_2)\subseteq E \times V$ \shortversion{that satisfies}\longversion{with} $\weightfun(R_1,R_2) \leq k$?
\end{minipage}
}}

\smallskip
\noindent
\centerline{\fbox{\begin{minipage}{.96\textwidth}
\textbf{Problem name: }\textsc{Roman Edge Cover}, or \textsc{REC} for short\\
\textbf{Given: } A graph $G=(V,{E})$ and $k\in \mathbb{N}$\\
\textbf{Question: } Is there a  Roman edge cover $(R_1,R_2)\subseteq V \times E$ \shortversion{that satisfies}\longversion{with} $\weightfun(R_1,R_2) \leq k$?
\end{minipage}
}}


\begin{lemma}
    Let $G=(V,E)$ be a graph. For each Roman edge cover $(R_1,R_2)\leq (V, E)$, $\weightfun(R_1,R_2)\leq \vert V\vert $.
\end{lemma}

\begin{pf}
    Let $G=(V,E)$ be a graph. Let $(R_1,R_2)$ be a Roman edge cover of~$G$. Define $V(R_2)\coloneqq\bigcup_{e\in R_2} e $. Each $e\in E$ belongs to  $\mathbf{E}(v)$ for exactly two $v\in V$. Therefore,
    $$\vert V(R_2)\vert = \left\vert \bigcup_{e\in R_2} e\right\vert \leq \sum_{e\in R_2}\vert e\vert  = 2\cdot\vert R_2\vert.$$
     As $V = R_1 \cup V(R_2)$ for a Roman edge cover, $$ \weightfun(R_1,R_2) = \vert R_1\vert + 2\cdot\vert R_2\vert \geq \vert R_1\vert + \vert V(R_2)\vert  \geq  \vert V\vert. $$
     This shows the claim.
\end{pf}

As $(V,\emptyset)$ is a Roman edge cover for each graph $G=(V,E)$ with $\weightfun(V,\emptyset)=\vert V\vert$, this is also an optimal solution, so that  \textsc{Roman Edge Cover} becomes a rather trivial problem, we only have to check if the parameter~$k$ is less than $|V|$.

\begin{corollary}
    \textsc{Roman Edge Cover} is solvable in logarithmic space.
\end{corollary}

\begin{theorem}
    \textsc{Roman Vertex Cover} is \NP-complete.
\end{theorem}

\begin{pf}
    As each instance of \textsc{Roman Vertex Cover} is also an instance of \RHS, the \NP-membership follows directly. For the hardness, We will use the \textsc{Vertex Cover} problem. Let $G=(V,E)$ be a graph and $k\in \mathbb{N}$. Define the graph $G'= (V',E')$ with $ V'\coloneqq \{v,v'\mid v\in V\}, E' \coloneqq E\cup \{ \{v,v'\} \mid v\in V\}$. Trivially, this is a polynomial-time reduction, as $G'$ has $2\cdot \vert V\vert$ many vertices and $\vert V\vert + \vert E\vert$ many edges. This leaves to show that there is a vertex cover $C\subseteq V$ on $G$ with $\vert C \vert \leq k$ \iffl there is a Roman vertex cover $(R_1,R_2)$ on $G'$ with $\weightfun(R_1,R_2) \leq  k + \vert V \vert$.
    
    Let $C \subseteq V$ be a vertex cover on $G$ with $\vert C \vert \leq  k$. Then define $(R_1, C)$ with $R_1\coloneqq \{\{v,v'\} \mid v\notin C\}$. Since $C$ is a vertex cover on~$G$, for each $e\in E$, $C \cap e$ is not empty. For $v\in C$, $v\in \{v,v'\} \cap C$ and for $v \notin C$, $\{v,v'\} \in R_1$. Therefore, $(R_1, C)$ is Roman vertex cover with $\weightfun(R_1, C) = \vert V \setminus C \vert + 2\cdot (\vert C \vert) \leq  k + \vert V \vert$. 

    Let $(R_1,R_2)$ be Roman vertex cover on $G'$ with $\weightfun(R_1,R_2)\leq k +\vert V \vert$. If there is a $v\in V$ with $v'\in R_2$, then $(R_1\cup \{\{v,v'\}\}, R_2 \setminus \{v'\})$ is a Roman vertex cover (as $v'$ is only in the edge $\{v,v'\}$) with a smaller weight. Therefore, we can assume $R_2 \cap \{v'\mid v\in V\}=\emptyset$. To cover $\{v,v'\}$ for each $v\in V$, this edge has to be in $R_1$ or $v\in R_2$. Furthermore, we can assume that both cases cannot hold at same time. Otherwise, $(R_1\setminus \{\{v,v'\}\}, R_2)$ would be a Roman vertex cover with smaller weight.   Let $\{v,u\}\in R_1\cap E$. If $v\in R_2$ or $u\in R_2$, then  $(R_1\setminus \{\{v,u\}\}, R_2)$ would be a Roman vertex cover with smaller weight. Thus, we can assume $v,u\notin R_2$. Hence, $\{v,v'\} \in R_1$. The tuple $(R_1',R_2')=(R_1\setminus \{\{v,u\},\{v,v'\}\}, R_2\cup \{v\})$ is also a Roman vertex cover, as each edge besides $\{v,u\},\{v,v'\}$ are covered in the same way as by $(R_1,R_2)$ and the other edges are cover by $v$. Furthermore, $\weightfun(R_1,R_2) = \weightfun(R_1',R_2')$. Therefore, we can assume that $R_1\cap E= \emptyset$. This implies that $R_2\subseteq V$ is a vertex cover of $G$ and $\vert R_1\vert + \vert R_2\vert= \vert V\vert = \vert R_1\vert + 2 \cdot \vert R_2\vert - k$. Therefore, $R_2$ is a vertex cover with $\vert R_2\vert =k$.   
\end{pf}

As we have seen that \textsc{Roman Vertex Cover} is \NP-complete, we will now show an \FPT-time algorithm with respect to the solution size.

\begin{algorithm}[ht]
\caption{RVC FPT time solver}\label{alg:rvc_fpt}
\begin{algorithmic}[1]
\Procedure{RVC Solver}{$G=(V,E),k$}\newline
 \textbf{Input:} Graph $G$, $k\in \mathbb{N}$.\newline
 \textbf{Output:} Is there a Roman vertex cover $(R_1,R_2)$ with $\weightfunction{R_1,R_2}\leq k$?
 \If {$E=\emptyset$ or $\vert E\vert=k=1$ }
 \State\Return{\yes} 
 \EndIf
 \If { $k = 0 < \vert E\vert $ or $k = 1 < \vert E\vert$}
 \State \Return{\no} 
 \EndIf
 \State Let $e=\{v,u\}\in E$\label{alg:edge_choice}
\If { \textsc{RVC Solver}$((V\setminus \{v\},\{e\in E\mid v\notin e\}),k-2)$}
 \State \Return{\yes} 
 \EndIf
 \If { \textsc{RVC Solver}\textsc{RVC Solver}$((V\setminus \{u\},\{e\in E\mid u\notin e\}),k-2)$}
 \State \Return{\yes} 
 \EndIf
 \State \Return{\textsc{RVC Solver}$((V,E\setminus \{ e\}),k-1)$}

\EndProcedure
\end{algorithmic}
\end{algorithm}
\begin{theorem}
    Algorithm \ref{alg:rvc_fpt} solves \textsc{Roman Vertex Cover} in \FPT-time if parameterized by an upper bound on the weight of a solution.
\end{theorem}

\begin{pf}
We will prove this theorem by induction on $k$. For $k=0$ it is trivially true as $(\emptyset, \emptyset)$ is the only tuple with weight 0. This tuple can only be a Roman vertex cover if there is no edge to cover. The only tuples $(R_1, R_2)$ with $\weightfunction{R_1,R_2}=1$ are the tuples with $\vert R_1\vert=1$ and $R_2=\emptyset$. These tuples can only cover one edge. Hence, our case distinction in the first two if-blocks is correct. 

Assume the the Algorithm runs correct for some $k-1$ and $k\in \mathbb{N}$. 

Let $e$ be the edge in Line \ref{alg:edge_choice}. This edge has to be covered. This can be done in three ways: $v\in R_2 \text{ or } u\in R_2 \text{ or } e\in R_1 $. In the first two cases, we the weight budget reduces by two and all edges including $v$ or $u$ are covered. In the remaining case, the weight budget reduces by one to only cover~$e$.
Therefore, the algorithm runs correctly. Since this branching is a $(1,2,2)$ branching and the first two cases can be calculated in linear time, this algorithm runs in $\mathcal{O}(2^k\cdot \vert E\vert)$.
\end{pf}

\begin{remark}
A very similar algorithm can be given to \emph{enumerate} all minimal Roman vertex covers of size at most~$k$, within the same time bound of $\mathcal{O}^*(2^k)$: instead of answering \yes, the algorithm would simply output the solution. Moreover, we can re-interpret the branching as: either put $v$ into $R_2$ or not; if not: either put $u$  into $R_2$ or not. In the very last case, the edge is put into $R_1$ by \autoref{rr:empty_hyperedge}, which is also valid in this discussion. This reasoning shows that no solution would be enumerated twice. Also, one can enforce polynomial delay (on top of polynomial space) in the enumeration process by including extension tests as in \autoref{thm:rhs-enumeration}, so we get an \FPT-time polynomial-delay and polynomial-space algorithm for enumerating all minimal Roman vertex covers of size at most~$k$. 
\autoref{thm:rhs-tight-example} can be re-interpreted as a collection of $P_2$, i.e., isolated edges, in this context. This example hence shows that in general, we can face instances where we need to enumeration time $\mathcal{O}^*(2^k)$.  

All this should be compared to what is known about parameterized enumeration for minimal vertex covers, as discussed in \cite{Dam2006,Fer02a}. Interestingly, the bad enumeration-time examples coincide with our Roman setting. However, the minimal vertex covers enumeration is worse that the Roman case with respect to polynomial delay, as the extension question related to \textsc{Vertex Cover} is \NP-complete, even in fairly restricted graph classes, see~\cite{CasFKMS2019} (a question mentioned first by Damaschke~\cite{Dam2006}).
\end{remark}

\begin{remark}
Both \textsc{Roman Domination} and \textsc{Roman Vertex Cover} could be motivated in a less martial way by claiming that it is less demanding to be `responsible' for a single vertex or a single edge, respectively, so that the corresponding price one has to pay is less than for a vertex that is `responsible' for more than one vertex or for more than one edge. If one neglects this difference, i.e., all prices are the same, then we move from \textsc{Roman Domination} to \textsc{Dominating Set} and from \textsc{Roman Vertex Cover} to what Damaschke~\cite{Dam2009a} called \textsc{Vertex Cover with Missed Edges}.
However, the latter problem is `uninteresting' in the sense that as an optimization problem with one parameter, it clearly boils down to \textsc{Vertex Cover}. Therefore, Damaschke considers the 2-parameter variant in which one can delete at most~$m$ edges to produce a graph whose edges can be covered by at most~$k$ vertices. This approach then really resembles  \textsc{Roman Vertex Cover}.
\end{remark}

\begin{remark}
With the idea of modeling costs instead of counting armies, but still with a kind of defense scenario in the background, one could also think of a different interpretation of a function $f:V\to\{0,1,2\}$ for a graph $G=(V,E)$, namely, a vertex $x\in V$ can be defended at a cost of~$1$ if there is some vertex $y\in N[x]$ with $f(y)=1$ and it can be defended at a cost of~$2$ if there is some vertex $y\in N[N[x]]$ with $f(y)=2$, because it is simply more costly to move to a location that is farther away. Let us call $f:V\to\{0,1,2\}$ a \emph{defense function} if every vertex of the graph is defended. Now, concerning the notion of minimality, only the pointwise extension of the ordering $0<1<2$ makes sense. Now, one can ask to decide if, given a graph $G=(V,E)$ and some function~$f:V\to\{0,1,2\}$, if there exists some minimal defense function $g:V\to\{0,1,2\}$ with $f\leq g$, or one can also ask to enumerate all minimal defense functions. Both questions can be answered efficiently (i.e., with  with polynomial time or with polynomial delay, respectively), by connecting these questions to Roman Hitting Set. 

First, observe a simple reduction rule: If $G$ contains true twins $u,v$ with $N[u]=N[v]$, then we can remove one of them.
Namely, if $u$ is dominated with weight~1, then $v$ is dominated with weight~1, as well.
If $u$ is dominated by some vertex~$y$ at distance~2, then there is some vertex~$x\in N(u)\cap N(y)$. Now, $x\neq v$ follows, as if $y\in N(x)$, then $y\in N(u)$, as $u$ and $v$ are true twins, contradicting our assumption that the distance between $u$ and $y$ is two. Hence, $v$ is dominated by~$y$ at distance~2. Similar arguments apply for discussing what happens if some vertex is dominated by, say, $v$ at distance~1 or~2. 
Also then, $u$ can take over the role of~$v$.
Therefore, we can assume in the following that $G$ does not contain true twins.

Because $G$ has no true twins, we can index all sets $N[N[x]]$ by $N[x]$, as we have a bijection~$N$ between $V$ and $I\coloneqq\{N[x]\mid x\in V\}$. Hence, we arrive at the hypergraph $H=(V,(N[i])_{i\in I})$. Now, observe that a rhs $(R_1,R_2)$ on~$H$ corresponds to a defense function~$f$ of~$G$ with $\omega(f)=\omega(R_1,R_2)$.

A less martial interpretation of a defense function would be the design of the coverage of a country with firefighter stations, where you have two types of such stations: one with and one without helicopters. The one without helicopters is cheaper (cost~1) than the one with helicopters (cost~2). But with helicopters, you can also reach places ``at distance~2'' in our model graph. Here, one could even see another motivation why it is interesting to list all minimal solutions: maybe, after you have come up with this model, other restrictions and conditions appear, for instance, because not every type of accident can be fought by cars, you like to see a helicopter at distance at most~4 for any vertex. Such additional constraints can be filtered by looking at all minimal defense functions. If some fire stations with helicopters are already built, we face another type of extension problem.
\end{remark}

The previous remark also shows that the topics studied in this paper are not only of purely theoretical interest, but there are a number of scenarios where they could apply. Also, it proves that the hypergraph models introduced in this paper can be useful for modeling concrete situations that might arise in practical settings.

\section{Conclusions}

We have generalized the notion of Roman domination in two ways towards hypergraphs. While \textsc{Roman Hitting Set} is a problem that behaves quite like \textsc{Roman Domination}, also having a polynomial-time decidable extension version and hence a polynomial-delay enumeration algorithm, for \textsc{Roman Hitting Function}, the crucial property that maintains these nice properties is the surjectivity of the correspondence function. This can be seen as a technical answer to our question what causes Roman domination to behave different from classical domination with respect to polynomial-delay enumerability. When the correspondence is not surjective, \textsc{Roman Hitting Function} rather behaves like \textsc{Dominating Set}; in particular, its extension problem is $\W{3}$-complete when parameterized by the given pre-solution's weight, and we observe in this case that all minimal solutions cannot be enumerated assuming the \emph{Hitting Set Transversal Thesis} (HSTT).


The main open problems in the context of this paper 
are the following ones:
\begin{itemize}
\item Can we base HSTT on other, better known or classical computational complexity assumptions? This might be related to another natural question:
\item How tight is the (in)feasibility of enumeration with polynomial delay linked to the (in)feasibility (in the sense of $\PTIME$ vs. $\NP$) of a related extension problem? In the line of the studies in this paper, these links were pretty tight. But in general, only one direction is clear: if extensibility can be decided in polynomial time, then enumeration is possible with polynomial delay. \footnote{More general questions relating enumeration complexity and classical complexity were also raised (and partially answered) in \cite{CapStr2019,CreKPSV2019,Mar2013a,Str2019}.}
\item We also do not know if the polynomial-delay enumerability questions that we discussed are really equivalent to the polynomial-delay enumerability of minimal hitting sets.
    \item A concrete technical problem might be to close the gap between the approximation algorithms and the inapproximability results for all Roman optimization problems described in this paper.
    \item We mentioned in the introduction that \textsc{Roman Domination} is in \FPT, when parameterized in a dual way, meaning, in this case, by $n-k$, where $n$ is the number of vertices of the graph and $k$ is an upper-bound on the weight of the Roman domination function. It might be interesting to have similar results for the two generalizations of Roman domination introduced in this paper. However, now it is not very clear what the `dual' of the standard parameterization should be.
\end{itemize}

We are currently looking for non-trivial graph-classes where bounded-\textsc{Extension Roman Domination} is 
solvable in polynomial time.

\bibliographystyle{splncs04}
\bibliography{ab,hen}

\end{document}